\documentclass[11pt]{article}
\usepackage{fix-cm}
\usepackage[utf8]{inputenc}
\usepackage{rotating}
\usepackage{multirow}
\usepackage{color}
\usepackage[english]{babel}
\usepackage{blindtext}
\usepackage{hyperref}
\usepackage{geometry}
\usepackage{rotating}
\usepackage{lscape}

\usepackage{subcaption,graphicx}
\usepackage{longtable}

\usepackage{float}

\def\offlinecapacityFSP{1.9}
\def\onlinecapacityFSP{1.1}
\def\totalcapacityFSP{2.2}

\def\onlinecapacityMSV{1.9}
\def\totalcapacityMSV{4.3}

\def\fractionexternalFSP{36}
\def\fractiontierzFSP{64}

\def\fractionexternalMSV{19}
\def\fractiontierzMSV{81}

\def\fractiontierzMSVs{63}

\begin{document}
\begin{titlepage}

\title{Conceptual Design Report for FAIR Computing}
\author{Mohammad~Al-Turany, \and Volker~Friese, \and Thorsten~Kollegger, \and Bastian~Loeher,  \and Jochen~Markert, \and \underline{Johan~Messchendorp}, \and Andrew~Mistry, \and Thomas~Neff, \and 
Adrian~Oeftiger, \and Michael~Papenbrock, \and Stephane~Pietri, \and Shahab~Sanjari, \and Tobias~Stockmanns}

\maketitle

\begin{abstract}

This Conceptual Design Report (CDR) presents the plans of the computing infrastructure for research at FAIR, Darmstadt, Germany. It presents the computing requirements of the various research groups, the policies for the computing and storage infrastructure, the foreseen FAIR computing model including the open data, software and services policies and architecture for the periods starting in 2028 with the "first science (plus)" phase to the modularized start version of FAIR. The overall ambition is to create a federated and centrally-orchestrated infrastructure serving the large diversity of the research lines present with sufficient scalability and flexibility to cope with future data challenges that will be present at FAIR.
\end{abstract}
\end{titlepage}

\tableofcontents
\newpage
\section{Executive summary}




\subsubsection*{FAIR research and the role of scientific computing}

The Facility for Antiproton and Ion Research (FAIR) will be the next-generation accelerator-driven infrastructure serving a large community of researchers from a variety of disciplines
with an internationally-oriented character. The broad spectrum in physics communities, from atomic to particle physics, actively 
involved at FAIR, leads to a challenging diversity in the computational activities that need to be performed to reach the ambitious physics goals.

The research activities that are foreseen at FAIR have a large spectrum of requirements as a consequence of the differences in 
computational complexity. The monolithic CBM and PANDA experiments will face data rates $\sim$500~GB/s with online data acquisitions that will run in streaming mode without a conventional hardware trigger, while
APPA and NUSTAR will operate at lower data-production rates, but with modular setups and various sub-communities imposing challenges with respect
to the data management and organisation. Moreover, for the preparation of the experiments and interpretation of the harvested data, 
large-scale simulation studies are required together with forefront model-based or, even, first-principle-based, calculations supported by the theory department.
In this respect, the computing infrastructure for FAIR needs to act as a high-throughput (HTC) as well as a high-performance 
computing center (HPC). 


\subsubsection*{The next-generation of scientific computing for FAIR research}

The computing requirements at FAIR need to be placed in the context of the developments that are taking place world-wide. 
Most noticeably are the recent paradigm shifts on the hard- and software side in processing data and extracting physics information.
Besides the standard x86-based computers based on CISC architectures (Intel, AMD), general-purpose computing on graphics processing
units (GPGPUs) have become popular to process the massive streaming data from the experiment, imposing as well challenges in deploying
algorithms that were originally developed on x86 architectures. RISC based architectures like ARM and RISC-V have delivered reduced energy consumption in consumer devices and promise new perspectives for the research community as well. From the software and algorithm perspectives, 
promising activities are ongoing in the direction of exploiting machine learning and artificial intelligence to address the complexity of the data, extracting more
valuable information, expanding the physics reach, and to develop new methodologies in boosting up the computational performances. Although,
these developments are not yet in full production mode, one may expect that it will become a common standard within the upcoming decade
when FAIR will become operational. Besides the many new advances in hard- and software, the open-science cultural aspects
and overall policies for the design of the computing infrastructure at FAIR are of utmost importance as well. FAIR has unambiguously
committed itself to follow-up the commonly accepted F.A.I.R. principles, requiring its data and services to 
be Findable, Accessible, Interoperable and Reusable. This imposes new challenges in data management, data processing, and access, but 
at the same time provides opportunities in the scientific interest of the research facility, {\it e.g.} long-term knowledge preservation, 
synergetic data analysis, enabling resource sharing, promoting transparency, and highlighting the research output from FAIR to the physics community and society in general. etc..  


FAIR scientific computing will use a centrally orchestrated infrastructure build upon the presently available Green-IT Cube (or in short GreenCube) as Tier0. The GreenCube is a successfully operating computing center located at the GSI campus and internationally recognized by its ecofriendly design, thereby an excellent basis for future FAIR computing.  
The objective is to optimise the sharing of the computing resources among the various research pillars and between the online and offline computations,
thereby minimizing the costs and power consumption while respecting and incorporating the needs of the various research groups.
Conceptually, no dedicated online clusters for experiments are foreseen, but rather shared in a central system.
Additional large-scale computing infrastructures associated to the international partners of FAIR need to be connected to Tier0 in a so-called
federated mode of operation. Their local autonomy will be respected with an agreed mixture of common policies that allow a transparent 
exchange of data and services following the F.A.I.R. principle.


\subsubsection*{Computing requirements for FAIR}

This document describes the conceptual design of the foreseen FAIR scientific computing infrastructure. It provides a perspective view 
of FAIR computing supporting the research divisions starting from the time when FAIR becomes operational with the SIS100 accelerator,
``First Science(+)'' (FS+) towards the completion of the modularized start version ``MSVc''. The presented plan is primarily based on requirements and computing estimates of the various FAIR research lines. These requirements and estimates are preliminary, reflect the plans of today, and subjective to changes in the course of time. To track and verify the resource needs and utilization of the FAIR research lines, we provide biannual updates steered by the FAIR computing coordinator and monitored by an external committee (``Compute Resource Assessment Committee''). 

To obtain a transparent overview as the basis to the proposed
computing model, we classified the various types of computational, storage, and bandwidth requirements according to a common template. 
In the case of CBM and PANDA, the estimates were primarily based on the results of detailed Monte Carlo simulations.
We extracted research-line specific information from ongoing computational activities at the Virgo HPC cluster
in the context of ``FAIR Phase Zero'' projects conducted at GSI and used this knowledge as a reference to extrapolate towards the computing 
and storage needs for FAIR. Moreover, with our participation in open-science projects, such as PUNCH4NFDI and ESCAPE, 
and as observer in EOSC, we are not only well informed about the respective developments, but also able to steer
towards the needs of a landmarked European research facility such as FAIR. 
Similarly, to ensure that the design of the computing infrastructure for FAIR 
will be aligned with the common plans of other partner facilities and research activities, some of the authors of this report are actively taken part in the computing-related discussions and write-ups for the long-range plan of NuPECC and the position paper of the joint 
ECFA-NuPECC-ApPEC (JENA) community. 


The present GreenCube HPC at GSI has about 1.1~MHEPSpec06~\footnote{To get an impression of this benchmark performance metric: 1 physical core of Intel E5-2680v4@2.4GHz processor corresponds to $\sim$22~HEPSpec06. The following link gives an overview of the benchmarks for other hardware architectures: \href{https://www.gridpp.ac.uk/wiki/HEPSPEC06}{https://www.gridpp.ac.uk/wiki/HEPSPEC06}.} of compute capacity ($\sim$50~kcores of AMD/Intel-based computers and $\sim$500 AMD GPUs) and $\sim$30~PBytes of storage capacity. An analysis of the past two years (1/2021-3/2023) of computing usage at this center shows that a large fraction of compute ($\sim$0.5~MHEPSpec06) and storage ($\sim$10~PB) resources is used for the FAIR Phase Zero program with significant contributions of all the research pillars. The remaining resources are primarily dedicated to GSI activities, {\it e.g.} Alice-Tier2, which are not part of FAIR and, therefore, not accounted for in this report. 

\begin{figure}[h]
\begin{center}
\vspace*{-0.2cm}
\includegraphics[width=1\textwidth]{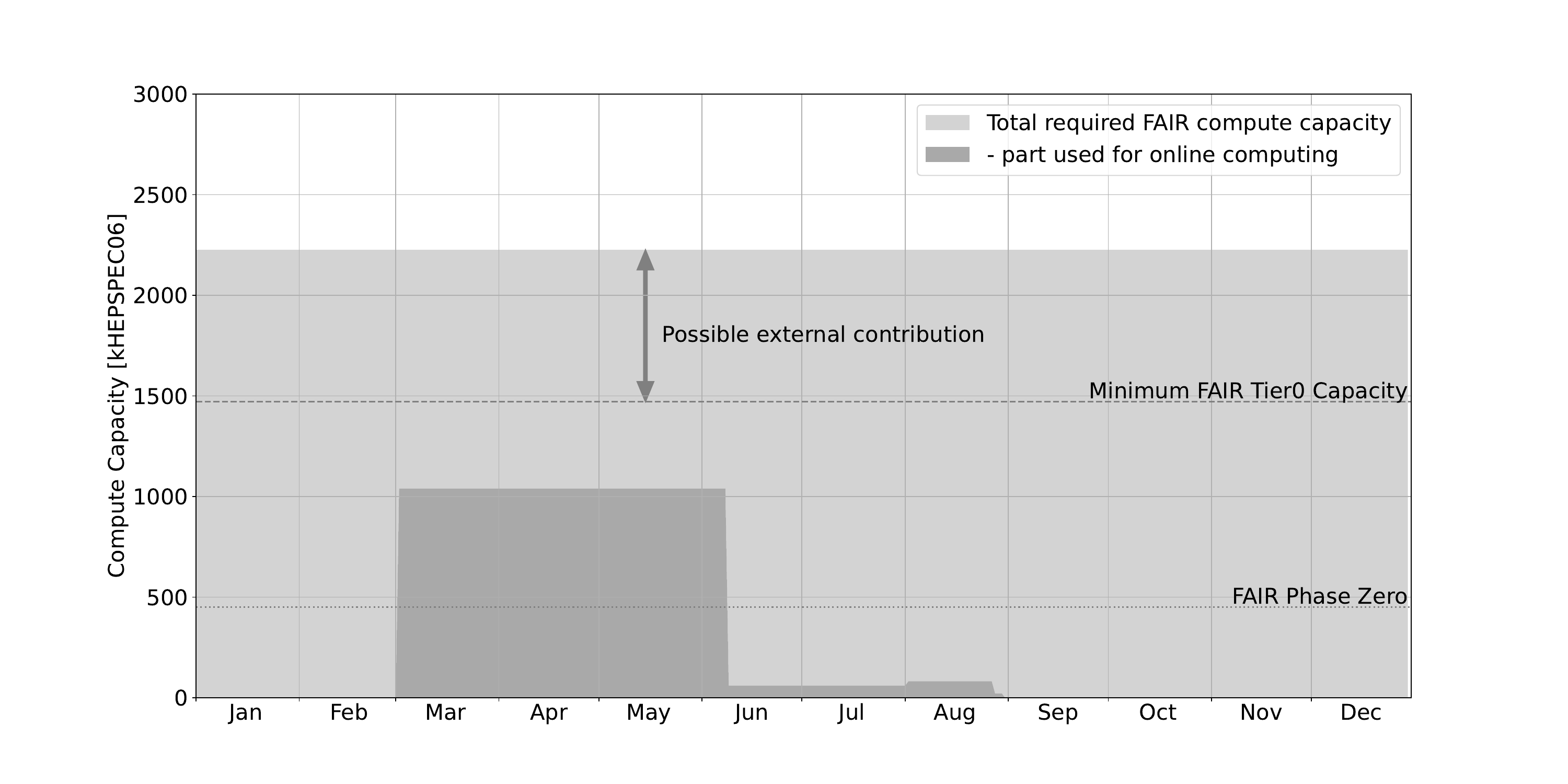} 
\vspace*{-1cm}
\caption{Sketch of the required compute capacity for a nominal FS+ year. The light-grey area depicts the total required shared compute capacity for online and offline computations whereby the online part is averaged out over the year. The dark-grey area indicated the required online computing capacity taking into account CBM (100 days), NUSTAR (180 days), HADES (30 days), and APPA (180 days). The dotted line represents the presently used compute capacity at the GreenCube for FAIR Phase Zero activities. The dashed line depicts the minimum required capacity at FAIR Tier0 which includes the maximum online compute capacity plus data intensive tasks.}
\label{fig:pretty_plot_fsp}
\end{center}
\end{figure}

With SIS100 becoming operational in 2028, the centrally-organized compute facility will need to provide about \totalcapacityFSP~MHEPSpec06 of resources to accommodate both online and offline computations in the FS+ scenario. The shared compute resources solely dedicated for data taking by itself will already add up to a capacity requirement of \onlinecapacityFSP~MHEPSpec06. For the offline computational activities - primarily for the production of derived data, higher-level data analyses, Monte Carlo studies, and theoretical calculations - we estimated a continuously available compute capacity of about \offlinecapacityFSP~MHEPSpec06 whereby all FAIR pillars require a significant share. Extrapolating towards the compute needs for MSVc, we expect an increase of less than a factor of two in capacity with respect to FS+. More precisely, we estimated a total shared compute capacity including both online and offline activities of about \totalcapacityMSV~MHEPSpec06 for which \onlinecapacityMSV~MHEPSpec06 is instantly required for online data taking assuming that CBM and PANDA can operate in parallel. Since the experiments will not continuously request the online capacity to be available, it is foreseen in our model to effectively share online with offline resources among all research lines, thereby optimising for cost and energy efficiency. To realise such model, it is of utmost importance to provide a strong local compute center at the FAIR campus hosting a significant fraction of the compute and storage resources for the FAIR communities.  

\subsubsection*{Compute and storage capacities during FAIR operations}

Figure~\ref{fig:pretty_plot_fsp} illustrates an overview of the required compute capacity for a nominal FS+ year. It presents both the total required compute capacity and the fraction required for online computing. We estimated the minimum compute capacity that would be required at Tier0, {\it i.e.} at the local infrastructure at the FAIR campus. For this, the maximum online compute requirements are supplemented with a minimum required fraction of additional computational needs that strongly depends on processing large volumes of raw or partly-derived data harvested by the experiments at FAIR. The remaining compute resources ($\sim$\fractionexternalFSP\% of the total), indicated by ``possible external contribution'', can be located at another computing center outside the FAIR campus. Those tasks include computations that do not depend on the availability of the data and allow for an asynchronous data transfer, such as Monte Carlo simulations and theoretical calculations that support the experiments. For MSVc, the fraction of possible external contributions reduces to about \fractionexternalMSV\% of the total required capacity.

\begin{figure}[h]
\begin{center}
\vspace*{-0.2cm}
\includegraphics[width=0.9\textwidth]{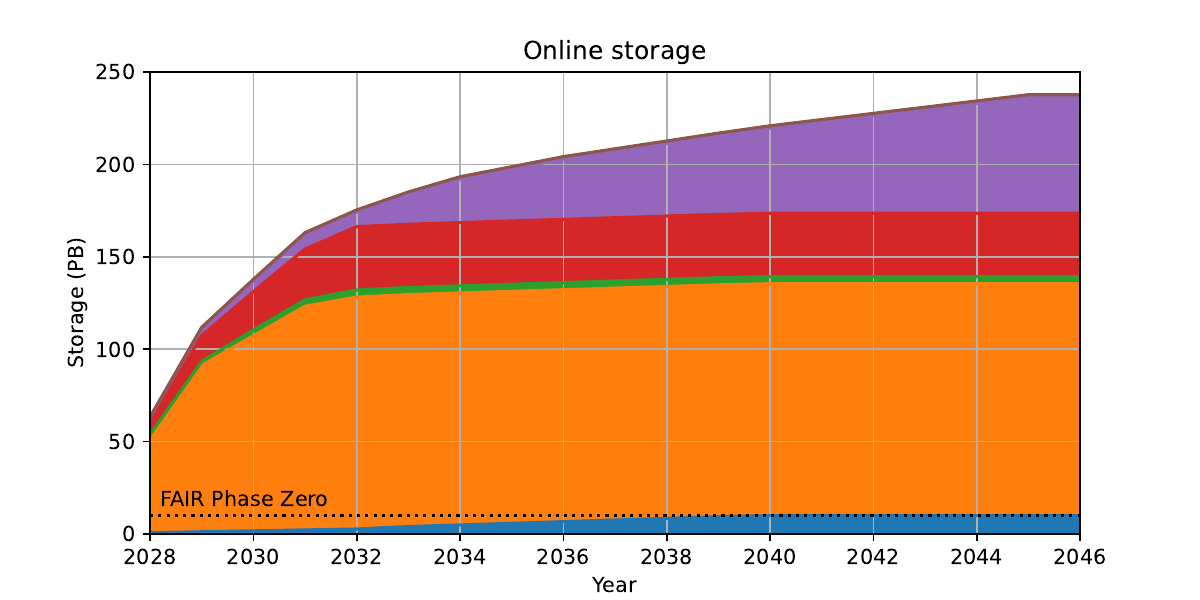}\\
\vspace*{-0.7cm}
\includegraphics[width=0.9\textwidth]{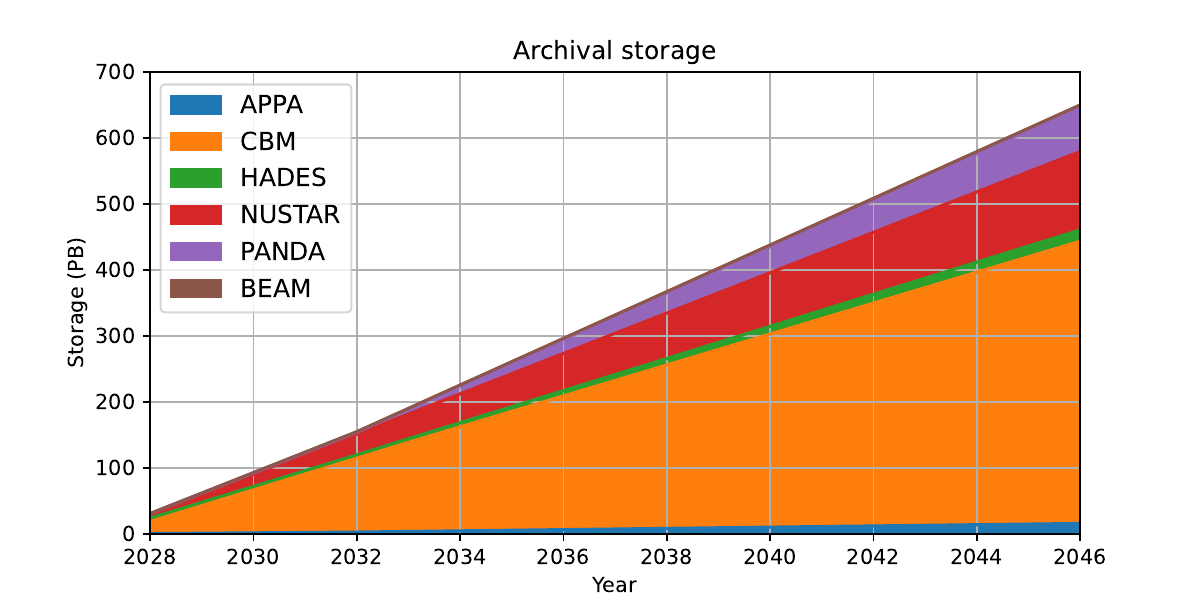}
\vspace*{-0.3cm}
\caption{The required amount of storage as a function of year with FS+ starting in 2028 and MSVc in 2032. The top panel depicts the requested disk space for fast access, whereas the bottom panel presents the needed long-term storage space (archive). The contributions of the various research lines are indicated by different colors. The dashed line shows the storage used on the Lustre filesystem for FAIR Phase Zero activities. Copies of the raw data at other FAIR facilities are not included, but imposed a requirement established by an external audit}.
\label{fig:storage}
\end{center}
\end{figure}

The immense compute requirements of the research lines go hand-in-hand with a significant increase of data volumes compared for the current needs at GSI. This is driven by the high data rates of the large-scale experiments producing up to $\sim$25~PB/year of raw data in FS+. Moreover, the amount of associated Monte Carlo and, other, simulated data that need to be produced to determine efficiencies, to control systematic uncertainties, and to interpret the harvested data, are of comparable order of magnitude ($\sim$15~PB/year). The top panel in Fig.~\ref{fig:storage} summarizes the required disk storage capacity as a function of year assuming FS+ starting in 2028 and MSVc in 2032 and taking into account the requested amount of time the data need to be available for fast access. The contributions of each FAIR pillar are indicated by the various filled and colored areas. For FS+, CBM and NUSTAR require the dominant fraction of storage capacity. PANDA will become an important disk user starting in 2032 assuming MSVc. The requested storage capacity saturates to $\sim$160 (210)~PB for FS+ (MSVc). The bottom panel in Fig.~\ref{fig:storage} depicts the permanently stored archived data (mostly raw data, but also partly including derived data) as function of year, which approximately shows a linear increase with a slope of $\sim$30~PB/year. 


\subsubsection*{FAIR data and F.A.I.R. principles}

FAIR has the ambition to serve an international research community with a broad spectrum of disciplines, and to that end commits to supporting its researchers in adopting Open Science practices. The physics will be extracted by researchers of this community from the data harvested and processed from the various experiments and derived from associated simulations and calculations. Most of the data will be centrally produced. To optimize and ease the link between the user and compute/data and to ensure long-term data and knowledge preservation, we follow the Findable Accessible Interoperable Reusable (F.A.I.R.) principles. The implementation of these principles will be tuned towards the basic needs. This includes the creation of a data management plan for each research project.  In general, the open-science policies are aligned with the presently formulated policies of GSI. To summarize and to be specific towards the conceptual design of the FAIR research IT infrastructure, the following aspects are foreseen to be implemented for FAIR:

\begin{itemize}
    \item{{\bf F}indable:} Centrally orchestrated storage and access of data essential to enable the (meta)data and software to become findable. Consistent usage of Persistent IDentifiers (PID) for example Digital Object Identifiers (DOI). Expand the GATE environment available at GSI. The open-source scientific software and service repository (OSSR), a successful deliverable of ESCAPE project with a large GSI contribution, will be promoted and supported.
    \item{{\bf A}ccessible:} Data and software produced and dedicated for FAIR communities and publications will be centrally stored. The usage of a common and “user-friendly” interface to store and retrieve data will be promoted. 
    We foresee a federated data infrastructure based on open-source protocols commonly used in HEP and Nuclear Physics.
    \item{{\bf I}nteroperable:} Participate in community-wide open-science initiatives, projects \& programs on institutional, national, and European levels. Follow-up the developments of a ``Datalake'' concept as introduced and pioneered by ESCAPE at the European level and/or within ErUM-Data at the German national level. The requirement is that it should enable interoperability among similar research disciplines as FAIR with minimum overhead. To what extent the data obtained at FAIR has to be interoperable and with whom depends on the specific needs of each (sub)collaboration and its policy falls outside the scope of this report. 
    \item{{\bf R}eusable:} 
    Data and software available under a suitable open licence. Ensure that all data and software are accompanied by comprehensive and structured documentation.  An important element is to establish domain-specific common metadata vocabularies and to employ collaborative platforms to facilitate reusable data and software.
\end{itemize}


\noindent FAIR aims to position itself within the open-science community via ESCAPE/EOSC and within the German-national PUNCH4NFDI activities. Particularly, our interest is to invest in science-cluster specific activities driven by the direct and common needs of our pillars and research facility. It will be left up to the specific needs of the (sub)collaborations to develop a finer granularity in their open-science policies.


\subsubsection*{Role of FAIR-research IT}

One may tentatively divide up the user community at FAIR for scientific IT in two categories. The first category involves large-scale monolithic experiments, {\it e.g.} CBM and PANDA, that rely heavily on online data processing and make use of extensive offline computations. Their computing activities are centrally organized according to the regulations of the respective collaboration with well defined standards for the data processing and analysis schemes with corresponding services and software. The second category relates to a broad spectrum of various types of smaller-scale experiments with specific use cases and a diverse user community, such as APPA, for a large part NUSTAR, but also researchers at the theory department supporting the experimental activities. These types of research activities require significantly less intensive computations during the online data processing, but do often require HPC for their simulations or calculations. Their computing activities exploit a large variety of specific software and services for which the standards are defined by smaller sub-collaborations and communities and, therefore, less top-down orchestrated. The challenges for FAIR IT are, on one hand, to provide a hard- and software infrastructure that addresses the requirements of both types of categories, and, on the other hand, to ensure that the computing resources are being used cost efficiently, avoiding as much as possible idle machines and minimizing the overhead in providing specific services.  
The responsibilities and role of the FAIR IT department would primarily concentrate towards providing a common infrastructure starting ``at the end of the fibres coming from the experiments''. Part of this is to define and setup interfaces between the experiment/user and compute/storage and to promote as much as possible common interfaces and software frameworks. To optimise the usage of the compute resources, {\it i.e.} minimizing the number of ``idle'' machines, while taking into account the community-specific use cases, we aim to strongly promote the usage of virtualization technologies, such as containerized approaches, virtual machines, and cloud services, discouraging the deployment of private computing clusters. Commonly-used services and frameworks, {\it e.g.} Fairroot, FairMQ, CDash, Gitlab, will (remain) supported and maintained by the FAIR research IT department. To optimize the synergy between the research communities and FAIR IT, the local scientific IT team will be expanded for FAIR operations. This team will partly be integrated in the research groups and will form a network or interface towards expert communities outside the FAIR campus, such as ROOT and GEANT4 collaborations. The research groups are responsible for the development of research-specific software and should make use of the common infrastructure. Moreover, they are accountable for the management of their data. 


\subsubsection*{R\&D activities}

The R\&D aspects within FAIR computing will be devoted to investigate new methodologies to reduce the costs and/or to enhance the physics output from the harvested data. A promising development in this context is the application of machine learning (ML) and artificial intelligence (AI) techniques. Particularly, the usage of ML algorithms to boost the performance of simulations, the ambitions to develop smart experiment and accelerator control to optimise the beam time usage and streaming readout processes, and the various techniques to improve tracking and particle identification may become opportune to incorporate in the computing models. At present, these developments are in a pioneering stage and not in full production mode yet. Therefore, it is difficult to predict the performance gain one may achieve with it. On the hardware side, one expects a significant gain in performance by further exploiting the deployment of data processing algorithms on alternative architectures, such as accelerator cards (GPU), and/or more data-acquisition specific, field-programmable gate arrays (FPGA). Exploratory activities in this direction are ongoing at GSI by various collaborations, such as by CBM in the context of the ongoing miniCBM experiment. 
The overall objective is to ensure that a large fraction of the computing resources for FAIR research will be based on hardware architectures meant for generic usage to be able to serve its diverse communities.
Further explorations within open-science driven and digital-innovative projects and programs on European and national levels, {\it} e.g. ESCAPE, PUNCH4NFDI, ErUM-Data, may enhance the physics reach in some of the cases. The focus will be towards supporting thematic clusters to make these efforts most beneficial for our purpose.   


\subsubsection*{The computing model for FAIR in a nutshell}

The backbone of the computing model is an effective resource sharing at FAIR Tier0, {\it i.e.} exploiting the available GreenCube infrastructure, accounting for all the online data processing and for a very large part for the offline computations. The online and offline activities will have a huge overlap in resources tending to conceptually merge eventually, both from a hardware perspective, but also software wise. Moreover, the various research pillars at FAIR will make use of the same compute center, thereby sharing the available resources in an effective manner. The key elements for a harmonized successful operation are

\begin{enumerate}
    \item Centralized storage of data and its management. This model is most suited to incorporate the F.A.I.R. principle, to introduce common interfaces, and to minimize the operational overhead.
    \item Containerized approaches and other virtualization methods for flexible compute operations serving diverse community.
    \item Data access using http protocol and AAI using widely accepted standards and token-based.
    \item Usage of hardware architectures largely dedicated for generic use combined with a significant fraction using technologies suited for online data processing, such as accelerator cards.  
\end{enumerate}

\noindent 
For the resource sharing and data distribution with other centers, we foresee to integrate a small number of large centers using a federated model.
Successful examples in this direction are the ongoing federation among the computing clusters at several universities in Frankfurt, Mainz, and the GreenCube at GSI/FAIR exploiting the Teralink network. A grid-like distributed computing infrastructure with many sites is considered not opportune given the composition of the collaborations involved at FAIR.

\newpage
\section{Introduction}
\label{preface}

\subsection{The Facility for Antiproton and Ion Research - FAIR}

The primary objective of this conceptual design report (CDR) is to provide an abstract description of the computing infrastructure that will enable the various research activities that are foreseen at the Facility for Antiprotons and Ion Research (FAIR) at Darmstadt, Germany. FAIR is an ESFRI landmark and a top priority for the European Nuclear Physics Community. FAIR will be the successor of the existing and operational GSI facility and it provides a unique laboratory serving a broad and international ($\sim$50 countries) community bringing together scientists ($\sim$3000) from the fields of applied, atomic, nuclear, and hadron physics. The research is driven by heavy-ion and proton beams, the production of secondary beams composed of rare isotopes or antiprotons, the storage rings dedicated to precision studies, and the diversity of monolithic and modular experimental setups. The foreseen high-intense beams and high-acceptance and granular experimental setups will give rise to computational complexities in various dimensions with impressive data rates ($\sim$TB/s), intelligent high-throughput data processing, and large storage requirements ($\sim$35~PB/year). Moreover, the volume and diversity in research communities with their international character, that go beyond the borders of Europe, will impose further challenges in the context of data preservation, access and management. The conceptual design of the computing infrastructure presented in this document is driven to address these challenges within an acceptable financial and managerial boundary conditions.


The experimental scientific program of FAIR embeds four pillars, namely 
\begin{itemize}
    \item Atomic, Plasma Physics and Applications (APPA) - an umbrella for several collaborations working on atomic physics, plasma and applied sciences. 
    \item Compressed Baryonic Matter (CBM) - a facility that exploits heavy-ion collisions and a modular fixed-target experiment to explore strongly interacting matter under extreme conditions, particularly at high nuclear densities.
    \item Nuclear Structure, Astrophysics and Reactions (NUSTAR) - a collaboration devoted to the study of nuclear structure using beams of radioactive species and various experimental setups, and within several sub-collaborations. 
    \item antiProton Annihilations at Darmstadt (PANDA) - a comprehensive physics program driven by antiprotons and a versatile detector installed at the high-energy storage ring.
\end{itemize}

Besides these four experimental pillars, the scientific ambitions of FAIR will rely on a local theory support and the beam physics department. Moreover, CBM is strongly linked to the existing HADES experiment that is presently installed in one of the caves connected to SIS18 at GSI. It is foreseen to move HADES at some stage to the same experimental area as CBM, thereby making use of the higher heavy-ion and proton beam energies available with SIS100.     


\begin{figure}[h]
\begin{center}
\includegraphics[width=1\textwidth]{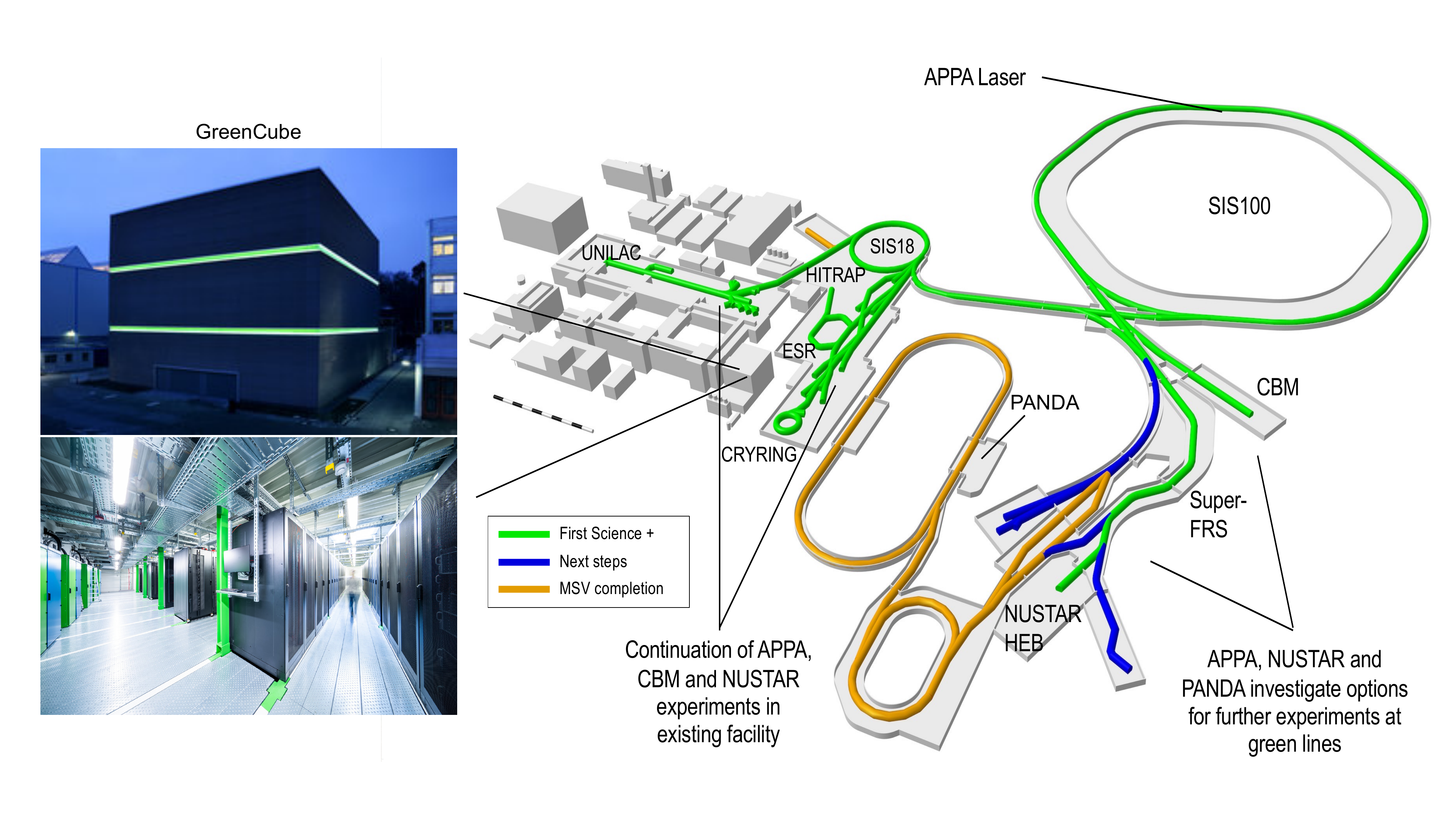} 
\vspace*{-1cm}
\caption{Schematic view of FAIR with its beam lines and experimental areas. The color code indicates the various stages of the construction relevant for this CDR. The left-hand side shows images of the existing GreenCube as seen from the outside (top) and inside (bottom).  }
\label{fig:fair_fsp_to_msv}
\end{center}
\end{figure}

\subsection{The timeline of FAIR}

During the writing of this CDR, FAIR is in its construction phase with the ambition to start its first experiments with the SIS100 accelerator in the second half of 2028, referred to as ``First Science (FS)". This will be the kick-off phase for parts of the NUSTAR and APPA physics programs. For NUSTAR the superconducting fragment separator (S-FRS) at the high-energy beam line (HEB) will become available while APPA can exploit laser-driven experiments in the SIS100 ring. The ambition is to enable as well the beam line and experimental equipments in the high-energy nuclear collision cave which would provide beams for the CBM collaboration and, potentially, for parts of the PANDA and APPA communities. This option is referred to ``First Science Plus (FS+)". The objective of FAIR is to complete the facility with the so-called Modularized Start Version (MSV) which includes additional beam lines and caves for the NUSTAR and APPA programs and the necessary infrastructure for the antiproton program. The latter includes the accumulation and HESR storage rings, antiproton production target, and p-linac, together with the PANDA detector at HESR. The ambition is to have the MSV ready by 2032. An overview of the available FAIR facilities starting at FS+ towards MSV is shown in Fig.~\ref{fig:fair_fsp_to_msv}. Prior to FS(+) phase, the SIS18 beam at GSI is presently the basis of the so-called Phase-0 program of FAIR in which experimental equipment designed for FAIR experiments is being used in existing setups or being tested. Moreover, part of the beam lines and the S-FRS at the new FAIR side (HEB) will become available already in 2026. This allows to perform some physics runs in 2027 within the NUSTAR collaboration using the S-FRS facility with SIS18 beams. This phase is referred to as ``Early Science (ES)". This CDR considers the computing requirements in the phases FS, FS+, and MSV, and hence, starting from 2028 on-wards. 

\subsection{The base of FAIR computing - Green-IT Cube}

The left-hand side of Fig.~\ref{fig:fair_fsp_to_msv} depicts the location of the Green-IT Cube, the main data center at GSI and core of the scientific data processing, together with a few photographs taken from the outside and inside of the center. 

It was build in 2014 to support the computing requirements for FAIR. The building has a total of 6 floors. Each floor can be equipped with 128 racks, providing for a total of 768 racks in the full data center.  The data center is designed for a total connect load of 12 MW with full redundancy from the high-voltage 25 kV lines down to the individual rack power supplies.

The technical installation partitions the building into three, largely independent, domains in the building. Two adjunct floors form a common domain, e.g. floors 5 and 6. Each domain has its own redundant power connection to the 25 kV lines and a common cooling system.

One of the major design goals of the Green-IT Cube have been low operating costs, which are one one hand driven by the overhead of the data center infrastructure compared to the electric load of the IT equipment, on the other hand by the operating conditions of the servers. To tackle this the cooling system of the Green-IT Cube is based on water cooled passive backdoors at each rack, with evaporative cooling towers as backcoolers. Operational history shows a typical PUE of 1.07 (i.e. the infrastructure overhead is around 7\% of the connected IT load), while at the same time keeping the operating conditions inside the ASHRAE TC9.9 recommended envelope (desing goal was the ASHRAE TC9.9 A2 envelope). The energy monitoring systems in the Green-IT Cube follow the DIN EN 50600-4-2 standard. Significantly going beyond the requirements of the standard, a large number of sensors have been deployed throughout the building infrastructure, which are fed into a common monitoring system with the data from the compute servers to allow for a holistic view on the system state and optimizations in operation.

In line with the FAIR requirements the initial installation of the Green-IT Cube consisted of 256 racks in 4 floors, with total electrical connection power of 4 MW. In 2023, this was extended by another two floors, which added another 128 racks and 4 MW of electrical connection power. This could be easily expanded by another 128 racks, the full infrastructure has already been prepared. 

The GreenCube has won many awards in the past decade~\cite{greencube:hpc,gsi:hpc}, thereby, demonstrated to be a sustainable infrastructure designed for long-term operations. With this available and fully-functional infrastructure, particularly designed for future operations, FAIR computing will heavily make use of the GreenCube facility as well serving as a Tier0 center.

The Green-IT Cube is supplemented by a second smaller data center RZ-1 on campus. This is a traditional air cooled data center, which provides only limited power (100 kW). The main purpose is to act as an independent location for mission-critical business services. It is however also used as location of the main tape library for FAIR and supports also the scientific use case. Originally designed in the 1970s, the data center has been remodelled for FAIR operation, now meeting todays requirements, including a similar monitoring infrastructure as the Green-IT Cube.




\subsection{The objectives and challenges of FAIR computing}

It is a no-brainer that computing at FAIR is an essential, but also a challenging, enterprise to enable to perform fore-front physics. The phrase ``big-data challenge" certainly applies for this facility as well as for other (future) accelerator-driven facilities in the world. With data rates up to about TB/s, a storage requirement of about 35~PB/year, in-situ streaming data processing, an international research community, commitment to open science, etc.. A unique and additional feature at FAIR relates to the diversity of computing requirements. This is for a large part a consequence of hosting a large variety of research communities, ranging from small-scale and modular experiments to large-scale monolithic setups, all within a broad spectrum of research fields and users. The experimental activities depend strongly on the support by the local theory and accelerator groups. Their computing requirements differ from experiments as well, but will (partly) make use of the same computing infrastructure. Moreover, the integration of the data harvest by these two groups with those by the experiments may become a vital element for the data processing, analysis and interpretation. The ambition is to create an eco-friendly computing infrastructure minimizing the (operational) costs while maintaining sufficient flexibility to support the research lines and their users such to reach their physics goals, to preserve the information for a long term, and to enable an open-science culture.     


This document describes various FAIR-focused computing aspects including an overview of the specific requirements of the research lines, the foreseen conceptual computing model for FAIR, and the planned and most promising technological developments that will be followed up. More specifically, this report will address the following list of questions:

\begin{itemize}
    \item What are the compute, storage, and bandwidth requirements of the various research lines, referred to as ``hard" requirements?
    \item What are the software- and policy-driven requirements of the research lines, referred to as ``soft" requirements?
    \item How will the data and software be stored and accessed? What policies will be considered in this context?
    \item How will the FAIR (Findable, Accessible, Interoperable, Reusable) principle be met?
    \item What R\&D activities in both soft- and hardware are being foreseen to prepare for future developments and to reduce the costs, improve the efficiency and physics reach?
    \item What will be the responsibility of the central FAIR-IT department and that of the research lines and corresponding collaborations?
\end{itemize}

The timeline for FAIR is tight, beginning with the Early Science (ES) phase in 2027, followed by First Science (Plus) (FS+) tentatively in 2028, and aiming for the completion of the Modularized Start Version (MSVc) in 2032. This schedule may suggest that achieving FAIR computing on time is highly challenging. However, it is important to note that FAIR is not starting from scratch. It builds on the robust foundation of GSI, which already operates a fully functional accelerator complex, the Green IT Cube computing infrastructure, a track record of successful experiments using the SIS18 accelerator, and various collaborative efforts on a global scale. Currently, experiments are continuously running as part of FAIR Phase Zero, including HADES with complex online data processing, mini-CBM as a prototype for free-streaming data processing, and numerous smaller-scale experiments within NuSTAR and APPA.
The core operations of future FAIR experiments do not depend critically on breakthroughs in IT technology, though emerging developments may offer cost-efficiency improvements.

Given the fast developments in computing and the large uncertainties in the realization of FAIR, it should be noted that some of the aspects described in this document may only be considered tentative and intentional. Hence, in this stage of the facility we therefore focus primarily on the conceptual computing aspects and that quantified numbers should be taken as guidelines in the design of the infrastructure. This document will be followed-up by a technical design report (TDR) quantifying in detail the various aspects of the proposed computing infrastructure.

To track and verify the resource needs and utilization of the FAIR research lines, we provide biannual updates steered by the FAIR computing coordinator and monitored by an external committee (``Compute Resource Assessment Committee''). The computing coordinators of the various FAIR pillars will be responsible for requesting and providing accounting of the offside resources while the local FAIR IT will collect those for Tier0 operations. 

The content presented in this CDR has been compiled by representatives from the various research lines at GSI/FAIR, from members of the GSI-IT department, and from the research data management \& open-science activities. 
\newpage
\section{Research-line specific requirements}
\label{sec:requirements}
\subsection{Introduction}

%
%

As input to the formulation of the computing model, we evaluated the requirements of the various pillars and departments that will rely strongly on the FAIR research IT. This section gives for each research line an overview of their requirements that would be needed for computing in order to reach their physics goals. Partly, the corresponding estimates are based on experiences during FAIR Phase Zero. We therefore studied the computational activities of the respective research lines using the GSI and FAIR computing systems for the past few years. Moreover, detailed Monte Carlo (MC) studies have been performed to provide estimates on expected compute times, bandwidths, data rates and sizes. Such detailed MC studies were performed for the two experiments that will dominate the need for computational resources needed for FAIR, {\it i.e.} CBM and PANDA. 

To derive to a strategy towards a model for the FAIR research IT, we have collected the various requirements and estimates in a structured and common manner. Particularly, we concentrated on the following types of requirements:

\begin{itemize}
\item{\bf Computing:} The amount of computational resources in units of HEPSpec06 for on- and offline data processing, Monte Carlo simulations, and theoretical models and simulations.
\item{\bf Data storage:} The expected data rates to storage devices in units of TB (or PB)/year for raw, derived, and simulated data. 
\item{\bf Bandwidths:} Particularly, the expected data rates (TB/s) from the experiments towards the Tier0 compute facility (GreenCube) is of interest, including average and peak values. 
\item{\bf ``Soft'' requirements:} Besides the ``hard'' estimates indicated above, we summarized the main wishes and constraints of the various FAIR research lines on software and services aspects inspired by the topics given in Ref.~\cite{Albrecht:2019}.  
\end{itemize}

The following scenarios of the FAIR facility have been considered in defining the requirements:

\begin{itemize}
\item{\bf First Science (Plus) (FS/FS+):} We refer to Sec.~\ref{preface} for description. In the context of computing resource requirements, the additional ``+'' includes CBM. FS(+) refers to a timeline starting in 2028 and ends when MSVc becomes operational.
\item{\bf Modularized Start Version (MSVc):} Also here, we refer to Sec.~\ref{preface} for further details. Relevant to note is that the MSVc phase includes the operation of PANDA at HESR. MSVc refers to a timeline starting in 2032.
\end{itemize}

\noindent We note that these refer to working assumptions pending on decisions by the shareholders of FAIR. Hence, timelines are at present tentatively. The steps beyond FS+ require additional funding.

The research lines that have taken part in the requirement survey include all the four experiment pillars, namely APPA, CBM, HADES~\footnote{HADES is presently connected to SIS18 and the secondary pion-beam facility. It is, however, foreseen that this setup will be reinstalled in the same cave as CBM. HADES serves researchers from several communities, including heavy-ion, nuclear, and hadron physics.}, NUSTAR, and PANDA, together with FAIR-related research activities of the theory department and those connected to beam diagnostics and operations. We also note that this CDR describes the FAIR {\it research} IT and does not address administrative IT aspects (enterprise IT).

\subsection{Classification scheme}

An important aspect as input to the FAIR computing model is to distinguish between the compute resources that will be part of the experiments (``experiment clusters'') and those that will be commonly used by FAIR users and designed/operated by FAIR IT (``shared compute clusters''). Moreover, it is of essence to get an overview for the ``shared compute clusters'' how much of those resources need to be reserved specifically during data taking, {\it e.g.} specifying the online versus offline computing needs. The resources for online computing require to take place at Tier0, whereas part of the offline computations could take place at another computing center. This fraction of computations will be an important guideline for the conceptual design of the FAIR computing model.   

To be more specific, the computing requirements are split up according to the following classifications:

\begin{enumerate}
\item[I.] Experiment Clusters

Integral part of the experiments; responsibility for the design/operation with the experiment

Subclasses:

\begin{enumerate}
\item at the experiment location; exclusively used by the experiment
\item at the experiment location; can be used part-time for general purpose computing
\item in the GreenCube; exclusively used by the experiment
\item in the GreenCube; can be used part-time for general purpose computing
\end{enumerate}

\item[II.] Shared Compute Clusters

Resources required on shared computing resources, responsibility for design/operation with FAIR/GSI Scientific IT

Subclasses:

\begin{enumerate}

\item can be located at any FAIR computing center; if required data accessible at necessary speed
\item must be located in the GreenCube, reserved guaranteed resources during data taking for e.g. online event selection

\end{enumerate}

\end{enumerate}

\subsection{Compute specifications}

The compute requirements of each experiment at FAIR are summarized in uniformly-organized tables as illustrated in Table~\ref{example_cr}. The last two columns in such table contain the compute requirements in units of HEPSpec06 for the various phases of FAIR operations (in the example FS(+) and MSVc). The compute resource requirements are subdivided according to the classification scheme as discussed above according to column ``Compute Class''.

\begin{table}[h]
\begin{center}
\caption{Example experiment: Compute Requirements (in HEPSpec06).}
\begin{tabular}{|c|c|c|c|}
\hline
Compute Class  &  FS / FS+ & MSVc  \\
\hline
I.a & 0 & 0 \\
\hline
I.b & 0 & 0 \\
\hline
I.c & 0 & 0 \\
\hline
I.d & 0 & 0 \\
\hline
II.a & 10,000 &  55,000 \\ 
\hline
II.b & 0 & 0  \\
\hline
\end{tabular}
\label{example_cr}
\end{center}
\end{table}%

\subsection{Storage specifications}

Throughout this chapter, the storage requirements of each experiment are summarized in structured/uniform tables as illustrated in Tabs.~\ref{example_sr1} and ~\ref{example_sr2}. 

Table~\ref{example_sr1} illustrates the storage requirements of an example experiment during data taking. The numbers in the last two columns reflect the requirements for the two operation scenarios of FAIR, namely first science (+) (FS/FS+) and modularized start version (MSVc). The first column summarizes the average bandwidth requirements from the experiment (DAQ) to the GreenCube including the required number of fiber bundles. The bandwidth to the permanent storage is specified for expected peak rates and average in units of MBytes/s. In this example, the peak bandwidth is the same as what was required for the bandwidth from experiment to GreenCube. In the case of online data selection, this might differ. The expected data volume in units of TBytes that will be stored permanently each year is specified in the next row in the table. In the last row of the table, the permanent disk storage that is needed for data taking is specified. Such disk storage is used as cache or scratch during online data taking.

Table~\ref{example_sr2} shows an example for storage requirements during offline data processing stage. Similar to Tab.~\ref{example_sr1}, estimates are provided for the two FAIR operation stages FS(+) and MSVc. Moreover, the table classifies the information in three different data types: raw data, simulated data and derived (processed) data. For each data type and FAIR operation stage, we specify the volume of data being produced in the processing stage each year in units of TB/s, the number of years the data are being kept on (online) disk (\#years), and the bandwidth required during processing stage in units of MBytes/s.

\begin{table}[h]
\begin{center}
\caption{Example experiment: Storage Requirements (I \textemdash Data Taking).}
\begin{tabular}{|c|c|c|c|}
\hline
&   &  \multirow {2} {2 cm}{FS/FS+}   &   \multirow {2} {2cm}{MSVc} \\
& & & \\
\hline
   \multirow {2} {3.5 cm}{ Experiment to GreenCube/RZ1} & \#fibers & 0 & 0 \\
\cline{2-4}
&     Bandwidth (MB/s) & 500 & 500  \\
\cline{2-4}
\cline{1-2}
\multirow {2} {*} {Bandwidth to permanent storage} & Peak (MB/s) & 500 & 500 \\
\cline{2-4}
&   Average (MB/s) & 200 & 200 \\
\hline
\multicolumn {2} { |c| }{  Permanent storage/year (TB/year)} & 500 &  500\\
\hline
\multicolumn {2} { |c| } {Additional disk storage (TB)} & 0 & 0 \\

\hline
\end{tabular}
\label{example_sr1}
\end{center}
\end{table}%

\begin{table}[h]
\begin{center}
\caption{Example experiment: Storage Requirements (II \textemdash Processing).}
\begin{tabular}{|c|c|c|c|}
\hline
&   &  \multirow {2} {2 cm}{ FS/FS+}   &   \multirow {2} {2cm}{MSVc} \\
& & & \\
\hline
 \multirow {3} {3.5 cm}{ Raw Data} & TB/year & 350 & 350 \\
\cline{2-4}
&    \#years &5 & 5 \\
\cline{2-4}
&  Bandwidth (MB/s) & 200  & 500 \\
\hline
\multirow {3} {3.5 cm}{ Simulation} & TB/year & 300 & 300  \\
\cline{2-4}
&    \#years &3 & 3\\
\cline{2-4}
&  Bandwidth (MB/s) & 6,000& 6,000\\
\hline

\multirow {4} {3.5 cm}{ Derived data} & TB/year & 400 & 400 \\
\cline{2-4}
&    \#years & 3 & 3  \\
\cline{2-4}
&  {Bandwidth (MB/s)} & 6,000,  & 6,000 \\

\hline
\end{tabular}
\label{example_sr2}
\end{center}
\end{table}%
\newpage

\subsection{FAIR Phase Zero}
\label{fair_phase_zero}

In this section we summarize the computing activities at the GSI and FAIR compute systems within the FAIR Phase Zero program at GSI. Particularly, we evaluated the used computing resources over the past few years by the FAIR research lines to, partly, extrapolate towards their needs once SIS100 becomes operational.  

Operational details of the HPC cluster at GSI, called Virgo and located in the Green-IT Cube are described in~\cite{gsi:hpc}. The system provides Virtual Application environments (VAEs) based on Linux container technology \cite{apptainer:2021} to enable multiple different user environments on a single host platform. Besides the VAEs supported by IT, it is possible for users to launch custom build containers as well. Slurm is used as workload management system. The capabilities of the workload management system have been adapted to allow flexible partitioning of resources. This is particularly useful in order to dynamically orchestrate groups of compute nodes upon requests for online-computing resources by different scientific experiments. To optimize the interconnectivity (bandwidth and latency) among the nodes, an InfiniBand network is used. Lustre is deployed as distributed storage system in order to cope with many PB of data.

\begin{figure}[h]
\begin{center}
\vspace*{-0.2cm}
\includegraphics[width=0.9\textwidth]{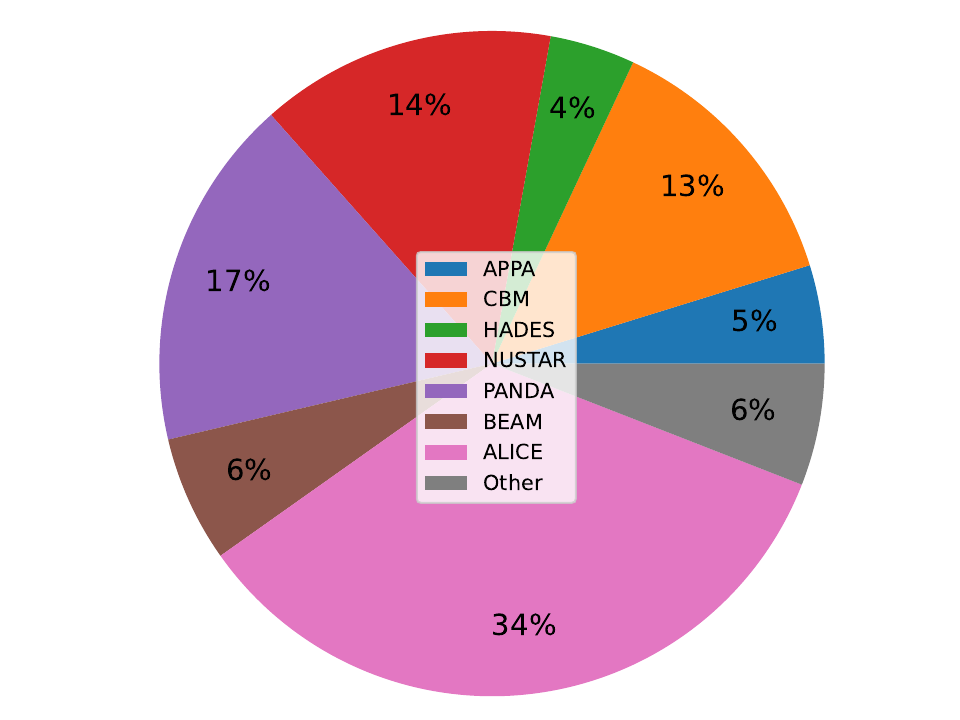}
\vspace*{-0.3cm}
\caption{An overview of the compute usage at the Virgo cluster at GSI classified according to the various research lines. The presented data correspond to the period 1/2021-3/2023 and amounts to about 82~kcore~years.}
\label{fig:fp0_compute}
\end{center}
\end{figure}

In the context of FAIR Phase Zero, various experiments collected and processed data using FAIR detector elements during the past years. Moreover, large-scale Monte Carlo simulations, theoretical calculations and other preparatory activities towards future FAIR operations have taken place. The Virgo cluster at the GreenCube HPC plays a central role in serving all these activities. We analyzed the compute and storage usage of the Virgo cluster classified according to user and group names for the years 2021 till March 2023 representing a snapshot of FAIR Phase Zero. 

Figure~\ref{fig:fp0_compute} summarizes the distribution of CPU usage at the Virgo cluster among the various research lines. The total CPU usage for 27 months is about 82 kcores years. The Virgo cluster embeds more than 50k of physical cores (AMD/Intel) corresponding $\sim$1.1~MHEPSpec06 of total compute resources. The largest (34\%) group of users is associated with Alice activities with a majority related to Grid computing~\footnote{GSI is a Tier2 side for CERN-Alice within WLCG}. These activities are not part of FAIR Phase Zero. The remaining part of the chart relates to FAIR Phase Zero activities, whereby all pillars consume a significant fraction of the available resources. Each pillar also contains a large contribution of theoretical calculations directly related to support the associated experiment and its physics program. We estimated that the contribution of model calculations is about 1/3 of the total requested compute capacity. Based on the analyzed administrative data, we estimated a total of 0.5~MHEPSpec06 of compute resources dedicated to FAIR Phase Zero program.

\begin{figure}[h]
\begin{center}
\vspace*{-0.2cm}
\includegraphics[width=1\textwidth]{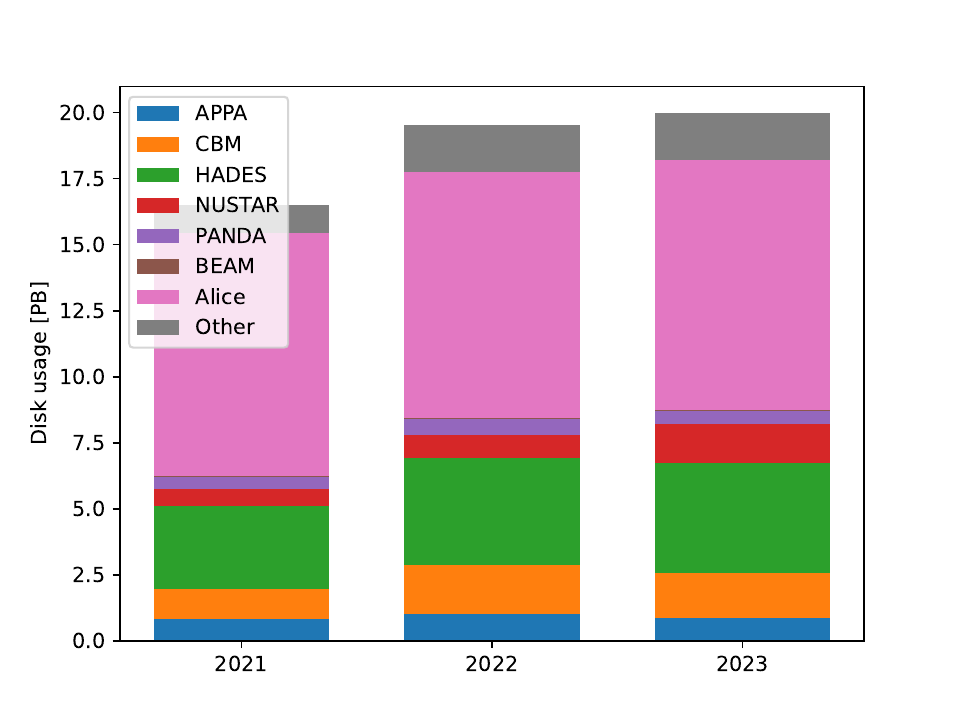}
\vspace*{-1.2cm}
\caption{A stacked overview of the maximum disk usage at the Virgo cluster (Lustre) at GSI classified according to the various research lines and for the past three years.}
\label{fig:fp0_storage}
\end{center}
\end{figure}

The maximum disk usage over the past years are depicted in Fig.~\ref{fig:fp0_storage}. Also here, we distinguish between the various contributions from the research pillars. The total used disk capacity has slightly grown over the past years to a level of about 20~PB, whereby the non-FAIR-related Alice activities allocated about 10~PB (of which 50\% relates to Grid activities and 50\% to local Alice usage). In the context of FAIR Phase Zero, HADES, an experiment using SIS18 with presently the highest data rates, has allocated the largest fraction of disk space ($\sim$4~PB). The total available disk capacity is about 30~PB (July 2023).



In summary, the computing capacity at Virgo within the context of FAIR Phase Zero amounts to about 0.5~MHEPSpec06 (compute) and 10~PB (storage). This estimate does not include the various Phase Zero activities taking place elsewhere at different laboratories and computing clusters. The purpose of presenting these estimates is many fold. First of all, it provides a reference value with respect to the request for FAIR operations. Moreover, it shows that all the research pillars of FAIR are already well connected to the computing infrastructure at the GreenCube (being the basis for future FAIR Tier0). Last, but not least, the consumed compute resources at FAIR Phase Zero are, in some cases, used to predict the required resources at the various stages of FAIR. Some of the FAIR experiments will only take place at MSVc, {\it i.e.} PANDA. However, preparation studies that already have been started in FAIR Phase Zero will be continued until the realization of MSVc. The statistics presented here will be used in such cases as educated guesses.

\newpage

\subsection{APPA}
\label{sec:appa}
\subsubsection{\label{atomic_physics}Atomic physics}
The atomic physics research programme of APPA at FAIR prepares experimental stations at different locations of the FAIR accelerator facility. These facilities are located at HESR, APPA CAVE, SIS100, CRYRING, ESR and HITRAP. 
Our computing requirements focus on long-term storage, highly reliable network infrastructure and shared cluster computing for data analysis and theory calculations.
In terms of data handling, the planned experiments can be characterised into three different classes.  

\begin{enumerate}
    \item The first describes the stand-alone and proprietary DAQ systems, which do not require any FAIR-related computing resources except for a small amount of long-term storage for the experiment data. They are not relevant for this discussion. 

    \item The second class represents the event-based DAQ systems, which consist of one or more nodes for the different detector systems, depending on the physics case. Examples are the lepton spectrometer at HESR; Ge(i) polarimeter systems at HESR, ESR and CRYRING; or the magnetic micro calorimeter systems. Data pre-processing will be performed close to the detector systems, resulting in an average raw data transfer rate (list mode) of 5-50 MB/s to the computing cluster and long-term storage. It is planned to perform cluster computing on particle detector data in the future, similar to the example below (Schottky detectors), but the requirements will be very moderate, almost negligible.

    \item The last group are the Schottky detector DAQ systems, which either continuously or shot by shot sample the pickup signal from resonant Schottky detectors, monitoring the electromagnetic wave driven by circulating ions or ion bunches in a storage ring. The sensitive data analysis in this case depends on absolute dead time free signal sampling at a data rate of approximately 5 Gb/s. To avoid any data loss due to network outages or failures, a temporary local storage solution with a 2 x 10~Gb/s connection is attached to the DAQ. Subsequently, the raw experimental data of about 50~TB/week will have to be transferred to the long-term storage at a reasonable transfer rate. Future detector development at the ESR would increase the amount of data required. Such detectors would include position sensitive resonant pickups to increase mass and lifetime measurements using isochronous velocity correction.

    Time domain Schottky data are processed into intermediate result files in frequency domain, as well as as combinations and second order derivatives of these data. For this step parallel computing in Green Cube is used. For a given parameter set, up to 10000 jobs are used each with up to 100 GB peak memory usage. The amount of intermediate data are in the same order as the original data , i.e. approx. 1:1. Such calculations are repeated many times during the analysis time throughout the year. Intermediate data will not be stored on the long term storage, but may be needed on the LUSTRE for the duration of analysis.
    
\end{enumerate}





\begin{table}[H]
\begin{center}
\caption{APPA-Atomic physics: Compute Requirements (in HEPSpec06).}
\begin{tabular}{|c|c|c|c|}
\hline
Compute Class  &  FS / FS+(+) & MSVc  \\
\hline
I.a & 0 & 0 \\
\hline
I.b & 0 & 0 \\
\hline
I.c & 0 & 0 \\
\hline
I.d & 0 & 0 \\
\hline
II.a & 45,000 &  90,000 \\
\hline
II.b & 0 & 0  \\
\hline
\end{tabular}
\label{appa_ap_cr}
\end{center}
\end{table}%


\begin{table}[H]
\begin{center}
\caption{APPA-Atomic physics: Storage Requirements (I \textemdash Data Taking).}
\begin{tabular}{|c|c|c|c|}
\hline
&   &  \multirow {2} {2 cm}{FS / FS+(+)}   &   \multirow {2} {2cm}{MSVc} \\
& & & \\
\hline
   \multirow {2} {3.5 cm}{ Experiment to GreenCube/RZ1} & \#fibers & 0 & 0 \\
\cline{2-4}
&     Bandwidth (MB/s) & 50 & 100  \\
\cline{2-4}
\cline{1-2}
\multirow {2} {*} {Bandwidth to permanent storage} & Peak (MB/s) & 50 & 100 \\
\cline{2-4}
&   Average (MB/s) & 20 & 50 \\
\hline
\multicolumn {2} { |c| }{  Permanent storage/year (TB/year)} & 200 & 400 \\
\hline
\multicolumn {2} { |c| } {Additional disk storage (TB)} & 10 & 50 \\

\hline
\end{tabular}
\label{appa_ap_sr1}
\end{center}
\end{table}%


\begin{table}[H]
\begin{center}
\caption{APPA-Atomic physics: Storage Requirements (II \textemdash Processing).}
\begin{tabular}{|c|c|c|c|}
\hline
&   &  \multirow {2} {2 cm}{FS/FS+(+)}   &   \multirow {2} {2cm}{MSVc} \\
& & & \\
\hline
 \multirow {3} {3.5 cm}{ Raw Data} & TB/year & 200 & 400 \\
\cline{2-4}
&    \#years & 1 & 1 \\
\cline{2-4}
\hline
\multirow {3} {3.5 cm}{ Simulation} & TB/year & 5 &  10  \\
\cline{2-4}
&    \#years & 1 & 1 \\
\cline{2-4}
\hline
\multirow {3} {3.5 cm}{ Derived data} & TB/year & 10 & 20 \\
\cline{2-4}
&    \#years & 2 & 2  \\
\cline{2-4}
\hline
\end{tabular}
\label{appa_ap_sr2}
\end{center}
\end{table}%


\clearpage
\subsubsection{Biology}
Experiments in biophysics differ from those in nuclear physics in two important ways. They do not generate large amounts of online or list mode data that need to be stored and processed immediately. 
For the foreseeable future, online control and data processing will be performed as close to the irradiation experiment as possible.
However, for data such as 3D or 4D tomographic and dosimetric data, as well as images, e.g. from various microscopy devices, an adequate amount of reliable long-term storage is required.
Computing requirements can be met using off-the-shelf or in-house hardware, integrated into the existing IT environment.
In the future, real-time control of sample irradiation will play an important role.



\begin{table}[ht]
\begin{center}
\caption{APPA-Biology: Compute Requirements (in HEPSpec06).}
\begin{tabular}{|c|c|c|}
\hline
Compute Class  &  FS / FS+(+)  &  MSVc\\
\hline
I.a & 0 & 0 \\
\hline
I.b & 0 & 0  \\
\hline
I.c & 0 & 0  \\
\hline
I.d & 0 & 0  \\
\hline
II.a & 4,500 &  9,000 \\
\hline
II.b & 0 & 0  \\
\hline
\end{tabular}
\label{appa_bio_cr}
\end{center}
\end{table}%


\begin{table}[ht]
\begin{center}
\caption{APPA-Biology: Storage Requirements (I \textemdash Data Taking).}
\begin{tabular}{|c|c|c|c|}
\hline
&   &  \multirow {2} {2 cm}{FS / FS+(+)}   &   \multirow {2} {2cm}{MSVc} \\
& & & \\
\hline
   \multirow {2} {3.5 cm}{ Experiment to GreenCube/RZ1} & \#fibers & 0 & 0 \\
\cline{2-4}
&     Bandwidth (MB/s) & 0 & 0  \\
\cline{2-4}
\cline{1-2}
\multirow {2} {*} {Bandwidth to permanent storage} & Peak (MB/s) & 0 & 0 \\
\cline{2-4}
&   Average (MB/s) & 0 & 0 \\
\hline
\multicolumn {2} { |c| }{  Permanent storage/year (TB/year)} & 100 & 150 \\
\hline
\multicolumn {2} { |c| } {Additional disk storage (TB)} & 0 & 0 \\

\hline
\end{tabular}
\label{appa_bio_sr1}
\end{center}
\end{table}%

\begin{table}[ht]
\begin{center}
\caption{APPA-Biology: Storage Requirements (II \textemdash Processing).}
\begin{tabular}{|c|c|c|c|}
\hline
&   &  \multirow {2} {2 cm}{FS / FS+(+)}   &   \multirow {2} {2cm}{MSVc} \\
& & & \\
\hline
\multirow {3} {3.5 cm}{ Derived data} & TB/year & 100 & 150 \\
\cline{2-4}
&    \#years & 2 & 2  \\
\hline
\end{tabular}
\label{appa_bio_sr2}
\end{center}
\end{table}%


\clearpage
\subsubsection{Material Science}

The amount of long term storage of raw data needed for material science experimetns is rather negligible for current discussion. But the main computing requirements of material science is focused on simulations and the resulting intermediate files. The requirements increase moderately during the MSVc period.

The computational resources necessary for materials research at GSI can be divided into three major parts, on and offline experimental data and simulations:

On-line: Online data are acquired during beamtimes and data rates are typically modest and can be handled by local computers. Long term, secure storage of course is necessary, but the total amount of data is about 1 TB/year.

Off-line: For many of the experiments, the beamtime is only one step in a long series of offline procedures and experiments that are performed on each sample. Here, accurate and easy to use metadata management is critical. For many experiments, a sample ID is given, which is kept by the scientist/student throughout the chain of off-line experiments. However, to date there is no solution for a fully digital, traceable logging system for samples, which ideally would be hosted in a central location and could also be used by other departments. The total amount of data here is higher than for online data, especially because of some microscopy techniques like scanning and transmission electron microscopy and X-ray diffraction data from synchrotron experiments. Still, the storage requirements are expected to be less than 10 TB/year. 

Simulations: For simulations on ion-materials interactions and detector development, access to clusters is required. Currently, Monte Carlo simulation of ion impacts for experiment design and ion acoustic detector development are run on the Virgo cluster. The current average is at 2000 Threads running about 5-10 days a month, future requirements are expected to be at a similar level, amounting to up to 50\% of the time.



\begin{table}[h]
\begin{center}
\caption{APPA-Material science: Compute Requirements (in HEPSpec06).}
\begin{tabular}{|c|c|c|}
\hline
Compute Class  & FS / FS+(+) &  MSVc\\
\hline
I.a & 0 & 0 \\
\hline
I.b & 0 & 0  \\
\hline
I.c & 0 & 0  \\
\hline
I.d & 0 & 0  \\
\hline
II.a & 9,000 &  18,000 \\
\hline
II.b & 0 & 0  \\
\hline
\end{tabular}
\label{appa_mat_cr}
\end{center}
\end{table}%

\begin{table}[h]
\begin{center}
\caption{APPA-Material science: Storage Requirements (I \textemdash Data Taking).}
\begin{tabular}{|c|c|c|c|}
\hline
&   &  \multirow {2} {2 cm}{FS/FS+(+)}   &   \multirow {2} {2cm}{MSVc} \\
& & & \\
\hline
   \multirow {2} {3.5 cm}{ Experiment to GreenCube/RZ1} & \#fibers & 0 & 0 \\
\cline{2-4}
&     Bandwidth (MB/s) & 0 & 0  \\
\cline{2-4}
\cline{1-2}
\multirow {2} {*} {Bandwidth to permanent storage} & Peak (MB/s) & 0 & 0 \\
\cline{2-4}
&   Average (MB/s) & 0 & 0 \\
\hline
\multicolumn {2} { |c| }{  Permanent storage/year (TB/year)} & 80 & 100 \\
\hline
\multicolumn {2} { |c| } {Additional disk storage (TB)} & 0 & 0 \\

\hline
\end{tabular}
\label{appa_mat_sr1}
\end{center}
\end{table}%

\begin{table}[h]
\begin{center}
\caption{APPA-Material science: Storage Requirements (II \textemdash Processing).}
\begin{tabular}{|c|c|c|c|}
\hline
&   &  \multirow {2} {2 cm}{FS/FS+(+)}   &   \multirow {2} {2cm}{MSVc} \\
& & & \\

\hline
\multirow {3} {3.5 cm}{ Derived data} & TB/year & 80 & 100 \\
\cline{2-4}
&    \#years & 2 & 2  \\
\hline
\end{tabular}
\label{appa_mat_sr2}
\end{center}
\end{table}%


\clearpage
\subsubsection{Plasma physics}

The computational requirements associated with data acquisition for plasma physics experiments at FAIR are modest and negligible. We aim to work mostly on single-core machines, taking modest amounts of data from single-shot experiments. For computational tasks, which are mostly simulations to support data analysis or make predictions, the highest demand comes from particle-in-cell simulations, which require MPI machine-optimised computing power. In the coming years, we expect to run 3D PIC simulations for the Plasma Physics programme on the Green IT Cube, with up to 15,000 cores per simulation. 

Storage requirement for data taken from the experiments in the APPA cave is also modest due to the nature of the experiments performed by the plasma physics collaboration.

The storage requirements associated with simulation activities, on the other hand, are very large. We plan to use the Green Cube for numerical plasma physics experiments that produce very large (\textgreater than 100 GB) files. These files will need to be stored locally for some time to await analysis.  In general, these files also require local post-processing before they can be exported and the physics data extracted. Post-processed data files also need to be stored long enough for analysis. Typical data sets for a simulation will be close to 10 GB and the transfer rate must be planned accordingly.


In the future, it is planned to use GPUs for simulations. For now, the planned requirements are negligible.


\begin{table}[ht]
\begin{center}
\caption{APPA-Plasma physics: Compute Requirements (in HEPSpec06).}
\begin{tabular}{|c|c|c|}
\hline
Compute Class  &  FS / FS+(+)  &  MSVc  \\
\hline
I.a & 0 & 0 \\
\hline
I.b & 0 & 0  \\
\hline
I.c & 0 & 0  \\
\hline
I.d & 0 & 0  \\
\hline
II.a & 60,000 &  130,000 \\
\hline
II.b & 0 & 0  \\
\hline
\end{tabular}
\label{appa_pp_cr}
\end{center}
\end{table}%


\begin{table}[ht]
\begin{center}
\caption{APPA-Plasma physics: Storage Requirements (I \textemdash Data Taking).}
\begin{tabular}{|c|c|c|c|}
\hline
&   &  \multirow {2} {2 cm}{FS / FS+(+)}   &   \multirow {2} {2cm}{MSVc} \\
& & & \\
\hline
   \multirow {2} {3.5 cm}{ Experiment to GreenCube/RZ1} & \#fibers & 0 & 0 \\
\cline{2-4}
&     Bandwidth (MB/s) & 0 & 0  \\
\cline{2-4}
\cline{1-2}
\multirow {2} {*} {Bandwidth to permanent storage} & Peak (MB/s) & 20 & 50 \\
\cline{2-4}
&   Average (MB/s) & 10-20 & 20-50 \\
\hline
\multicolumn {2} { |c| }{  Permanent storage/year (TB/year)} & 4 & 10 \\
\hline
\multicolumn {2} { |c| } {Additional disk storage (TB)} & 1 & 2 \\

\hline
\end{tabular}
\label{appa_pp_sr1}
\end{center}
\end{table}%


\begin{table}[ht]
\begin{center}
\caption{APPA-Plasma physics: Storage Requirements (II \textemdash Processing). We note that simulated data are stored in relatively large files, whereas derived data are stored in smaller files.}
\begin{tabular}{|c|c|c|c|}
\hline
&   &  \multirow {2} {2 cm}{FS / FS+(+)}   &   \multirow {2} {2cm}{MSVc} \\
& & & \\
\hline
 \multirow {3} {3.5 cm}{ Raw Data} & TB/year & 2 & 5 \\
\cline{2-4}
&    \#years &1 & 1 \\
\cline{2-4}
&  Bandwidth (MB/s) & 10  & 50 \\
\hline
\multirow {3} {3.5 cm}{ Simulation} & TB/year & 300 & 600  \\
\cline{2-4}
&    \#years & 10 & 10 \\
\cline{2-4}
&  Bandwidth (MB/s) & 2,000 & 2,000\\
\hline

\multirow {3} {3.5 cm}{ Derived data} & TB/year & 3 & 3 \\
\cline{2-4}
&    \#years & 10 & 10  \\
\cline{2-4}
&  Bandwidth (MB/s) & 100 & 100 \\
\hline
\end{tabular}
\label{appa_pp_sr2}
\end{center}
\end{table}%

\clearpage
\newpage

\subsection{CBM}

In the following, we derive estimates of the requirements of the CBM experiment for online and offline computing, bandwidths, and storage capacity.
It must be noted that several years before the start of operation,
many issues are yet quite uncertain, both concerning CBM design features as well as operation 
conditions not directly influenceable by CBM.
The numbers have to be interpreted accordingly as estimates only, which are to be substantiated
by more involved investigations in the years to come, in particular with experience gathered
from the miniCBM full-system test.

\subsubsection{Operation of the CBM experiment}
\label{sec:operation}
\paragraph{Annual run time and beam conditions}

For the assessment of the CBM computing requirements, the following assumptions on the operating conditions are made:
\begin{enumerate}
\item The SIS-100 machine is planned to be operated 6,000 hours (250 days) per year
Accelerator-wise, slow extraction to fixed-target experiments can be efficiently combined
with fast extraction to storage-ring experiments in a super-cycle.
We thus assume a beam into the CBM cave for 50\,\% of the machine operation time.
HADES request about 30 days of operation per year.
The remaining beam time for CBM is hence 2,280~h or $8.2 \cdot 10^6$~s.

\item We assume 100\% operational efficiency of the experiment during CBM beamtimes. It is understood that it will be lower in the first years of data taking with a steep learning curve. Experiences with past and running experiments suggest that the actual experiment duty cycle stabilizes at values above 90\% after some years of operations.


\item 
Assuming CBM to be run in parallel to an experiment using fast extraction,
an effective duty cycle of 75\,\% seems a realistic operation scenario.

\item There are no predictions for the in-spill beam intensity variations to be expected from SIS-100.
CBM and HADES require the variations to be less than three (peak/average)
For the resource estimates in this document, we assume a peak-to-average ratio of two.
It should be noted that this means a substantial improvement of the
beam quality with respect to SIS-18.
\end{enumerate}

\paragraph{Interaction rates and CBM setups}
\label{subsec:rates_setups}
For the interaction or data rates, we employ the following terminology:
\begin{itemize}
\item Peak rate: the maximum instantaneous rate, averaged over a time window of 10~$\mu$s;
\item Average rate: the mean in-spill rate, averaged over beam intensity variations on a time scale of ms;
\item Sustained rate: the mean data taking rate during operation, averaged over the machine duty cycle on a time scale of 10~s.
\end{itemize}
The CBM detector systems, with the exception of the Micro-Vertex Detector (MVD) and the 
Forward Spectator Detector (FSD), are designed for a peak interaction rate of $10^7$~/~s for minimum-bias Au+Au collisions at $p_\textrm{beam} = 12A$~GeV/c.
In Au+Au collisions, the MVD is limited by the load from delta electrons, produced by beam particles in the target, to average beam intensities of $10^7$~/~s, corresponding to an average interaction rate of $10^5$~/~s for a 1\,\% target. 
The FSD is limited 
by its readout
to a peak interaction rate of $10^6$~/~s.

CBM is designed to operate with flexible combinations of detector systems (``setups'')
serving different physics objectives.
For the purpose of this document, we consider the following setups,
which are decisive for the computing requirements:

\begin{itemize}
\item Hadron setup: it comprises the Silicon Tracking System (STS), 
the Transition-Radiation Detector (TRD) and the Time-of-Flight System (TOF).
This setup gives access to hadronic probes like e.g., multi-strange hyperons
or hyper-nuclei. It can be operated at highest interaction rates.
Adding the FSD for the determination of the event plane (e.g., for flow measurements) limits
the operation to a peak interaction rate of $10^6$~/~s.
\item Electron setup: it comprises the Hadron setup plus the MVD, the Ring-Imaging Cherenkov Detector (RICH) and the Forward Spectator Detector (FSD). In this setup, both hadronic probes and electron pairs can be measured simultaneously. The MVD limits the operation of this setup to average interaction rates of $10^5$~/~s.
\item Muon setup: it comprises the STS, the Muon Detection System (MUCH), the TRD, and TOF. 
This setup is exclusively used to measure muon pairs. It can be operated at highest interaction rates.
\end{itemize}
The CBM beam times will be split between the setups according to the physics priorities, which will be agreed on 
before each data taking campaign. Switching between the setups necessitates moving or exchanging of detector
sub-systems.

\paragraph{Online data flow}
\label{subsec:dataflow}

Raw data delivered from the detector front-end electronics are aggregated by the DAQ hardware and streamed
to the Entry Cluster, a number of commodity-hardware computers equipped with custom FPGA interface cards.
The experiment DAQ is designed to cope with a maximum instantaneous interaction rate of $10^7$/s (Au+Au at $p_\textrm{beam} = 12A$~GeV/c).
The DAQ hardware and/or the Entry Cluster will buffer in-spill beam intensity variations;
it is not yet determined whether the average over the duty cycle will be established in the Entry Cluster or in the Compute Cluster.
The size of the Entry Cluster is determined by bandwidth and connectivity to the experiment rather than by compute power.
It will be located close to the experiment; the responsibility for its design and operation is with CBM.
Within the FAIR computing model, it belongs to the class I.a.

The Entry Cluster is connected via long-range Infiniband to the Compute Cluster located in the Green Cube.
The FLESnet~\cite{ref:flesnet} software running on entry and compute nodes aggregates raw data into time-slices, which are delivered to the compute nodes for processing in real-time. 
The role of online data processing is the selection of a subset of the raw data for permanent storage.
The Compute Cluster belongs to class II.b of the FAIR computing model; the corresponding reserved resources must be
guaranteed and exclusive during CBM data taking.
The responsibility for design and operation of the Compute Cluster has to be agreed on between FAIR IT and CBM.

\paragraph{Online data processing concept}
\label{subsec:online_processing}

The operation modes of CBM comprise both data taking at maximum interaction rate with highly selective software triggers
and minimum-bias data taking without online data selection.
For Au+Au collisions at $p_\textrm{beam} = 12A$~GeV/c, the maximum instantaneous rate is $10^7$ events/s, corresponding to a sustained rate of  $3.75 \cdot 10^6$ events/s, averaged over the machine duty cycle.
The minimum-bias interaction rate is determined by the saturation of the archival bandwidth and differs between the experimental setups.

Time-slices are processed independently on the compute nodes, which constitutes a first, trivial data-level parallelism 
on the time-slice / node level. 
The size (duration) of a time-slice is adjusted to the hardware architecture of the compute nodes.

Selective triggers necessitate partial reconstruction of raw data to track and event level. 
This real-time reconstruction has to make optimal use of the computing capacities of the compute nodes,
in particular by vectorisation (SIMD) and inter-node concurrency. 
Parallel features may be used on the algorithm level or on the framework level. 
The result of real-time reconstruction and event selection is a portion of the raw data stream
presumably corresponding to one collision. This raw data container will be forwarded to file storage
and provides the interface to offline reconstruction and analysis.

\begin{figure}[bhtp]
\begin{center}
\includegraphics[width=0.7\textwidth]{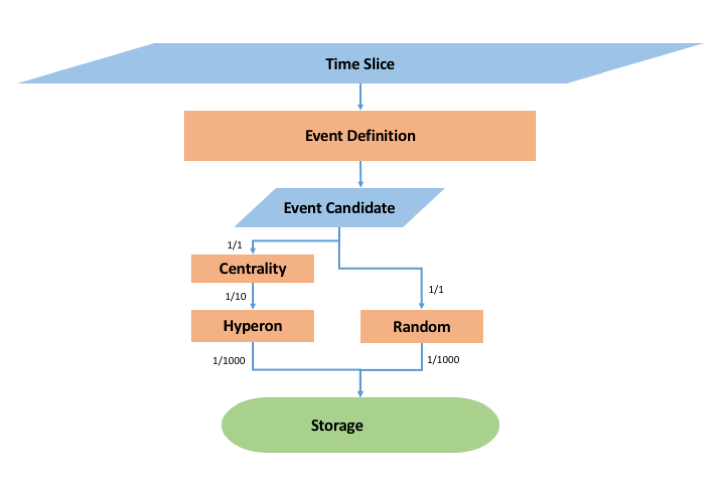} 
\caption{Schematic view of a software trigger set operating on time-slices. The physics trigger comprises 
the selection of central events (data reduction of 10) followed by a trigger on multi-strange hyperons with a
selectivity of 100. In parallel, a minimum-bias trigger is run by randomly selecting events
such that the output data volume is similar to that of the the physics trigger.}
\label{fig:trigger}
\end{center}
\end{figure}

Figure~\ref{fig:trigger} schematically shows the data flow for online data selection. 
Starting from a time-slice as input, event candidates are defined which are then subjected to selection
criteria following the physics objectives. 
The event definition from the raw data stream may happen based on the time information only by defining
regions-of-interest or may already involve track reconstruction. 
In minimum-bias data taking, events are forwarded to storage without further processing. 
In trigger mode, software triggers decide whether to store or discard an event candidate.
Trigger decisions can proceed in several steps with subsequent reduction of the data volume. Similarly, several physics triggers can be run in parallel, sharing the archival bandwidth.
Physics triggers not selective enough to reduce the data volume to the targeted archival bandwidth can be
run with additional random data suppression; an example is a minimum-bias trigger run in parallel to a highly selective trigger 
by randomly suppressing events.

\paragraph{Event reconstruction}
\label{subsec:reco_time}
Event reconstruction from raw data comprises the following steps:
\begin{itemize}
    \item Local reconstruction: reconstructing the detector response to a particle crossing from the digis. For a tracking detector, the reconstructed object is typically a space-point (hit); for e.g., the RICH detector, it is a ring.
    \item Tracking: Associating hits and other measurements through the detector setup to a trajectory (''track``) and determining its parameters.
    \item Vertexing: Finding the primary (event) vertex as a 4-d point in space and time from a number of tracks.
\end{itemize}
Event reconstruction is used both online for the trigger decision and offline for the production of the input for high-level analysis. Because of the usage in real-time data processing, high-performance software must be employed.

The most complex and thus compute-intensive part in the reconstruction process is track finding in the STS.
For this task, an algorithm based on the Cellular Automaton (``L1 track finder'') was developed.
It is highly optimized with respect to speed and parallelised
using multi-threading paradigms.
The L1 CPU time was measured for simulated minimum-bias Au+Au collisions at $p_\textrm{beam} = 25A$~GeV/c on Intel Xeon E7-4860 
processors to be 8.5~ms per event and physical core
The throughput in this machine hosting
in total 40 physical cores is thus about 4,700 events/s. Hyper-threading was found not to improve the
performance significantly.

Several components in the reconstruction process, in particular tracking in the downstream detectors, are not yet optimized with respect to performance, such that resilient numbers on the total reconstruction speed cannot be derived from the current software status. 
Highly optimized parts of the reconstruction chain are, besides the L1 track finder, the Kalman Filter (KF) package for track fitting, the KFParticle package for vertexing, and the KFParticleFinder package for the simultaneous reconstruction of particle decays.
It was demonstrated that the KFParticleFinder, employing the functionality of KFParticle and KF, consumes less than 20\,\% of the CPU time of the L1 CA track finder.
In addition, local reconstruction in the STS, TRD and TOF detectors was lately strongly optimized.

\paragraph{Observables and trigger conditions}
\label{subsec:observables}

The observables relevant for CBM can be divided into classes according to the required detector setup and
the presence of trigger signatures:

\begin{itemize}
\item Electron pairs, bulk hadrons, and fluctuations: these observables do not have a trigger signature occurring less
frequently than once per event on average. 
For electron pairs, the absence of a trigger signature is due to the abundant background electrons from $\pi^0$ decays, $\gamma$ conversions and secondary interactions.
They will be measured with the electron setup limited to $10^5$ interactions/s (average) because of the MVD.
\item Rare hadrons (e.g., multi-strange anti-hyperons, hyper-nuclei, exotic strange objects): these observables
necessitate high interaction rates and come with highly selective trigger signatures, e.g., on the decay topology.
The will be measured with the hadron setup. In case the PSD is required for event-plane determination (e.g., for flow studies),
the instantaneous interaction rates are limited to $10^6$/s.
\item Charmed hadrons: selective trigger signatures exist on the decay topology, but require the usage of the MVD, which limits the average interaction rates to $10^5$/s. They will be measured in the electron setup simultaneously with electron pairs.
\item Muon pairs: they will be measured with the muon setup at maximal interaction rate. For charmonia, a highly selective trigger signature exists; for low-mass muon pairs, a less selective trigger with data suppression factors of the order of several tens is possible.
\end{itemize}

\paragraph{Raw data event size}
\label{subsec:eventsize}
The raw data event sizes were evaluated by simulation of minimum-bias Au+Au collisions at $p_\textrm{beam} = 12A$~GeV/c generated by UrQMD. 
The simulations were performed with CbmRoot r.12882 (27 April 2018) using the default geometries and digitizers.
The simulation yields the number of raw detector hits (``messages'') per event for the various detector systems.
This number is weighted with the single-message size within a First Level Event Selection (FLES) time-slice, according to the current design
of the front-end ASICS.
The overhead introduced by the micro-slice and time-slice containers is neglected.

Tables~\ref{tab:eventsize_hadron}--\ref{tab:eventsize_muon} show the resulting numbers. 
The event sizes for the STS vary slightly in the different setups because of shadowing from delta electrons by the MVD detector and because of backward scattered particles from the MUCH absorbers, when these systems are present in the setup. TOF shows a higher rate in the electron setup compared to the hadron setup, caused by secondary particles
produced in the RICH and TRD detector materials. 
The comparison of the TOF and TRD rates in the muon setup and the electron setup reflects the 
effect of the MUCH absorbers on the produced particles.

\begin{table}[htp]
\caption{Raw data average event sizes for minimum-bias Au+Au collisions at $p_\textrm{beam} = 12A$~GeV/c in the hadron setup}
\begin{center}
\small
\begin{tabular}{|c|r|r|r|}
\hline
System  &  Messages  &  Bytes / message  &  Event size [kB]  \\
\hline \hline
STS &  5,395  &  4   &  21.6 \\
TRD & 1,810  &  12  & 21.7 \\
TOF &     670  &  8   &  5.4 \\
\hline
Total  &  &  & 48.7 \\
\hline
\end{tabular}

\end{center}
\label{tab:eventsize_hadron}
\end{table}

\begin{table}[htp]
\caption{Raw data average event sizes for minimum-bias Au+Au collisions at $p_\textrm{beam} = 12A$~GeV/c in the electron setup}
\begin{center}
\small
\begin{tabular}{|c|r|r|r|}
\hline
System  &  Messages  &  Bytes / message  &  Event size [kB]  \\
\hline \hline
MVD   &    3,156      &   4   &  12.6 \\
STS    &    4,779      &   4   &  19.1 \\
RICH  &       425      &   4   &  1.7   \\
TRD   &     2,487     &  12   &  29.8   \\
TOF    &    1,079     &    8   &   8.6 \\
PSD    &       403     &    4  &   1.6 \\
\hline
Total  &  &  & 73.4 \\
\hline
\end{tabular}
\end{center}
\label{tab:eventsize_electron}
\end{table}

\begin{table}[htp]
\caption{Raw data average event sizes for minimum-bias Au+Au collisions at $p_\textrm{beam} = 12A$~GeV/c in the muon setup}
\begin{center}
\small
\begin{tabular}{|c|r|r|r|}
\hline
System  &  Messages  &  Bytes / message  &  Event size [kB]  \\
\hline \hline
STS      &  5,668      &  4   &  22.6 \\
MUCH  &     926      &  4   &  3.7   \\
TRD      &     55  &  12   &  0.7   \\
TOF      &     126   &  8   &  1.0 \\
\hline
Total  &  &  & 28.0 \\
\hline
\end{tabular}
\end{center}
\label{tab:eventsize_muon}
\end{table}

The event sizes given in Tables~\ref{tab:eventsize_hadron}--\ref{tab:eventsize_muon} correspond 
to the largest collision system for CBM at SIS-100 (Au+Au at $p_\textrm{beam} = 12A$~GeV/c).
Lower collision energies or smaller collisions systems will entail smaller raw data events.
For instance, the event size for Au+Au collisions at $p_\textrm{beam} = 4A$~GeV/c
is 70\,\% of that at top SIS-100 energy.

\paragraph{Delayed event filtering}
\label{subsec:delayed_filter}
Software triggers on the decay topology in the STS (multi-strange hyperons, hyper-nuclei) rely on precise
knowledge of the detector geometry, i.e., on precision alignment. 
CBM currently has no solid knowledge on how frequently the
alignment has to be renewed.
Possible scenarios are to perform alignment with data from special runs (e.g., field-off) before the actual data
taking, or to perform alignment online during data taking.

The LHCb experiment follows a concept of delayed event filtering, where raw data taken with approximate
alignment only are transiently stored to file after a modest first-level online data reduction.
Track-based alignment is then performed in parallel to the data taking on the transiently stored data.
After having obtained the final alignment parameters, the transiently stored data are then subjected
to a second-level filtering before permanent storage.

Because of the current absence of an alignment concept for CBM, we foresee the possibility for a similar scheme,
which comes with an additional requirement for transient disk storage capacity and bandwidth.

\subsubsection{Offline data processing}
\label{sec:offline}
Offline computing for CBM comprises
\begin{itemize}
    \item Processing of experiment raw data;
    \item Simulation and processing of simulated data;
    \item User-level physics analysis of experiment and simulated data.
\end{itemize}

Offline computing can be performed at any computing site, provided the required input data are
resident or accessible at the necessary speed. 
The corresponding compute resources thus belong to category II.a in the FAIR computing model.

\paragraph{Processing of experiment data}
\label{subsec:data_formats}
A basic design choice of CBM is to permanently store raw data only as result of the experiment operation, irrespective to what information is obtained during online data inspection and selection. 
The motivation for this choice is based on two facts. First, we find the raw data format to be less storage-consumptive than any derived, reconstruction-level format. Second, we are developing highly optimised and fast reconstruction procedures for online data inspection. We find these algorithms to provide a level of accuracy also satisfying offline needs (possibly with minor modifications and additions), such that offline reconstruction can be performed at high speed, without the need for persistent storage of reconstruction-level (ESD) data. 

The storage entity is a set of raw data corresponding to one interaction (see section~\ref{subsec:online_processing}), which provides a level of data compression similar to the time-slice container.

Alignment and calibration parameters required for online data inspection will be obtained from special calibration runs before the actual data taking. 
For offline analysis, parameters with better accuracy will be obtained from the taken raw data. Typically, only a subset of the data is required for that purpose.

After suitable alignment and calibration parameters are obtained, Analysis Object Data (AOD) will be centrally created.
These serve as input for the user analysis. 
AODs typically comprise for each event a list of reconstructed tracks with all parameters (momentum, vertex, PID information) required for the physics-level analysis.
The size of the AOD is by a factor of ten or more smaller than the
corresponding raw data. Different AODs may be produced for the various physics analysis objectives, but currently, we use a single format called \texttt{AnalysisTree} for most analyses.
We estimate the total amount of AOD storage (single copy) to be  $<20$\,\% of the raw data volume. AODs can be distributed to various computer centres outside FAIR, serving as analysis centres for a regional subset of the collaboration.

Foreseeing a small number of passes over a complete raw data set for the production of AODs, the default planning foresees to keep the
raw data on disk storage for fast access for the year of data taking and the following year. Alternatively a data carousel like model could be considered, where the raw data is stored on tape and recalled for reconstruction passes in organized campaigns.
Once the final calibration and alignment parameters are obtained and AODs are produced, the raw data usually
need not be accessed any longer. The raw data will be put on long-term storage (tape or equivalent) with two copies in different locations as a requirement imposed by an external audit.

AODs will be kept available on fast (disk) access for user-level analysis for five years. Since they can be reproduced from raw data if required, there is no need for long-term archiving of AODs. We currently do not foresee to keep several versions of the same AOD data (with different calibration constants), meaning that older AODs from the same raw data set will be removed after newer ones have been produced.
 
\paragraph{Simulation}
\label{subsec:simulation}
Simulations for CBM comprise two steps: first, a transport engine (GEANT4) is run on
input events generated by a suitable generator (e.g., UrQMD), taking into account the detailed
detector geometry and the magnetic field. The output of this step (MC) are the simulated
intersection points of particles with sensitive detector elements and the corresponding energy deposits.
Table~\ref{tab:mctime} show the processing time and MC event sizes for the transport
simulation of minimum-bias Au+Au collisions at $p_\textrm{beam} = 12A$~GeV/c. 
The PSD was excluded from this evaluation since for this device, a fast simulation with
parametrised response will be introduced already on the MC level.

When comparing the MC event sizes to the raw data event sizes in Tables~\ref{tab:eventsize_hadron}--~\ref{tab:eventsize_muon}, it should be noted that MC information is logically equivalent to the hit (space-point) level. The raw data format is strongly optimised w.r.t.\ size, and we find it less storage-consumptive than any reconstruction-level format.
It should be noted further that the size of the MC data was not yet optimised e.g., w.r.t.\ the usage of single versus double precision floats, and can thus be expected to shrink
in future software versions. 

\begin{table}[htp]
\caption{Average processing time (Intel E5-2680 v4 @ 2.40~GHz) and MC size per event for the 
transport simulation of minimum-bias Au+Au collisions at $p_\textrm{beam} = 12A$~GeV/c}
\begin{center}
\small
\begin{tabular}{|c|c|c|}
\hline
setup   &  wall time & Event size \\
\hline \hline
hadron  &  1.1 s  & 140 kB \\
electron  &  3.6 s  &  250 kB \\
muon  &  2.7 s  &  140 kB \\
\hline
\end{tabular}
\end{center}
\label{tab:mctime}
\end{table}

CBM makes use of the VirtualMC concept introduced by ROOT, which allows to run the same experiment software for both Geant3 and Geant4. A simple flag in the simulation settings allows to switch between the two transport engines. While Geant4 is the default option of CBM, the Geant3 option is kept for simulation of specific setups and for verification purposes. This flexibility comes at negligible costs in terms of code efficiency and maintenance. We find the execution time of both engines comparable (within 20\%) with the same geometrical setup and equivalent physics settings.


MC data serve as input for the detector response simulation (``digitisation'').
This step of simulation produces simulated raw data in the same format as experiment data,
i.e., as a time-stream of data not associated to events. 
It is considerably faster than the transport simulation (about 0.3 s/event).

Rare signals will be subjected to transport simulations separately from the background events
and embedded at the digitisation stage. This means that the same transport simulation
of background events can be re-used for different physics purposes and also for
different digitisation conditions (e.g., interaction rates). Thus, MC data are usually saved to file.

For large-scale simulations, we foresee digitisation and reconstruction of the simulated raw data
to be performed on-the-fly, either subjected directly to analysis or to the production of AODs similar
to those for real data.

From the point of view of memory consumption, simulation runs can be performed independently on single cores; 
we currently thus do not see the need to use concurrency features here.

\subsubsection{Resource estimates: computing}
\label{sec:resources}

\paragraph{Entry Cluster (category I.a)}
The number of nodes of the Entry Cluster will be determined by bandwidth from and connectivity to the experiment. 
Our current estimate is that about 100 nodes will be required for full CBM operation.
These resources fall into the category I.a of the FAIR computing model.
In principle, these nodes are available for general-purpose computing between CBM data-taking periods.
However, the computing capacities of the Entry Cluster can be neglected in comparison to that of the
Compute Cluster.

\paragraph{Compute Cluster (category II.b)}
The most compute-demanding software triggers employed in CBM will be those involving full track reconstruction in the STS, e.g., triggers on the decay topology of anti-hyperons or hyper-nuclei. Trigger on muon pairs can be, at least at the first trigger level, derived from data of the most downstream MUCH stations and do not require prior STS track reconstruction. It has been shown that a di-muon software trigger can be obtained at very high speed and thus with moderate compute capacities.
The demands for online computing are thus determined by decay topology triggers, requiring STS, TOF and TRD information, the latter for the identification of multiply-charged composites.

The estimate of the compute capacities needed for real-time-reconstruction and data selection is based on the
L1 track finding in the STS detector as the most compute-intensive part. This was measured to be 8.5~ms
per core and event on a Intel Xeon E7-4860 processor (see section~\ref{subsec:reco_time}) for Au+Au collisions
at $p_\textrm{beam} = 25A$~GeV/c. We estimate that the total reconstruction time, including cluster finding, hit finding and
global tracking, is a factor of three larger than the L1 time consumption, i.e. 25.5~ms. The difference
in beam momentum (12$A$~GeV/c instead of 25$A$~GeV/c) amounts to a factor 1.5 in processing speed, which gives 17~ms per event and core. The throughput on the 40-core-machine is thus 2,350 events/s. 

For the very similar machine Xeon E7-4870, a HepSpec06-calibration exists, yielding 654 in full parallel
operation.
Both machines differ only in the clock frequency (2.40 MHz for E7-4870
versus 2.26 MHz for E7-4860) and, slightly, in the cache size. We assume that the HepSpec06 number
scales by the ratio of the clock frequency and is hence 616 for the E7-4860 used for the L1 performance
measurement.

In order to process the targeted $3.75 \cdot 10^6$ events per second in real-time, about 1,590 of the above mentioned
computers are required (63,750 cores), which amounts to 979 kHepSpec06. These resources correspond to 
category II.b in the FAIR computing model.

\textit{Uncertainties}: The derived number scales linearly with the sustained interaction rate 
($3.75 \cdot 10^6$ / s). It depends on the assumption that a similar level of optimisation and
concurrency as achieved in L1 track finding can be established in all parts of the reconstruction process.
The results reported in
were obtained under idealised conditions, e.g., no mis-alignment and
no detector noise. It can be expected that the inclusion of such effects will deteriorate the performance.
On the other hand, future improvements can be expected in particular by vectorisation on larger
registers. The requirement thus has to be regarded as a current and preliminary estimate only.

\paragraph{Offline computing (category II.a)}

Offline compute resources will be used for simulation, reconstruction of experiment 
and simulated data (production of AODs) and data analysis.

The current CPU consumption for simulation with CBM is moderate in comparison to LHC experiments
(see Table~\ref{tab:mctime}).
It is to be expected that the processing time will increase in the future, triggered by more detailed
description of detector geometries and by using GEANT4 instead of the
currently used GEANT3.
For the estimate of the associated compute resources, we assume a processing time of 4~s per event 
(Intel E5-2680 v4 @ 2.40~GHz, 22 HepSpec06 / physical core) 
for transport simulations and digitisation of the electron setup measuring bulk and untriggered observables.
The demand that the statistical error originating from simulations should be smaller than that originating from the measured data leads to the requirement that at least as many events as recorded by the experiment (see Table~\ref{tab:storage}) have to be simulated. 
The corresponding compute requirements for the electron setup amount to 22,500 cores or 500,000 HepSpec06 used continuously over one year.

The transport of rare probes measured with the hadron or muon setup will be performed separately and then embedded at digitization stage into real raw data.
Since in this case, only a small number of tracks has to be transported, the required processing time is negligible compared to the electron setup.



Reconstruction and production of AOD from both real and simulated data requires very moderate compute resources compared to the simulation. This is obvious from the comparison of the event processing time of 17 ms per core to the simulation time of 4 s. Assuming this number, $1.1 \times 10^{11}$ events taken or simulated with the electron setup can be reconstructed within one week on less than 500 cores. Even taking into account that the offline processing time might be larger than the one achieved online for better accuracy or additional processing steps, we require at maximum 1,000 cores (22,000 HepSpec06) for this task, allowing for multiple passes over the data with refined calibration within a short period.

The estimate for user-level analysis on AODs is, at the current stage, difficult. Typically, many passes over the data will be performed, shaping the analysis and addressing a multitude of physics objectives.
Experience from LHC experiments show that simulation is the most
time-consuming task for offline computing. 
For instance, ALICE typically spend about 70\,\% of their total
CPU usage for simulation, including digitization and reconstruction.
Taking this a a guidance,
we tentatively estimate the user-level analysis of both experiment and simulated
data to consume 50 \% of CPU resources used by the transport simulation. The total requirement 
for offline computing resources thus amounts to 780 kHepSpec06.

\textit{Uncertainties}: The derived number depends on far-reaching assumptions in particular 
concerning the relation of data analysis and simulation, the uncertainty of which can at present
not be quantified.

\subsubsection{Resource estimates: data storage}

\paragraph{Raw data storage volume}

In order to estimate the total raw data storage volume per year, we assume that the available annual beam time
of $6.6 \cdot 10^6$~s will be evenly split between the setups and run modes listed in Table~\ref{tab:datarates}.
Both the hadron and the muon setup are assumed to be operated with a physics trigger with a data reduction of 200
in parallel to a random minimum-bias trigger with the same output bandwidth. 
The electron setup is assumed to be operated without a physics trigger.

\renewcommand{\arraystretch}{1.5}
\begin{table}[htp]
\caption{
Annual raw data storage volume assuming an equal distribution of the available beam time among the hadron,
electron, and muon setups (see text). 
The equivalent event numbers are calculated assuming a trigger efficiency of 100\,\%. 
}
\begin{center}
\small
\begin{tabular}{|c|c|c|c|c|c|r|c|}
\hline
\multirow{2}{*}{setup}  & \multirow{2}{*}{ run time}  &  sustained  & \multirow{2}{*}{ trigger}  &  \multirow{2}{*}{selectivity}  & \multirow{2}{*}{ rand.~red.}   &  \multirow{2}{*}{storage}   &  \# equiv.  \\
     &   & data rate   &   &   &   &   & events \\
\hline \hline
\multirow{2}{*}{hadron}  &  \multirow{2}{*}{$2.75 \cdot 10^6$~s}  & \multirow{2}{*}{183 GB/s} &  physics  &  
200  &  1  &  {2.5 PB}  & {$1.0 \cdot 10^{12}$} \\
  &  &  &  min.~bias  &  1  &  200  &  {2.5 PB}  &  {$0.5 \cdot 10^{12}$} \\
\hline
electron&  {$2.75 \cdot 10^6$~s}  & 5.5 GB/s  &  min. bias  &  
1  &  1  &  {15 PB}  & {$2.0 \cdot 10^{11}$} \\
  \hline
\multirow{2}{*}{muon}  &  \multirow{2}{*}{{$2.75 \cdot 10^6$~s}}  & \multirow{2}{*}{105 GB/s}  &  physics  & 
 200  &  1  &  {1.4 PB}  &  {$1.0 \cdot 10^{12}$} \\
  &  &  &  min.~bias  &  1  &  200  &  {1.4 PB}  &  {$0.5 \cdot 10^{12}$} \\
  \hline \hline
  Sum & {$8.2 \cdot 10^6$ s} & & & & & {22.8 PB} &\\
  \hline
\end{tabular}
\end{center}
\label{tab:storage}
\end{table}%
\renewcommand{\arraystretch}{1}

The resulting numbers are presented in Table~\ref{tab:storage}. 
The above outlined running scenario requires an annual raw data storage capacity of about 18 PB.

This number is to be understood before data compression. Our raw data format is already intrinsically highly compressed; consequently, we see only a moderate gain ($< 20 \%$) when using e.g., \texttt{zstd}. We plan to investigate other compression algorithms as well. 

\textit{Uncertainties}: The storage volume scales linearly with the annual experiment run time (95 days),
the experiment duty cycle (80\,\%), the sustained interaction rate ($3.75 \cdot 10^6$ / s in the hadron
and muon setups, $7.5 \cdot 10^4$ / s in the electron setup), and the raw data event sizes, in particular
that of the electron setup (see Table~\ref{tab:eventsize_electron}). It further depends on the splitting
of the beam time between the setups, which will finally be decided on the base of physics objectives.

It is obvious that un-triggered data taking with the electron setup is the most consumptive case.
We are currently investigating the possibility of a trigger for intermediate-mass electron pairs ($1 \, \mathrm{GeV} < m_{\mathrm{inv}} < 3 \, \mathrm{GeV}$), where the background level is lower. If such a trigger signature is identified, the possibility for dedicated runs for this physics objective will open up. The storage requirements, however, will not change significantly, since such runs would be taken at higher interactions rates, compensating the gain by online data reduction, such that better event statistics are obtained with the same amount of storage resources.

The requested event statistics for di-electron measurements is  $8.6 \cdot 10^{10}$ events per
beam energy,
which would be accommodated in the above scenario, measuring one beam energy
per year. Similar estimates for other physics cases are not yet available. 

In the sense that the above estimate was calculated for the largest collision system for CBM at SIS-100, and that lower collision
energies or smaller systems consume less disk space, it can
be regarded as an upper limit.
A better assessment necessitates
the evaluation of required event statistics for our main physics objectives, which we target for
in the near future.

\paragraph{Additional disk storage during data taking}

For the highly selective triggers sensitive to the decay topology (hadron setup), 
delayed event filtering with intermediate transient file storage must not be excluded (see section~\ref{subsec:delayed_filter}). 
Assuming a first-level data reduction by a factor of ten before intermediate storage
and a storage capacity for seven days, we require an additional disk storage capacity of {14~PB}.

\paragraph{Derived data}
There will be no ESD-type data for permanent storage (see section~\ref{subsec:data_formats}). 
The AOD data volume (single copy) is estimated to be 20\,\% of the raw data volume; it thus amounts to
{4.5 PB} per year of operation. 
AODs have to be kept on disk at least five years after data taking (see section~\ref{subsec:data_formats}.

\paragraph{Simulation data}
CBM aims at simulating as many events as recorded (triggered) by the experiment
(see Table~\ref{tab:storage}). For the majority of these simulated data, AODs will be stored and
distributed for user analysis. The requested storage capacity for simulated AODs is hence
equivalent to that of experiment data ({4.5~PB} per year). No simulated raw data will be stored
(see section~\ref{subsec:simulation}). Simulated AODs have to be kept on disk for the same 
period as the corresponding experiment AODs, i.e., for at least five years.

For 10\,\% of the simulated event statistics, the MC data will be stored for the detailed analysis
of efficiencies and backgrounds using the full MC truth information. 
With the MC event sizes given in Table~\ref{tab:mctime}, the corresponding storage volume is
{0.7~PB} for the hadron setup, {5.0~PB} for the electron setup and {0.7~PB} for the muon setup. The
total amount is {6.4~PB} per year.
MC data have to be kept on disk for {four} years after their production. 

Because of reproducibility, there is no need for long-term archiving of simulated data, neither MC nor AOD.

\subsubsection{Resource estimates: bandwidths}

\paragraph{Data rate from experiment to Green Cube}

Table~\ref{tab:datarates} show the data rates from the experiment (Entry Cluster) to the Green Cube (Compute Cluster) for the various setups and run conditions. The numbers follow from the considerations in section~\ref{sec:operation}.
It is assumed that averaging over the machine duty cycle will not happen before the Green Cube, such that the average in-spill interaction rates have to be applied.  

\renewcommand{\arraystretch}{1.5}
\begin{table}[htp]
\caption{Raw data rates from the experiment to the Green Cube}
\begin{center}
\small
\begin{tabular}{|c|c|r|}
\hline
Setup  &  Average interaction rate &  Data rate to GC \\
\hline \hline
hadron        &  $5 \cdot 10^6$ / s      & 244 GB/s  \\
electron      &  $1 \cdot 10^5$ / s      &     7 GB/s  \\
muon          & $5 \cdot 10^6$ / s      &   140 GB/s  \\
\hline
\end{tabular}
\end{center}
\label{tab:datarates}
\end{table}
\renewcommand{\arraystretch}{1}

The bandwidth requirement is hence set by the hadron setup. The dark rate in the STS is estimated to be
9.6~GB/s at a threshold-to-noise ratio of four, which is less than 10\,\% of the STS event data rate.
Assuming similar conditions for TRD and TOF, an upper limit for the total data rate to the Green Cube
is 270~GB/s.

\textit{Uncertainties}: The data rate scales linearly with the average interaction rate ($5 \cdot 10^6$/s in the hadron setup) and with the event raw data size (see Table~\ref{tab:eventsize_hadron}).

Applying a contingency factor of 1.5, we arrive at a bandwidth requirement of 400 GB/s.

Taking into account an average utilisation factor of $<$ 1 and a sufficient safety factor, the current
FLES design foresees about 120 optical fibres (Infiniband HDR) from the experiment to the Green Cube.

\paragraph{Bandwidth to permanent storage}
Since for the hadron setup and the muon setup, selective triggers reducing the raw data rate by factors of 100 or more will be applied, the limiting case for the archival bandwidth is given by the electron setup,
where no physics trigger will be applied.
The sustained data rate to storage is 5.5~GB/s. 
This number refers to the average over the machine duty cycle, which will be established in the Compute
Cluster (see section~\ref{subsec:dataflow}). Since no reconstruction will be run online, variations in the online processing time are estimated to be small, too. 

\textit{Uncertainties}: The number scales linearly with the average interaction rate ($10^5$ / s), 
the machine duty cycle (75\,\%) and the event raw data size 
(see Table~\ref{tab:eventsize_electron}).

Applying a  contingency factor of 1.5, we require a peak bandwidth to permanent storage
of  8~GB/s.

\paragraph{Bandwidth to transient storage}
If the concept of delayed event filtering involving transient disk storage is followed
(see section~\ref{subsec:delayed_filter}), sufficient bandwidth to the storage cluster
must be foreseen.
Relevant for this consideration is the sustained data rate from the hadron setup after an online first-level data reduction
by a factor of ten, which amounts to 18.3~GB/s.
It must be noted that in this concept, the filtering of data from transient storage to permanent
storage proceeds in parallel to the experiment and at the same speed, such that a similar
read bandwidth is necessary.

\subsubsection{Summary}

Based on the present knowledge on operation conditions, experiment design and data processing concept, the CBM collaboration estimate to require for the operation of the experiment at SIS-100 the resources summarised in Tables~\ref{cbm_cr}, \ref{cbm_sr1} and \ref{cbm_sr2}.


\begin{table}[h]
\begin{center}
\caption{CBM: Compute Requirements (in HEPSpec06)}
\begin{tabular}{|c|c|c|}
\hline
Compute Class  &  FS+  &  MSVc  \\
\hline
I.a & 6,000 & 6,000 \\
\hline
I.b & 0 & 0  \\
\hline
I.c & 0 & 0  \\
\hline
I.d & 0 & 0  \\
\hline
II.a & {780,000} &  {780,000} \\
\hline
II.b & 980,000& 980,000   \\
\hline
\end{tabular}
\label{cbm_cr}
\end{center}
\end{table}%

\begin{table}[h]
\begin{center}
\caption{CBM: Storage Requirements (I \textemdash Data Taking).}
\begin{tabular}{|c|c|c|c|}
\hline
&   &  \multirow {2} {2 cm}{FS+}   &   \multirow {2} {2cm}{MSVc} \\
& & & \\
\hline
   \multirow {2} {3.5 cm}{ Experiment to GreenCube/RZ1} & \#fibers & 120 & 120 \\
\cline{2-4}
&     Bandwidth (GB/s) & 400 & 400  \\
\cline{2-4}
\cline{1-2}
\multirow {2} {*} {Bandwidth to permanent storage} & Peak (GB/s) & 8 & 8 \\
\cline{2-4}
&   Average (GB/s) & 5.4 & 5.4 \\
\hline
\multicolumn {2} { |c| }{  Permanent storage/year (TB/year)} & {22,500} &  {22,500}\\
\hline
\multicolumn {2} { |c| } {Additional disk storage (TB)} & {14,000} & {14,000} \\

\hline
\end{tabular}
\label{cbm_sr1}
\end{center}
\end{table}%

\begin{table}[h]
\begin{center}
\caption{CBM: Storage Requirements (II \textemdash Processing).}
\begin{tabular}{|c|c|c|c|}
\hline
&   &  \multirow {2} {2 cm}{FS+}   &   \multirow {2} {2cm}{MSVc} \\
& & & \\
\hline
 \multirow {3} {3.5 cm}{ Raw Data} & TB/year & {22,500} & {22,500} \\
\cline{2-4}
&    \#years &2 & 2 \\
\cline{2-4}
&  Bandwidth (MB/s) & 20,000  & 20,000 \\
\hline
\multirow {3} {3.5 cm}{ Simulation} & TB/year & {11,000} & {11,000}  \\
\cline{2-4}
&    \#years &4 & 4 \\
\cline{2-4}
&  Bandwidth (MB/s) & 10,000 & 10,000 \\
\hline

\multirow {3} {3.5 cm}{ Derived data} & TB/year & {4,500} & {4,500} \\
\cline{2-4}
&    \#years & 5 & 5  \\
\cline{2-4}
&  Bandwidth (MB/s) & 40,000 & 40,000 \\
\hline
\end{tabular}
\label{cbm_sr2}
\end{center}
\end{table}%

\subsubsection{Timeline}

The resource estimates given above were derived for full CBM operation at design specifications. Obviously, the requirements differ strongly between the setups: while the (untriggered) electron setup does not put high demands in terms of online compute power, it has the strongest requirements in terms of storage capacity. The situation for the hadron and muon setups with highest interaction rates and highly selective online triggers is reverse. An annual breakdown of the resource demands requires the definition of physics goals for a given run period, which cannot be detailed at present. The assumption that the annual beam time is equally divided between the three setups thus provides a long-term average for the resource requirements.

According to the current planning, CBM will start operations in 2028 by commissioning with beam. The first physics beam time is anticipated for 2029. In this first data taking year, the electron setup will be used with interaction rates of $10^5$ / s. High-rate operation of up to $10^7$ interactions / s will be gradually approached within the next two years, accounting for a learning curve in view of the challenging goal. Consequently, the full online compute power for real-time data reduction will be required from 2030 on, while the annual storage capacity is needed from the start of operations.

\newpage
\subsection{HADES}

Different operation scenarios have to be distinguished: 

\begin{enumerate}
\item Data Taking during beam times: HADES data are recorded on dedicated
servers ( $\sim$10) on experiment side. Data rates are in the order of 200-500 Mb/s.
The data are stored in binary list mode data format (hld) to a local disk array
and simultaneously written to the tape archive (tsm) and to the /lustre file system
on the common compute cluster to allow for a "real time" online DST production
with only a small delay. For the online DST production 1000 cores on the farm
are sufficient. The typical data volume on raw data can reach 350 TB during an 4 weeks
experiment.
\item Archiving of RAW data: Hld raw data are directly written to archive during beam time.
The tape archive has to provide the rate capability to archive close to real time.
The permanent longterm backup of experimental data is required by law. Backup media
are to be paid by HADES and include the costs for archiving infrastructure.
\item DST production: The DST production is CPU heavy with only marginal I/O per job
(~0.4Mb/s read, 0.3Mb/s write). Jobs on the batch farm are organized file by file
in parallel. To finish the production in a few days a decent amount of jobs 
(5000-6000) is needed.
\item Simulation of experimental data: Typical event generators as UrQMD etc. usually
are compute heavy with little I/O. DST production for simulation therefore share the
requirements with data DST production.
\item Archiving of DST data: The final DST which are used for publications are archived
to tape. This operation is not time critical.
\item Analysis of DST data by HADES users: The typical analysis of HADES DST data
is I/O heavy, mainly in reading the input DST data. The load in terms cores on the batch farm
is lower as compared to DST production (2000-3000).
\item Disk space: In HADES the DST production and simulations are organized in an common
approach and require the most diskspace. The space requirements will typically grow with 
each new experiment data. Analysis takes usally years and will lead to overlapping
campaigns on the disk. Simulations will take at least the same amount as the raw data.
\end{enumerate}


\begin{table}[h]
\begin{center}
\caption{HADES: Compute Requirements (in HEPSpec06).}
\begin{tabular}{|c|c|c|}
\hline
Compute Class  &  FS(+) configuration  &  MSVc configuration  \\
\hline
I.a & 2,200 & 0 \\
\hline
I.b & 0 & 0  \\
\hline
I.c & 0 & 0  \\
\hline
I.d & 0 & 0  \\
\hline
II.a & 36,000 &  36,000 \\
\hline
II.b & 22,000& 22,000   \\
\hline
\end{tabular}
\label{hades_cr}
\end{center}
\end{table}%

\begin{table}[h]
\begin{center}
\caption{HADES: Storage Requirements (I \textemdash Data Taking).}
\begin{tabular}{|c|c|c|c|}
\hline
&   &  \multirow {2} {2 cm}{FS(+)}   &   \multirow {2} {2cm}{MSVc configuration} \\
& & & \\
\hline
   \multirow {2} {3.5 cm}{ Experiment to GreenCube/RZ1} & \#fibers & 0 & 0 \\
\cline{2-4}
&     Bandwidth (MB/s) & 500 & 500  \\
\cline{2-4}
\cline{1-2}
\multirow {2} {*} {Bandwidth to permanent storage} & Peak (MB/s) & 500 & 500 \\
\cline{2-4}
&   Average (MB/s) & 200 & 200 \\
\hline
\multicolumn {2} { |c| }{  Permanent storage/year (TB/year)} & 500 &  500\\
\hline
\multicolumn {2} { |c| } {Additional disk storage (TB)} & 0 & 0 \\

\hline
\end{tabular}
\label{hades_sr1}
\end{center}
\end{table}%

\begin{table}[h]
\begin{center}
\caption{HADES: Storage Requirements (II \textemdash Processing).}
\begin{tabular}{|c|c|c|c|}
\hline
&   &  \multirow {2} {2 cm}{ Day 1 configuration}   &   \multirow {2} {2cm}{Full MSV configuration} \\
& & & \\
\hline
 \multirow {3} {3.5 cm}{ Raw Data} & TB/year & 350 & 350 \\
\cline{2-4}
&    \#years &5 & 5 \\
\cline{2-4}
&  Bandwidth (MB/s) & 200  & 500 \\
\hline
\multirow {3} {3.5 cm}{ Simulation} & TB/year & 300 & 300  \\
\cline{2-4}
&    \#years &3 & 3\\
\cline{2-4}
&  Bandwidth (MB/s) & 6,000& 6,000\\
\hline

\multirow {4} {3.5 cm}{ Derived data} & TB/year & 400 & 400 \\
\cline{2-4}
&    \#years & 3 & 3  \\
\cline{2-4}
&  \multirow {2}{*} {Bandwidth (MB/s)} & 6,000(DST),  & 6,000(DST), \\
&   &  $>$40,000(Analysis), &  $>$40,000(Analysis) \\

\hline
\end{tabular}
\label{hades_sr2}
\end{center}
\end{table}%

\newpage

\subsection{NUSTAR}

   As explain in the executive summary the NUSTAR experiments at FAIR are composed of modular setups serving difference scientific sub-communities. This creates different challenges to solve compared to more monolhitic approaches. We can subdivide the NUSTAR experiments in the FAIR facility in two different classes: 
   \begin{itemize}
       \item Those mounted alongside or directly at the end of the Super-FRS thus using the same data acquisition and data processing paradigms, this corresponds to the collaborations HISPEC/DESPEC, R3B and the Super-FRS EC (Experimental Collaboration). In this case the data flow of the experiment is directly merged with the one of the detectors of Super-FRS. To assure smooth operation, easily online durring experiment and simple compatiliby those experiments will follow the NUSTAR DAQ. 
       \item The experiment(s) mounted directly at rings. In the case of the MSVc this includes only ILIMA but could be expanded post MSVc to ELiSe and EXL (not part of the discussion in this IT TDR). In this case the data flow scheme and data acuiqisiton are not integrated in the NUSTAR DAQ so computer paradigm is a bit different
   \end{itemize}

   We present below the merged tables for the computer requirement for all NUSTAR operation as foreseen inside of the scope of the TDR, i.e. including NUSTAR at Rings and NUSTAR at Super-FRS. Then more details are given in the two types of experiments considered here, each with their own estimate for information. Those numbers are estimates and for the two last tables, the estimates from ILIMA being the size of the error bar of the estimates of the ``NUSTAR Experiments directly at Super-FRS'' the tables for full NUSTAR and ``NUSTAR experiments directly at Super-FRS'' show the same values.

\begin{table}[h]
\begin{center}
\caption{Full NUSTAR: Compute Requirements (in HEPSpec06).}
\begin{tabular}{|c|c|c|}
\hline
Compute Class  &  FS(+)  &  MSVc \\
\hline 
I.a & 12,000 & 24,000 \\
\hline
I.b & 0 & 0  \\
\hline
I.c & 0 & 0  \\
\hline
I.d & 0 & 0  \\
\hline
II.a & 70,000 & 205,000 \\
\hline
II.b & 60,000 & 150,000 \\
\hline
\end{tabular}
\label{nustar_cr_sfrs2}
\end{center}
\end{table}%

\begin{table}[h]
\begin{center}
\caption{Full NUSTAR: Storage Requirements (I \textemdash Data Taking).}
\begin{tabular}{|c|c|c|c|}
\hline 
&   &  \multirow {2} {2.5 cm}{FS(+)}   &   \multirow {2} {2.5 cm}{MSVc} \\
& & & \\
\hline
   \multirow {2} {3.5 cm}{ Experiment to GreenCube/RZ1} & \#fibers & 12 & 12 \\
\cline{2-4}
&     Bandwidth (MB/s) & 500 & 500 \\
\cline{2-4}
\cline{1-2}
\multirow {2} {*} {Bandwidth to permanent storage} & Peak (MB/s) & 1,000 & 1,000\\
\cline{2-4}
&   Average (MB/s) & 500 & 500 \\
\hline
\multicolumn {2} { |c| }{  Permanent storage/year (TB/year)} & 5,000 & 5,000 \\
\hline
\multicolumn {2} { |c| } {Additional disk storage (TB)} & 500 & 500 \\
\hline
\end{tabular}
\label{nustar_sr1}
\end{center}
\end{table}%
\begin{table}[hbt!]
\begin{center}
\caption{Full NUSTAR: Storage Requirements (II \textemdash Processing).}
\begin{tabular}{|c|c|c|c|}
\hline 
&   &  \multirow {2} {2.5 cm}{FS(+)}   &   \multirow {2} {2.5 cm}{MSVc} \\
& & & \\
\hline
 \multirow {3} {3.5 cm}{ Raw Data} & TB/year & 5,000 & 5,000 \\
\cline{2-4}
&    \#years & 5 & 5  \\
\cline{2-4}
&  Bandwidth (MB/s) &10,000  & 10,000 \\
\hline

\multirow {3} {3.5 cm}{ Simulation} & TB/year & 500 & 500  \\
\cline{2-4}
&    \#years & 5 & 5 \\
\cline{2-4}
&  Bandwidth (MB/s) & 1,000 & 1,000\\
\hline
\multirow {3} {3.5 cm}{ Derived data} & TB/year &1,250 & 1,250 \\
\cline{2-4}
&    \#years & 5 & 5  \\
\cline{2-4}
&  Bandwidth (MB/s) & 10,000 & 10,000 \\
\hline
\end{tabular}
\label{nustar_sr2}
\end{center}
\end{table}%

  Those numbers are estimates and for the two last tables, the estimates from ILIMA being the size of the error bar of the estimates of the "NUSTAR Experiments directly at Super-FRS" the tables for full NUSTAR and "NUSTAR experiments directly at Super-FRS" show the same values.

\subsubsection{NUSTAR experiment directly at Super-FRS}

  The three sub-collaborations (HISPEC/DESPEC, R3B and Super-FRS EC) have in common to each have their own suits of different modular setups, but they all operate several multi-detectors few $10^{4}$ channels. The experiments being operated directly in or at the end of the Super-FRS the three collaboration rely on a complete data flow merging between their detectors and the Super-FRS. This is of outmost importnace because durring a typical few weeks campaign using one given setup several smaller experiments of few days, with quite different beams (primary and secondary) will run. To assure the quality of the data taken a quick an efficient online is mandatory.  
  To tackle the problem of integrating those highly modular setups the NUSTAR data acquisition framework was developed to be used by the three collaborations and the Super-FRS spectrometer (NUSTAR DAQ \cite{ndaq:tdr}). In this scheme the raw time stamped data is sent to storage, local and archiving, A part of it is directly processed online for monitoring while a quick near line analysis is performed on the stored raw data during the experiments. During the running of an experiment some IT resources should be reserved for the online/near line analysis of the running experiment.

   In a typical NUSTAR approach the analysis of the data is done but requires only few times to process all the data. This can change between communities and setups but those experiments do not constantly process the stored data. Which means there is a request that processed data be somehow archived in medium term storage, or systematically backed up. This could be for ease of use (several people working on the same processed data, weeks could be lost if a mistake is made and the data is deleted) or for proper scientific practice to for example keep a level of  processed data which was used from publication.

  Because we can not predict which sub-collaboration with which detector will be running at a given time the numbers are estimated on an average deduced on the FAIR phase 0 operation, or previous campaigns (like AGATA~\cite{ref:agata} for HISPEC, WASA~\cite{ref:wasa} for Super-FRS EC or standard R3B setup). In the estimate and projection we end up with the same order of magnitude for the data flow size for each sub-collaboration within one year (supposing their more complex setup runs regularly).

\paragraph{Computing power}

The total computing power for operations and analysis of those experiments is expected to be approximately 5500 cores. Up to 10\% of these will be placed close to the setups. A decent part of the remaining resources need to be located on-site and should be reserved during runs for online and near line analysis. The roughly similar value for Type II.a and Type II.b comes from the fact that recently acquired data operated in II.b is processed more often that data stored since longer, which will still processed few times during the analysis.

Besides data processing, experiment control and data analysis, the requested computing power is also needed for modelling and simulation of the experiments. Thus the projected resource usage will be mostly uniform throughout the year.

\paragraph{Permanent storage}

Experiment and simulation data on disk are typically needed for a duration of five to ten years to assure proper analysis with an annual production of about 5000 TByte. The collected raw data must be stored on tape for long term storage.

It is required that access to data in long term storage is available for all involved scientists in a consistent and reliable way across all of NUSTAR experiments. Documentation for all related procedures must be freely available.

\paragraph{User access}

Users should be able to access the GSI/FAIR computing resources via centrally managed computers running the Linux operating systems, including (a few) desktop PCs extended by making use of virtual machines. The general availability of such terminals has proven very valuable in the past, and it is a must for operation of the spectrometer at the center of the NUSTAR experiments (FRS and Super-FRS). Indeed a strong link exist between the online analysis from detectors used in the experiment and the machine tuning of the spectrometers. To achieve this reliably standardized computers for operation are needed. Access to these computers will follow the general AAI policies as described in Sec.~\ref{sec:aai} to ease access for users, to centralize the administration, and to avoid any insecure connections from the outside.

For better understanding, in the below we present the estimated values for NUSTAR at Super-FRS only. Please note that these values are already included in tables \ref{nustar_cr_sfrs2}, \ref{nustar_sr1} and \ref{nustar_sr2}.

\newcommand{\redparagraph}[1]{\paragraph{\color{red} #1}}

\begin{table}[h]
\begin{center}
\caption{NUSTAR-Super-FRS (only): Compute Requirements (in HEPSpec06).}
\begin{tabular}{|c|c|c|}
\hline
Compute Class  &  FS(+) &  MSVc\\
\hline 
I.a & 12,000 & 24,000 \\
\hline
I.b & 0 & 0  \\
\hline
I.c & 0 & 0  \\
\hline
I.d & 0 & 0  \\
\hline
II.a & 60,000 & 150,000 \\
\hline
II.b & 60,000 & 150,000 \\
\hline
\end{tabular}
\label{nustar_cr_sfrs1}
\end{center}
\end{table}%

\begin{table}[h]
\begin{center}
\caption{NUSTAR-Super-FRS (only): Storage Requirements (I \textemdash Data Taking).}
\begin{tabular}{|c|c|c|c|}
\hline 
&   &  \multirow {2} {2.5 cm}{FS(+)}   &   \multirow {2} {2.5 cm}{MSVc} \\
& & & \\
\hline
   \multirow {2} {3.5 cm}{ Experiment to GreenCube/RZ1} & \#fibers & 12 & 12 \\
\cline{2-4}
&     Bandwidth (MB/s) & 500 & 500 \\
\cline{2-4}
\cline{1-2}
\multirow {2} {*} {Bandwidth to permanent storage} & Peak (MB/s) & 1,000 & 1,000\\
\cline{2-4}
&   Average (MB/s) & 500 & 500 \\
\hline
\multicolumn {2} { |c| }{  Permanent storage/year (TB/year)} & 5,000 & 5,000 \\
\hline
\multicolumn {2} { |c| } {Additional disk storage (TB)} & 500 & 500 \\
\hline
\end{tabular}
\label{nustar_sr3}
\end{center}
\end{table}%
\begin{table}[h]
\begin{center}
\caption{NUSTAR-Super-FRS (only): Storage Requirements (II \textemdash Processing).}
\begin{tabular}{|c|c|c|c|}
\hline 
&   &  \multirow {2} {2.5 cm}{FS(+)}   &   \multirow {2} {2.5 cm}{MSVc} \\
& & & \\
\hline
 \multirow {3} {3.5 cm}{ Raw Data} & TB/year & 5,000 & 5,000 \\
\cline{2-4}
&    \#years & 5 & 5  \\
\cline{2-4}
&  Bandwidth (MB/s) &10,000  & 10,000 \\
\hline

\multirow {3} {3.5 cm}{ Simulation} & TB/year & 500 & 500  \\
\cline{2-4}
&    \#years & 5 & 5 \\
\cline{2-4}
&  Bandwidth (MB/s) & 1,000 & 1,000\\
\hline
\multirow {3} {3.5 cm}{ Derived data} & TB/year &1,250 & 1,250 \\
\cline{2-4}
&    \#years & 5 & 5  \\
\cline{2-4}
&  Bandwidth (MB/s) & 10,000 & 10,000 \\
\hline
\end{tabular}
\label{nustar_sr4}
\end{center}
\end{table}%

\clearpage
\subsubsection{NUSTAR in Rings: ILIMA}

As a part of NUSTAR collaboration, ILIMA sub-collaboration focuses on measurements of mass and lifetimes of exotic nuclei at FAIR, with main experimental areas being ESR and later Super-FRS and Collector Ring (CR). Computing requirements focus on long-term storage, highly reliable network infrastructure and shared cluster computing for data analysis and theory calculations. In terms of data handling, the planned experiments can be characterised into two different classes.

\begin{enumerate}

    \item The first class represents the event-based DAQ systems, which consist of one or more nodes for the different detector systems, depending on the physics case. Examples are TOF and DSSSD detectors at ESR and CR. Data pre-processing will be performed close to the detector systems, resulting in an average raw data transfer rate (list mode) of 5-50 MB/s to the computing cluster and long-term storage. It is planned to perform cluster computing on particle detector data in the future, similar to the example below (Schottky detectors), but the requirements will be very moderate, almost negligible.

    \item The second group are the Schottky detector DAQ systems, which either continuously or shot by shot sample the pickup signal from resonant Schottky detectors, monitoring the electromagnetic wave driven by circulating ions or ion bunches in a storage ring. The sensitive data analysis in this case depends on absolute dead time free signal sampling at a data rate of approximately 5 Gb/s. To avoid any data loss due to network latency, a temporary local storage solution with a 2 x 10Gb/s connection is attached to the DAQ. Subsequently, the raw experimental data of about 50TB/week will have to be transferred to the long-term storage at a reasonable transfer rate. Future detector development at the ESR would increase the amount of data required. Such detectors would include position sensitive resonant pickups to increase mass and lifetime measurements using isochronous velocity correction.

    Time domain Schottky data are processed into intermediate result files in frequency domain, as well as as combinations and second order derivatives of these data. For this step parallel computing in Green Cube is used. For a given parameter set, up to 10000 jobs are used each with up to 100 GB peak memory usage. The amount of intermediate data are in the same order as the original data , i.e. approx. 1:1. Such calculations are repeated many times during the analysis time throughout the year. Intermediate data will not be stored on the long term storage, but may be needed on the LUSTRE for the duration of analysis.
    
\end{enumerate}

Due to huge overlap of experimental scenarios, the computing resources needed for the periods of FS and FS+(+) will be covered by atomic physics as explained in section \ref{atomic_physics}. In the MSVc period, requirements for additional detectors hugely increases the computation and resource requirements due to addition of several new Schottky detectors as well as TOF and pocket detectors.


For better understanding, in the below we present the estimated values for NUSTAR at storage rings i.e. ILIMA only. Please note that these values are already included in tables \ref{nustar_cr_sfrs2}, \ref{nustar_sr1} and \ref{nustar_sr2}.

\begin{table}[ht]
\begin{center}
\caption{NUSTAR-ILIMA (only): Compute Requirements (in HEPSpec06).}
\begin{tabular}{|c|c|c|c|}
\hline
Compute Class  &  FS / FS+(+) & MSVc  \\
\hline
I.a & 0 & 0 \\
\hline
I.b & 0 & 0 \\
\hline
I.c & 0 & 0 \\
\hline
I.d & 0 & 0 \\
\hline
II.a & 0 &  55,000 \\ 
\hline
II.b & 0 & 0  \\
\hline
\end{tabular}
\label{appa_ilima_cr}
\end{center}
\end{table}%

\begin{table}[ht]
\begin{center}
\caption{NUSTAR-ILIMA (only): Storage Requirements (I \textemdash Data Taking).}
\begin{tabular}{|c|c|c|c|}
\hline
&   &  \multirow {2} {2 cm}{FS / FS+(+)}   &   \multirow {2} {2cm}{MSVc} \\
& & & \\
\hline
   \multirow {2} {3.5 cm}{ Experiment to GreenCube/RZ1} & \#fibers & 0 & 0 \\
\cline{2-4}
&     Bandwidth (MB/s) & 0 & 100  \\
\cline{2-4}
\cline{1-2}
\multirow {2} {*} {Bandwidth to permanent storage} & Peak (MB/s) & 0 & 100 \\
\cline{2-4}
&   Average (MB/s) & 0 & 50 \\
\hline
\multicolumn {2} { |c| }{  Permanent storage/year (TB/year)} & 0 & 400 \\
\hline
\multicolumn {2} { |c| } {Additional disk storage (TB)} & 0 & 50 \\

\hline
\end{tabular}
\label{appa_ilima_sr1}
\end{center}
\end{table}%

\begin{table}[ht]
\begin{center}
\caption{NUSTAR-ILIMA (only): Storage Requirements (II \textemdash Processing).}
\begin{tabular}{|c|c|c|c|}
\hline
&   &  \multirow {2} {2 cm}{FS/FS+(+)}   &   \multirow {2} {2cm}{MSVc} \\
& & & \\
\hline
 \multirow {3} {3.5 cm}{ Raw Data} & TB/year & 0 & 400 \\
\cline{2-4}
&    \#years & 0 & 5 \\
\hline
\multirow {3} {3.5 cm}{ Simulation} & TB/year & 0 &  10  \\
\cline{2-4}
&    \#years &0 & 5 \\
\hline

\multirow {3} {3.5 cm}{ Derived data} & TB/year & 0 & 20 \\
\cline{2-4}
&    \#years & 0 & 5  \\
\hline
\end{tabular}
\label{appa_ilima_sr2}
\end{center}
\end{table}%

\clearpage

\newpage

\subsection{PANDA}

\subsubsection{General comments}

The antiproton branch of the FAIR facility, with PANDA as its main experiment, is part of the MSVc stage of the construction schedule of FAIR and therefore will not be available before 2032. Because of the long time scale and still open questions about the operation mode at the beginning of the experiment, all the numbers given for PANDA are subject to large uncertainties with the potential to significantly increase or decrease the required computing needs. All the numbers given inside this chapter are based on the current status of the software development and an assumed luminosity of $1\cdot 10^{31} cm^{-2}s^{-1}$.

In addition to the computing demands to run the experiment, once it is available, further resources are needed a couple of years before the start of PANDA to perform detailed simulation campaigns to prepare the data taking and speed up the generation of physics output. These requirements are listed in the column for FS(+).

\subsubsection{Compute Requirements}

To determine the compute requirements of the PANDA experiment different operation scenarios have to be distinguished: online processing of the raw detector data coming from the DAQ stage with the same rate as it is produced by the experiment, offline reprocessing of the raw data and simulation of the complete experiment. In these different scenarios different processing stages have to be performed. The different stages are part ofthe simulation and reconstruction software of the PANDA experiment, PandaRoot:

\begin{enumerate}
\item Event Generation: The primary interaction of the antiproton beam with the target material is simulated. This is done by the usage of external event generators which are different for different purposes e.g.: EvtGen for dedicated physics channels, DPM and FTF as generators for the most common background processes and particle guns for specific type of particles. The output of this stage are origin, type and momentum of the generated particle but no detector material nor magnetic fields are taken into account.
\item Simulation: Particles generated on the Event Generation are propagated through the detector, taking into account their interaction with the material of the detector or the magnetic fields and their possible decay into secondary particles. The propagation is done using the external software Geant3 or Geant4. The output of this stage is a set of Monte-Carlo tracks and Monte-Carlo points which have been created. A Monte-Carlo track contains the information about a generated particle (position, time, type and momentum). A Monte-Carlo point holds the information where and when a detector was hit by a particle and how much energy was deposited in the detector. Often Event Generation and Simulation are combined in one processing stages as it is done for the compilation of the compute requirements.
\item Digitization: From the Monte-Carlo points a detector response is calculated taking into account the physical processes in the detector material as well as the electronic processes necessary to amplify and digitize the data. The output is an individual data-set of digis for each sub-detector with hit information looking as it would be raw data from the experiment.
\item Local Reconstruction: Each sub-detector calculates from the digis the actual physical information measured by the detector. For a tracking detector this would be a space point, time and deposited energy.
\item Tracking: Combination of all data from the different tracking detectors. In the first stage the hits are sorted into groups which belong to the same track (track finding) including a rough track fitting. In the second stage the tracks are fitted with a Kalman based track fitter which respects the magnetic filed as well as the detector material to retrieve the precise track parameters: position on track and momentum vector of the track. The output of this stage is a reconstructed track with the hits belonging to the track.
\item Particle Identification (PID): The track data is combined with the hit information of the different PID detectors and calorimeters and the particle type is determined. The output of this stage is a charged or neutral particle candidate.
\item Event Building and Filtering: In the last stage the particle candidates are grouped into events which contain all particle candidates coming from the same antiproton-target interaction. In the online processing of the data the events are roughly analyzed and only those which seem to be interesting stored for further analysis.
For the offline processing and simulation stage the event data serves as the input into the physics analysis which tries to select interesting signal channels out of the huge background and to determine the properties of the selected particles.
\end{enumerate}

The individual times per processing stage are summarized in table \ref{panda_processing_times}. They have been achieved by the mean values of different simulation runs with various numbers of events for 15 GeV/c beam momentum and the FTF background generator. The processing was performed on a dedicated machine at GSI which a given HEPSpec06 value of 22 per core provided by the FAIR IT department. The most time of almost 3 seconds per event is needed for the propagation of the particles through the detector in the simulation stage. All other stages need about or less than 0.1 s per event. The only exception is the particle tracking with the Kalman filter with 0.64 seconds. The Kalman filter tries to determine the particle track with the highest resolution taking into account the detailed geometry of the detector and the magnetic fields. For online processing this time might be too much and the achieved precision might not be needed. Therefore also the tracking time without Kalman filter is given which is about a factor 10 less than with the filter.

\begin{table}[h]
\begin{center}
\caption{Time per processing stage in PandaRoot.}
\begin{tabular}{|c|c|c|c|c|c|c|}
\hline
Sim & Digi & Reco & Tracking& Tracking & PID & Event \\
& & & wo Kalman & with Kalman & & Filter \\
\hline
2.86 s & 0.07 s & 0.05 s & 0.06 s & 0.64 s & 0.12 s & 0.01 s \\
\hline
\end{tabular}
\label{panda_processing_times}
\end{center}
\end{table}%

Depending on the operation status three different scenarios are considered:

\begin{itemize}
\item Online: Processing of the data coming directly from the FPGA based DAQ system. The data has to be processed with highest speed and reduced precision for event selection. The processing stages are: local reconstruction, tracking without Kalman filter, particle identification and event filtering. It is assumed that the event rate was reduced by the FPGA-stage by a factor of 10.
\item Offline: Reprocessing of the experiment data with highest precision. The processing stages are: local reconstruction, tracking with Kalman filter, particle identification. It is assumed that the online processing has reduced the data rate by another factor of 10.
\item Simulation: Simulation of the experiment with highest precision. The processing stages are: event simulation, propagation, digitization, local reconstruction, tracking with Kalman filter, particle identification.
\end{itemize}

The combined processing times per event and the needed HEPSpec06 values for the different scenarios are given in table \ref{panda_time_per_event}. A complete simulation takes 3.7 s, the reprocessing of the raw data 0.81 s and the online processing 0.23 s per event.

\begin{table}[h]
\begin{center}
\caption{Computing time and corresponding HEPSpec06 values per event.}
\begin{tabular}{|c|c|c|}
\hline
& Time per event & HS06 per event \\
\hline
Online & 0.23 s & 5.06 \\
\hline
Offline & 0.81 s & 17.82 \\
\hline
Simulation & 3.7 s & 81.4 \\
\hline
\end{tabular}
\label{panda_time_per_event}
\end{center}
\end{table}

To calculate the needed compute power per year the following assumptions were made:
\begin{itemize}
\item PANDA has 100 effective days of data taking
\item 350 days of continuous CPU activity for offline/simulation
\item 100 days of continuous CPU activity for online processing
\item CPU efficiencies of 85 \% online/offline and 95 \% for simulation (describes the availability of cores)
\item Offline and simulation data is reprocessed four times a year
\end{itemize}

The result is shown in table \ref{panda_HS_year}. 

\begin{table}[h]
\begin{center}
\caption{PANDA: Compute Requirements per year or during experiment operation (in HEPSpec06).}
\begin{tabular}{|c|c|c|c|c|c|c|c|}
\hline
& \# events & HS06 & HS06 & CPU & HS06 & \# gen & HS06 \\
& per year & / event & / second & efficiency & per gen & & / year\\
\hline 
Online (80 kHz) & 7.0$\times$10$^{11} $  & 5.06 & 3.5$\times$10$^{12}$ & 85 \% & 480,000 & 1 & 480,000 \\
\hline
Offline (8 kHz) & 7.0$\times$10$^{10}$ & 17.82 & 1.2$\times$10$^{12}$ & 85 \% & 46,600 & 4 & 186,400\\
\hline
Sim (8 kHz) x 2 & 14$\times$10$^{10}$ & 81.4 & 11$\times$10$^{12}$ & 95 \% & 381,000 & 4 & 1,524,000\\
\hline
\end{tabular}
\label{panda_HS_year}
\end{center}
\end{table}

The largest compute resources are needed for the simulation of the experiment due to the large processing time per event of 3.7 s. The advantage is, that the time per simulation is very well known which is not true for the online case. Here the numbers calculated are connected with large uncertainties: The time needed for alignment and calibration and additional noise are not included, it is not clear if the suppression factor of 10 in the FPGA stage can be achieved and if it is possible to run online without a Kalman filter for tracking. Thus, a quite large safety factor is taken into account for the online processing of the data and 0.75 MHEPSpec06 are requested for the sum of online and offline processing and the same number for the simulation stage at the start of PANDA. A summary of the requested computing requirements divided into the different classes is given in table \ref{panda_cr}.


\begin{table}[h]
\begin{center}
\caption{PANDA: Compute Requirements (in HEPSpec06).}
\begin{tabular}{|c|c|c|}
\hline
Compute Class  &  FS+  &  MSVc  \\
\hline
I.a & 0 & 0 \\
\hline
I.b & 0 & 0  \\
\hline
I.c & 0 & 0  \\
\hline
I.d & 0 & 0  \\
\hline
II.a & 200,000 &  1,500,000 \\
\hline
II.b & 0 & 750,000   \\
\hline
\end{tabular}
\label{panda_cr}
\end{center}
\end{table}%

\subsubsection{Storage Requirements}

The storage requirements were calculated based on the simulation of FTF background events with an antiproton beam momentum of 15 GeV/c. For each processing stage the data size per event was calculated and based on this number the data size per second calculated. The results are summarized in table \ref{panda_fileSize}. For the simulation stage two numbers are given. The bigger one includes the Monte-Carlo data of both tracks and points while the smaller only contains track data. The difference between the two is with more than a factor 30 very large and the MC point information is in most cases not needed. Therefore it is foreseen to keep only the Monte-Carlo truth data of the tracks and not of the points. From the data sizes per second the amount of data per year is calculated based on the assumption of 100 days of effective data taking with an event rate to storage of 8 kHz.

\begin{table}[h]
\begin{center}
\caption{File size per event, second and operation year with 100 effective days.}
\begin{tabular}{|c|c|c|c|c|c|c|}
\hline 
File Size per & Sim & Sim & Digi & Reco & Tracking & PID \\
& track + point & only track & RAW & & & AOD \\
\hline
Event & 132.8 & 4.2 & 16.5 & 28.9 & 22.3 & 8.7 \\
$[$kByte$]$ & & & & & & \\
\hline
Second (8 kHz) & 1038 & 33 & 130 & 226 & 175 & 68 \\
$[$MByte$]$ & & & & & & \\
\hline
Year (100 d) & 8549 & 270 & 1063 & 1860 & 1435 & 560 \\
$[$TByte$]$ & & & & & & \\
\hline
\end{tabular}
\label{panda_fileSize}
\end{center}
\end{table}

To estimate the storage requirements for the experiment data it is assumed that after the online reconstruction only the raw data is stored. The size of this data corresponds to the digi column in table \ref{panda_fileSize} with 1.1 PByte/y. This data has to be stored in a secured manner (e.g. on tape storage) called RAW-COLD. In addition four consecutive years of raw data are kept in  a fast accessible storage space for reprocessing (in total 4.4 PByte) called RAW-HOT.
After the offline reconstruction of the raw data only the final output is kept for long term storage as input for future physics analysis. This data corresponds to the column PID in table \ref{panda_fileSize}, also labeled as Analysis Object Data (AOD). The AOD data of one year of data taking is reprocessed in the three following years with updated algorithms, calibration and alignment data and all four iterations are kept for further analysis. This causes a non-linear increase of the storage requirements for AOD data per year starting with 560 TByte in the first year, and rising up to 2240 TByte/y. The total amount of storage needed after 13 years sums up to 22.4 PByte for AOD data.

To estimate the amount of storage needed for simulated data it is assumed that twice as much events have to be simulated than recorded by the experiment.
The MC truth data for the tracks and the digi data is stored for four years for processing with a requirement of 2.7 PByte/y for four years. The AOD data coming from simulations is kept for the whole time with an increase of 1.2 PByte/y for ten years. In addition, 1 PByte of volatile storage space is requested for alignment, calibration and quality assurance.

The summary of the storage requirements is given in table \ref{panda_fileSizeSummary} and the requested storage requirements and bandwidth in tables \ref{panda_sr1} and \ref{panda_sr2}.

\begin{table}[h]
\begin{center}
\caption{PANDA: Storage estimate for different type of data.}
\begin{tabular}{|c|c|c|c|c|}
\hline 
  & min. increase & max. increase & years & total \\
  & per year & per year & &  \\
  & [PByte/y] & [PByte/y] & & [PByte] \\
\hline
RAW-COLD & 1.1 & 1.1 & 10 & 11 \\
\hline
RAW-HOT & 1.1 & 1.1 & 4 & 4.4 \\
\hline
AOD & 0.56 & 2.24 & 13 & 22.4 \\
\hline
\hline
SIM & 2.7 & 2.7 & 4 & 11 \\
\hline
AOD-SIM & 1.12 & 1.12 & 10 & 11.2 \\
\hline

\hline
\end{tabular}
\label{panda_fileSizeSummary}
\end{center}
\end{table}

\begin{table}[h]
\begin{center}
\caption{PANDA: Storage Requirements (I \textemdash Data Taking).}
\begin{tabular}{|c|c|c|c|}
\hline 
&   &  \multirow {2} {2 cm}{FS(+)}   &   \multirow {2} {2cm}{MSVc} \\
& & & \\
\hline
   \multirow {2} {3.5 cm}{ Experiment to GreenCube/RZ1} & \#fibers & 0 & 120 \\
\cline{2-4}
&     Bandwidth (GB/s) & 0 & 300 \\
\cline{2-4}
\cline{1-2}
\multirow {2} {*} {Bandwidth to permanent storage} & Peak (MB/s) & 0 & 2,000 \\
\cline{2-4}
&   Average (MB/s) & 0 & 1,000 \\
\hline
\multicolumn {2} { |c| }{  Permanent storage/year (TB/year)} & 0 & 1,200 \\
\hline
\multicolumn {2} { |c| } {Additional disk storage (TB)} & 0 & 1000 \\
\hline
\end{tabular}
\label{panda_sr1}
\end{center}
\end{table}%
\begin{table}[h]
\begin{center}
\caption{PANDA: Storage Requirements (II \textemdash Processing).}
\begin{tabular}{|c|c|c|c|}
\hline 
&   &  \multirow {2} {2 cm}{FS(+)}   &   \multirow {2} {2cm}{MSVc} \\
& & & \\
\hline

 \multirow {3} {3.5 cm}{ Raw Data} & TB/year & 0 & 1,200 \\
\cline{2-4}
&    \#years & 0 & 4  \\
\cline{2-4}
&  Bandwidth (MB/s) & 0  & 20,000 \\
\hline

\multirow {3} {3.5 cm}{ Simulation} & TB/year & 700 & 2,800  \\
\cline{2-4}
&    \#years &4 & 4 \\
\cline{2-4}
&  Bandwidth (MB/s) & 10,000& 20,000\\
\hline

\multirow {3} {3.5 cm}{ Derived data experiment} & TB/year & 0 & 2,240 \\
\cline{2-4}
&    \#years & 0 & 13  \\
\cline{2-4}
&  Bandwidth (MB/s) & 0 & 20,000 \\
\hline

\multirow {3} {3.5 cm}{ Derived data simulation} & TB/year & 560 & 1,120 \\
\cline{2-4}
&    \#years & 4 & 13  \\
\cline{2-4}
&  Bandwidth (MB/s) & 10,000 & 20,000 \\
\hline
\end{tabular}
\label{panda_sr2}
\end{center}
\end{table}%

\newpage

\clearpage

\subsection{THEORY}

\subsubsection{Theory activities and FAIR research lines}

This section summarizes the compute and storage requirements of the theory department. Particularly, we address those activities that strongly support FAIR experiments, {\it i.e.}, the CBM, NUSTAR, and PANDA pillars. An overview of the activities, organized according to the FAIR-physics topics, is given in Tab.~\ref{tab:theory}. It includes a description of the physics motivation and the type of calculations that are associated with this. The theoretical work associated with APPA is not included in the table and compute estimates given in this subsection since that is an integral part of the APPA research line and discussed in Sec.~\ref{sec:appa}.


\begin{landscape}
    
\begin{table}[ht]

\vspace*{-2.4cm}
\begin{center}
\caption{Overview of the planned theoretical activities in the context of FAIR research.}
\vspace*{-0.3cm}
\begin{tabular}{|p{0.2\textwidth}|p{0.75\textwidth}|p{0.75\textwidth}|}
\hline
\vspace{0.2cm}
\centerline{\bf Research area}  & 
\vspace{0.2cm}
\centerline{\bf Physics objectives}  &
\vspace{0.2cm}
\centerline{\bf Calculation types}  \\
\hline
\vspace{1.0cm}
Hot and & & \\
Dense QCD & & \\
Matter & 
\vspace*{-2.7cm}
{\small
\begin{itemize}
    \setlength\itemsep{-0.2em}
    \item Event-by-event transport and hydrodynamics simulations for the detailed dynamical description of heavy-ion collisions.
    \item Calculations for rare probes such as electromagnetic probes, multi-strange particles, and clusters.
    \item Evolution driven by Mean-fields and multi-particle interactions.
    \item Non-equilibrium dynamics around the phase transition between the hadron gas and the quark-gluon plasma.
\end{itemize}
}
&
\vspace*{-2.7cm}
{\small
\begin{itemize}
    \setlength\itemsep{-0.2em}
    \item Numerical solution of relativistic Boltzmann equation in some incarnation and with different methods (QMD, BUU), this corresponds to (test) particle based simulations with thousands of particles and their microscopic interactions.
    \item 3+1 D viscous relativistic hydrodynamic calculations implying calculations on a space- time grid with millions of cells over extended time and macroscopic information on energy density, pressure, temperature and density of the system.
    \item Event generators are used by experimental collaborations around the globe and at GSI for planning, design and analysis of experimental results.
\end{itemize}
}
\\
 \hline
 \vspace{1.2cm}
Hadron Physics & 
\vspace*{-0.5cm}
{\small
\begin{itemize}
    \setlength\itemsep{-0.2em}
    \item Derivation of effective hadron interactions from Lattice QCD and experimental data.
    \item Computation of hadron-hadron interactions from Lattice QCD and from effective field theory.
    \item Spectroscopy of strange baryons, charmed and charmonium(-like) hadrons.
\end{itemize}
}
&
\vspace*{-0.5cm}
{\small
\begin{itemize}
    \setlength\itemsep{-0.2em}
    \item Large-scale coupled-channel computations based on realistic interactions with long- range forces.
    \item Large-scale Markov chain Monte Carlo simulations of the strong interaction including dynamical up, down, and strange quarks.
    \item Calculations of the quark propagators needed to calculate hadron physics observables from lattice QCD simulations.
    \item Contraction of quark propagators and hadronic sources/sinks into hadronic correlation functions.
    \item Data analysis and fitting.
\end{itemize}
}
\\
\hline
 \vspace{2.5cm}
Nuclear & & \\ 
Astrophysics & & \\
and Structure & 
\vspace*{-4.2cm}
{\small
\begin{itemize}
    \setlength\itemsep{-0.2em}
     \item Determine nucleosynthesis output of astrophysical explosions including neutron star mergers and supernovae.
    \item Understanding observables of neutron star mergers and core-collapse supernovae including the gravitational wave signal, nucleosynthesis yields and electromagnetic counterparts by advanced dynamical simulations.
    \item Modelling of spectral energy distribution and bulk light curve properties of kilonovae.
    \item Investigate nuclear structure and decay properties of exotic nuclei with large scale shell-model calculations.
    \item Global calculations of astrophysical reaction rates (beta-decay, neutron capture, fission, ...).
    \item Nucleosynthesis networks.
    \item Radiation hydrodynamical simulations for neutron star mergers and supernova.
    \item Monte Carlo Radiation transport spectral modelling for kilonova.
\end{itemize}
}
&
\vspace*{-4.2cm}
{\small
\begin{itemize}
    \setlength\itemsep{-0.2em}
    \item Three-dimensional radiation hydrodynamical simulations for neutron star mergers and supernova using particle-based hydrodynamics schemes and grid-based solvers.
    \item Global calculations of astrophysical reaction rates.
    \item Nuclear reaction networks.
    \item Large scale shell-model calculations.
    \item Radiation transport spectral modelling for kilonovae based on three-dimensional Monte-Carlo schemes.
\end{itemize}
}
\\
\hline
\end{tabular}
\label{tab:theory}
\end{center}

\end{table}%

\end{landscape}

\subsubsection{Computational requirements}

The requirements of the theory groups are diverse and can be summarized as follows:
\begin{itemize}
    \item Massive sequential jobs (e.g. network calculations for whole sets of trajectories).
    \item Multithreaded jobs typically using OpenMP (e.g. nuclear shell model calculations, neutron star mergers).
    \item Parallel jobs using MPI and up to several thousand cores (e.g. neutron star merger simulations, nuclear structure calculations, lattice QCD calculations, kilonova modeling).
    \item Large parallel jobs (up to several thousand cores) currently using a client/server model with a plan to switch to a heterogeneous MPI setup (e.g. coupled channel hadron structure).
    \item Intermediate scale parallel jobs using a mixed OpenMP/MPI setup with typically a 100 to 1000 cores (lattice QCD calculations, neutron star mergers, kilonova modeling).
    \item Single node data analysis/fitting jobs (lattice QCD calculations).
    \item Sequential jobs (event generators) with trivial parallelization by running different events on different cores/nodes.
\end{itemize}

\noindent Tables~\ref{theory_cr} and \ref{theory_sr} present an overview of the compute and storage requirements. These estimates are primarily based on experiences obtained at FAIR Phase Zero (see Sec.~\ref{fair_phase_zero}) scaled with respect to additional activities foreseen during the time FAIR becomes operational. 

\begin{table}[h]
\begin{center}
\caption{THEORY: Compute Requirements (in HEPSpec06).}
\begin{tabular}{|c|c|c|}
\hline
Compute Class  &  FS(+)  &  MSVc  \\
\hline 
I.a & 0 & 0 \\
\hline
I.b & 0 & 0  \\
\hline
I.c & 0 & 0  \\
\hline
I.d & 0 & 0  \\
\hline
II.a & 750,000 & 1,000,000 \\
\hline
II.b & 0 & 0 \\
\hline
\end{tabular}
\label{theory_cr}
\end{center}
\end{table}%

\begin{table}[h]
\begin{center}
\caption{THEORY: Storage Requirements.}
\begin{tabular}{|c|c|c|c|}
\hline 
&   &  \multirow {2} {2.5 cm}{FS(+)}   &   \multirow {2} {2.5 cm}{MSVc} \\
& & & \\
\hline

\multirow {3} {3.5 cm}{ Simulation/Scratch} & TB & 3,000 & 4,000  \\
\cline{2-4}
&    \#years & permanent & permanent \\
\cline{2-4}
&  Bandwidth (MB/s) & 10,000 & 10,000 \\
\hline

\multirow {3} {3.5 cm}{ Derived data } & TB/year & 100 & 150 \\
\cline{2-4}
&    \#years & 10 & 10  \\
\hline
\end{tabular}
\label{theory_sr}
\end{center}
\end{table}%
\newpage

\subsection{BEAMS}

The computing activities in the accelerator sector are very diverse. The two departments Accelerator Physics and Storage Rings mainly require general cluster resources and offline storage for theoretical beam dynamics simulations. The accelerator control systems, on the other hand, necessitate dedicated computing hardware and pose higher demands for available online storage \& bandwidth capabilities. The quoted numbers in Tables \ref{beams_cr} and \ref{beams_br} include the accelerator device frontends, embedded systems and DAQ systems as well as the dedicated ASL cluster. A non-intrusive load monitoring is planned to support optimal network utilisation. Training data held for three months require a few TB of additional disk storage for the online services. Archiving services of accelerator state and beam parameters demand for a continuously provided bandwidth to storage in the Green IT Cube with a specified expected average of 35 MB/s (at maximum 300 MB/s peak short-term) bandwidth. In the future, towards the MSVc configuration, the online accelerator control systems are envisioned to make use of ML-targeting computing hardware such as dedicated GPUs in a low number of up to $\approx 10$ cards.

\subsubsection{Computational requirements}

\begin{table}[h]
\begin{center}
\caption{BEAMS: Compute Requirements (in HEPSpec06).}
\begin{tabular}{|c|c|c|}
\hline
Compute Class  &  FS+  &  MSVc  \\
\hline 
I.a & 285,000 & 460,000 \\
\hline
I.b & 15,000 & 20,000  \\
\hline
I.c & 190,000 & 210,000  \\
\hline
I.d & 0 & 0  \\
\hline
II.a & 25,000 & 25,000 \\
\hline
II.b & 0 & 0 \\
\hline
\end{tabular}
\label{beams_cr}
\end{center}
\end{table}%

\begin{table}[h]
\begin{center}
\caption{BEAMS: Online storage \& bandwidth.}
\begin{tabular}{|l|c|c|c|}
\hline
 & &  FS+  &  MSVc  \\
\hline 
\multirow{2}{*}{To Green IT Cube} & Fibres & 20 & 20 \\
 & Bandwidth (MB/s) & 700 & 4,000 \\
\hline
To permanent & Average (MB/s) & 100 & 100  \\
\hline
Permanent storage / year (TB/a) & & 10 & 15  \\
\hline
Additional disk storage (TB) & & 5 & 30  \\
\hline
\end{tabular}
\label{beams_br}
\end{center}
\end{table}%

\begin{table}[h]
\begin{center}
\caption{BEAMS: Offline storage.}
\begin{tabular}{|c|c|c|c|}
\hline 
 & &  FS+  &  MSVc  \\
\hline

\multirow{3}{*}{ Simulation/Scratch} & TB/year & 100 & 100  \\
\cline{2-4}
&    \#years & 2 & 2 \\
\hline

\multirow{3}{*}{Derived data} & TB/year & 200 & 200 \\
\cline{2-4}
&    \#years & 5 & 5  \\
\hline
\end{tabular}
\label{beams_sr}
\end{center}
\end{table}%
\newpage

\subsection{Overview requirements}
\label{summary_exp_req}

\subsubsection{Compute requirements}

\begin{figure}[h]
\begin{center}
\vspace*{-0.1cm}
\includegraphics[width=0.9\textwidth]{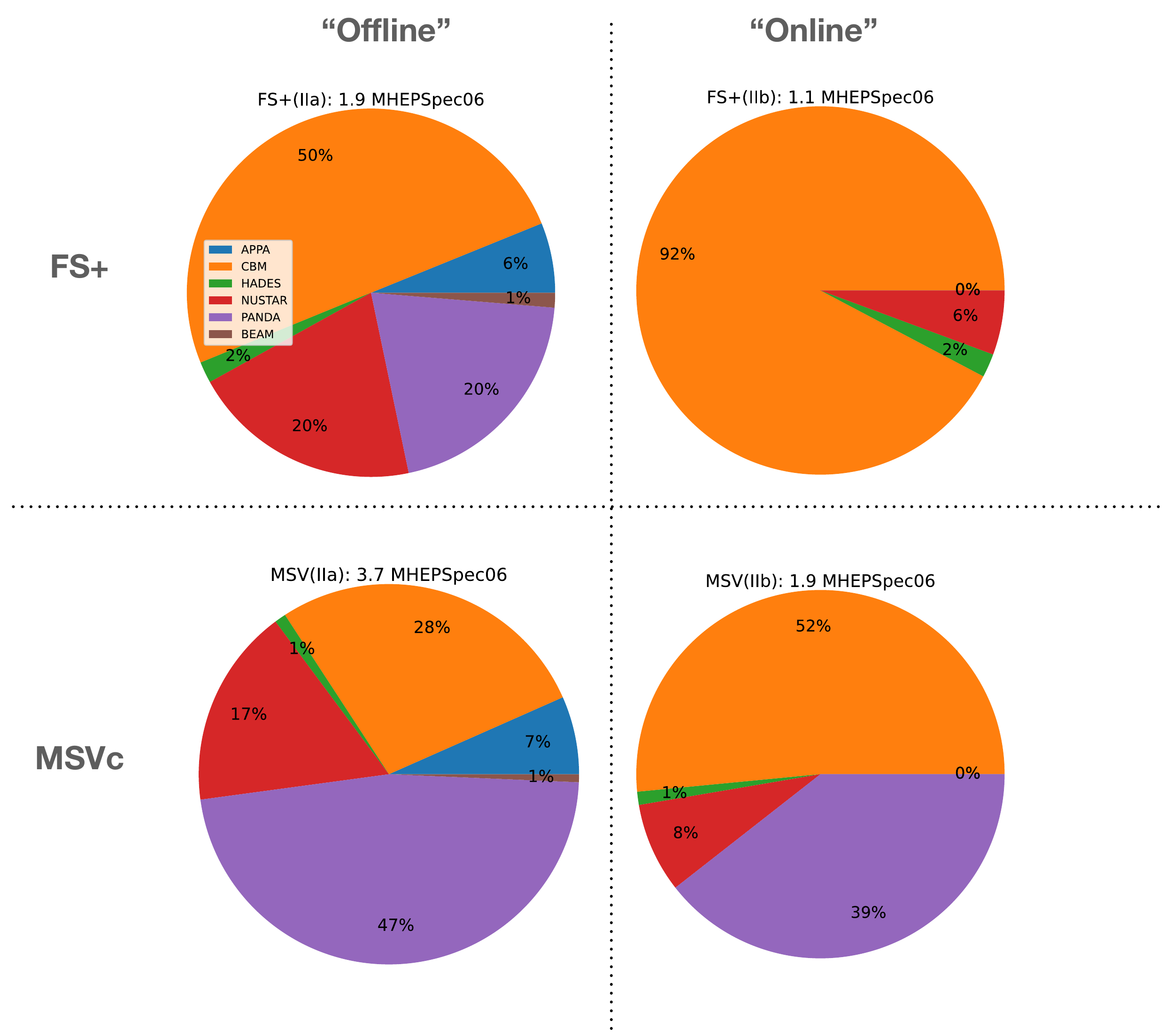}
\caption{Overview of the required compute resources for the FS+ (top panels) and MSV (bottom panels) scenarios. The left panels depict the resources for classes IIa (shared ``offline'') and the right panels those for classes IIb (shared ``online''). The title of each panel indicates the total integrated amount of compute resources in units of MHEPSpec06 and the color codes with percentages correspond to the contributions of the various FAIR research lines.}
\label{fig:CompFair}
\end{center}
\end{figure}

\noindent Figure~\ref{fig:CompFair} summarizes the compute requirements for class II type resources, {\it e.g.} commonly shared computing, of the various FAIR research lines and for the different running scenarios being considered (FS+ and MSVc). Other non-FAIR activities, such as ALICE, are excluded. We note that the experiment-supporting theory contribution has been included and redistributed among the respective FAIR research lines according to the fraction as observed during FAIR Phase Zero. For FS+, the compute resources during ``online'' operations (data taking, type IIb) are dominated by the requirements of CBM with only 8\% contribution from other research lines, particularly NUSTAR. The online compute capacity will reach a value of about \onlinecapacityFSP~MHEPSpec06. For ``offline'' computations, type IIa, the computing resources need to reach a capacity of about \offlinecapacityFSP~MHEPSpec06 and it includes contributions from all the FAIR research lines.  For MSVc, both ``online'' and ``offline'' capacities increase by a factor of two with respect to FS+. In this case, the three CBM, NUSTAR, and PANDA pillars ask for a significant fraction of computing resources both for ``online'' and ``offline'' operations. 

Considering the types of computational requirements of the various FAIR pillars, our analysis indicates that for FS+, approximately 60\% of the total requested compute capacity aligns with a High Performance Computing (HPC) infrastructure, while the remaining 40\% is better suited for a High Throughput Computing (HTC) architecture. Similarly, for MSVc, around 51\% of the compute capacity is optimal for HPC, with the remaining 49\% fitting an HTC environment.

\begin{figure}[t]
\begin{center}
\vspace*{-0.2cm}
\includegraphics[width=0.85\textwidth]{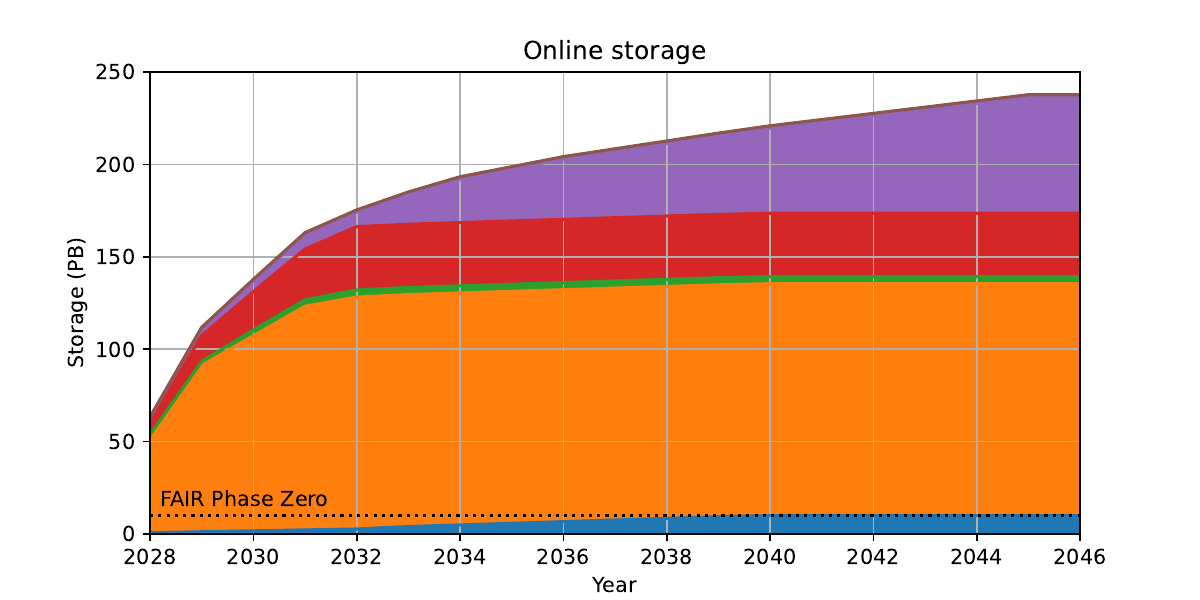}\\
\vspace*{-0.7cm}
\includegraphics[width=0.85\textwidth]{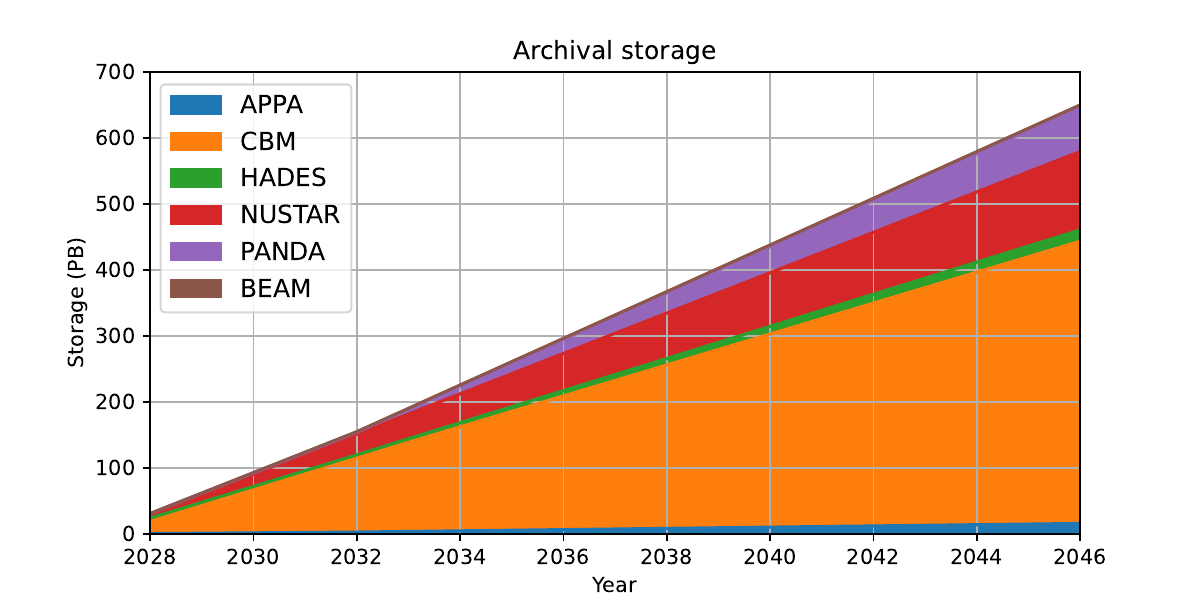}
\vspace*{-0.3cm}
\caption{The required amount of storage as a function of year with FS+ starting in 2028 and MSVc in 2032. The top panel depicts the requested disk space for fast access, whereas the bottom panel presents the needed long-term storage space (archive). The contributions of the various research lines are indicated by different colors. The dashed line shows the used storage on the Lustre filesystem for FAIR Phase Zero activities. Copies of the raw data at other FAIR facilities are not included, but imposed a requirement established by an external audit}.
\label{fig:storage2}
\end{center}
\end{figure}

\subsubsection{Storage requirements}

The time evolution of the required disk storage and archive is shown in Fig.~\ref{fig:storage2}. It includes the FS+ and MSVc scenarios with tentative starts in 2028 and 2032, respectively. For FS+ (MSVc) the total available storage saturates to about 160 (210)~PB accounting for the requested number of years the data needs to be directly accessible. The largest storage usage is by the CBM collaboration, followed by NUSTAR, and with the start of MSVc by PANDA. All other research lines only contribute marginally to the required storage capacity. Also here, we do not include non-FAIR related activities, such as ALICE.

\newpage

\newpage
\newpage
\section{FAIR computing model \& IT support}
\label{comp_model}

\subsection{Introduction}

This section describes the conceptual computing model for research IT at FAIR. The model is based on the requirements given by the research lines, presented in Sec.~\ref{sec:requirements}. The overall strategy is to optimize the sharing of computing resources among the various research groups and activities, to federate with partner institutes world-wide, and to minimize the amount of overhead in maintenance of the infrastructure with unambiguously-formulated responsibilities of the FAIR-IT research department.
The follow-up section~\ref{sec:open-science} reflects on how the model will be incorporated in our data access strategy and in light of respecting the overall F.A.I.R. principles. Moreover, it will be the basis of the focus on R\&D activities as described in Sec.~\ref{sec:randd}. 

\subsection{Computing model at FAIR}

\begin{figure}[h]
\begin{center}
\vspace*{-0.7cm}
\includegraphics[width=1\textwidth]{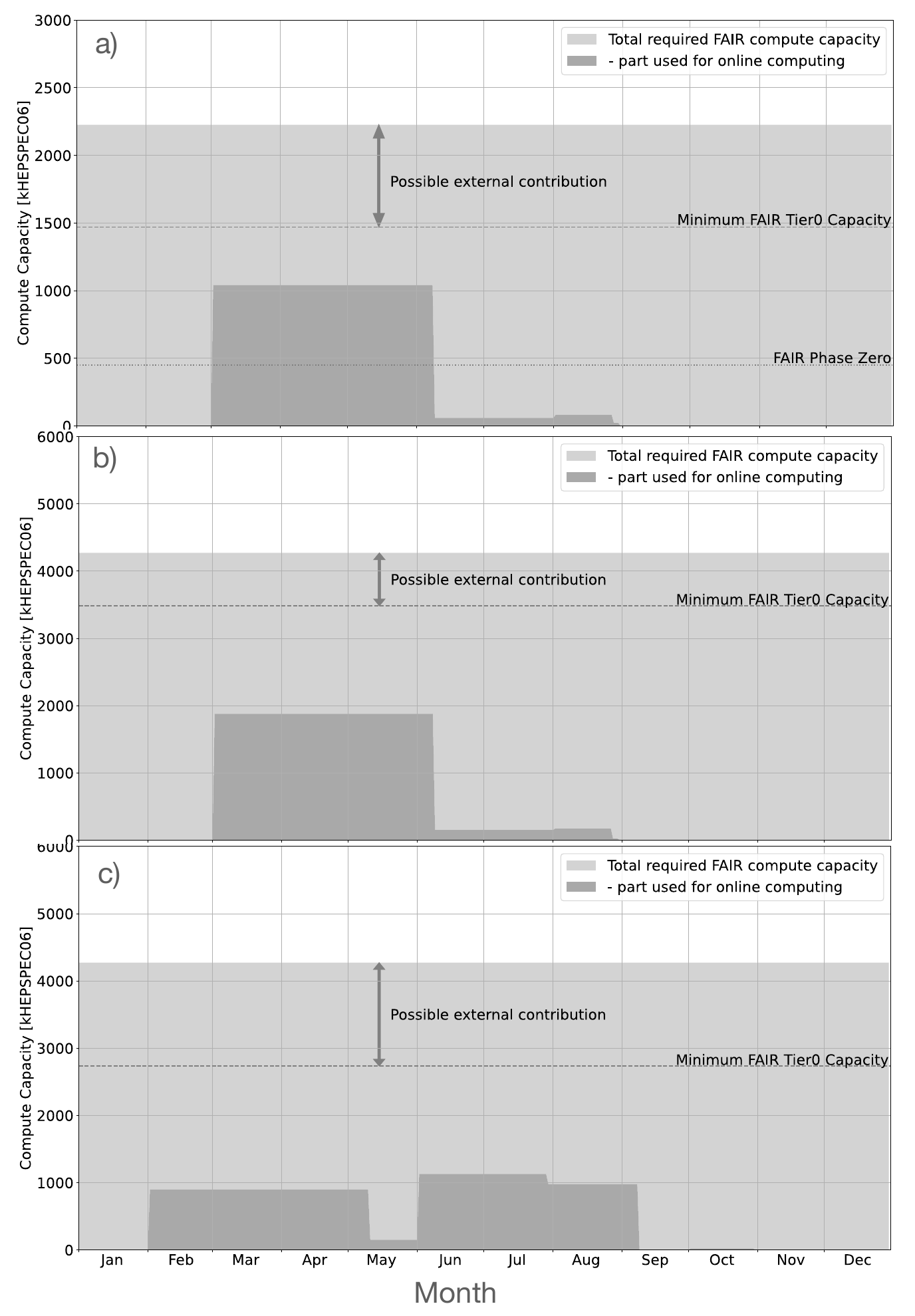}
\vspace*{-0.4cm}
\caption{Compute capacities for a nominal FS+ and MSVc year: a) FS+; b) MSVc with PANDA and CBM running in parallel; c) MSVc with PANDA and CBM running sequentially. The minimum FAIR Tier0 capacity is computed by summing up the maximum online capacity with the fraction of class IIa-type computing that is IO/data intensive. }
\label{fig:pretty_plot_all}
\end{center}
\end{figure}

Figure~\ref{fig:pretty_plot_all} illustrates the required compute capacity for a typical year of data taking and processing for three scenarios, namely FS+, MSVc assuming parallel operation of PANDA and CBM, and MSVc with a sequential operation of PANDA and CBM. 
The light-grey area depicts the total required compute capacity including both ``online'' and ``offline'' contributions, whereby the ``online'' contribution accounts for the fraction of days of beam time and that for the ``offline'' part one assumes the resources are fully used over the whole year. The part used for ``online'' computing is indicated by dark-grey areas. 

In general, the ``offline'' computations, summarized in Fig.~\ref{fig:CompFair}, contain both data-dependent (such as the production of derived data) and data-independent (such as Monte Carlo simulations and theoretical calculations) contributions. It is highly advantageous to provide computing resources at FAIR Tier0 to account fully for ``online'' requirements together with the part of ``offline'' computations that strongly depend on accessing large chunks of FAIR data. We estimated based on the requirements from the experiments, the fraction of ``offline'' computations that depend strongly on availability of FAIR data. Together with the maximally-required ``online'' contribution, we indicated the minimum FAIR Tier0 capacity (dashed lines) and, thereby, the contribution of compute resources that can be provided by external (FAIR-related) compute facilities.

The total compute capacity that is required throughout the year amounts to \totalcapacityFSP~ (\totalcapacityMSV) MHEPSpec06 for FS+ (MSVc). Moreover, we estimated that for FS+ at least \fractiontierzFSP\% of the total compute capacity needs to be located at FAIR Tier0. For MSVc in parallel (sequential) operation, this fraction becomes about \fractiontierzMSV\% (\fractiontierzMSVs\%). We note that the parallel operation of CBM and PANDA would be in general the most cost effective and efficient mode of operation from the perspective of running the accelerator. For the formulation of the FAIR computing model, we therefore bias our conceptual design towards a scenario with MSVc in parallel operation of CBM and PANDA.

\subsection{Federated computing}

The majority of the required compute resources will be provided by FAIR Tier0 as indicated above. The part of computing that can be offloaded to external facilities will be those jobs that do not require massive IO operations on FAIR data. Moreover, the data produced of these jobs can be transmitted from/to the central storage at Tier0 in an asynchronous mode. Typical compute tasks that fall in this category are theoretical model calculations and (Monte Carlo) simulations. 

\begin{landscape}
\begin{figure}[h]
\begin{center}
\vspace*{-0.2cm}
\includegraphics[width=1.5\textwidth]{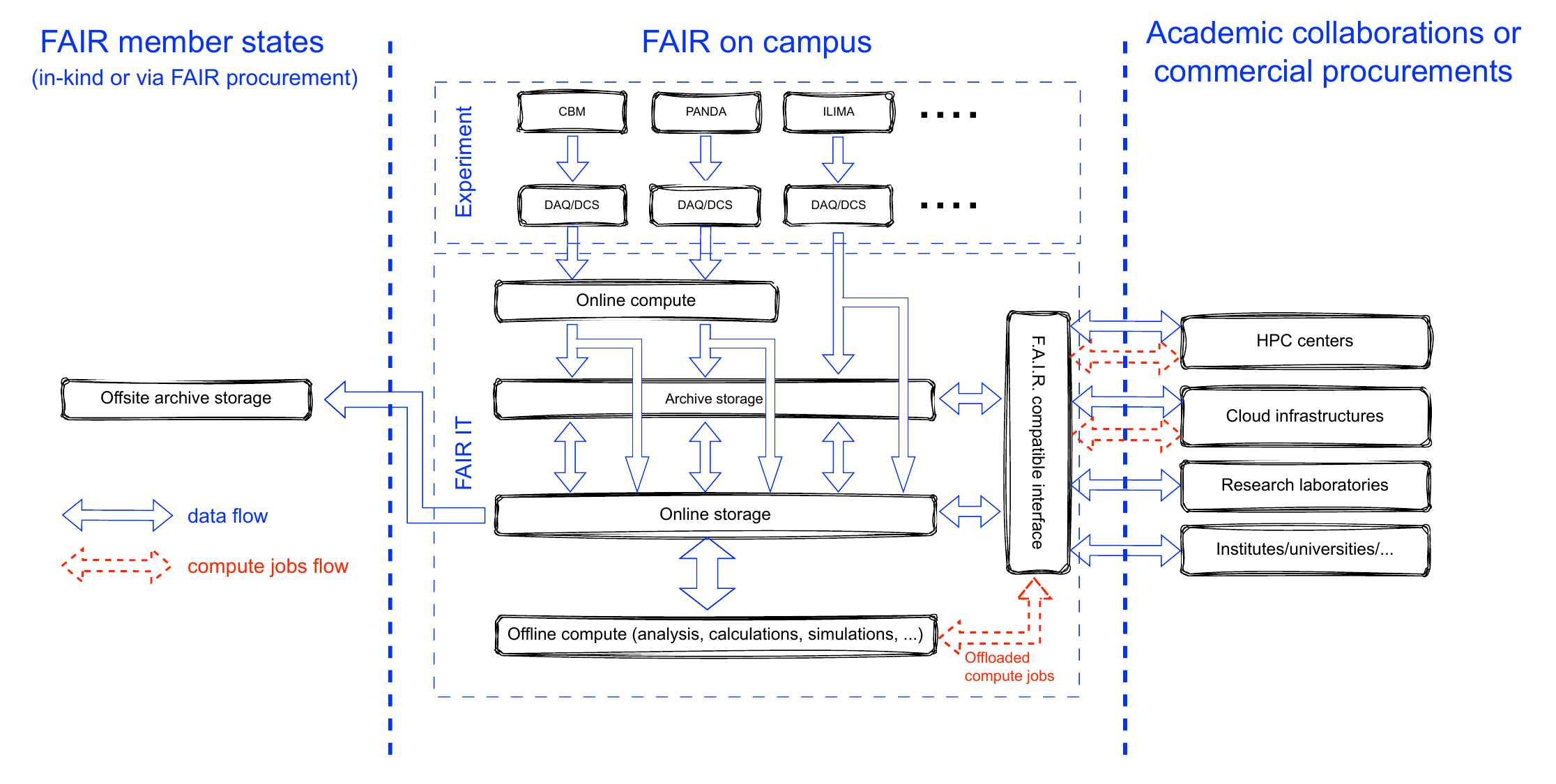}\\
\caption{Sketch of the computing model proposed for FAIR. Indicated are the various experiment, storage, and compute elements that will be integrated within the FAIR campus (middle), the archive copy to external storage (left), and the data and compute links with outside facilities (right). The data flows are indicated by solid, blue arrows and the compute jobs that possibly will be offloaded to outside facility by the red, dashed arrows. The data and compute flows to and from outside facilities are linked via an interface compatible with the F.A.I.R. principles.}
\label{fig:comp_model}
\end{center}
\end{figure}
\end{landscape}

\subsubsection{Conceptual design}

Figure~\ref{fig:comp_model} shows a conceptual sketch of the FAIR computing model from the perspectives of the data/processing flows and the connections with the ``outside'' (non-Tier0) facilities. The flow of data obtained by the DAQ/DCS of the various FAIR experiments (including accelerator data as well), from/to offline computations (processed data, simulated data, etc.), and from/to outside centers (offsite storage, HPC centers, cloud infrastructures, etc.) are indicated by blue arrows. The part of compute jobs that can be outsourced is presented by red (dashed) arrows. The backbone of ``FAIR IT'' will the Green-IT Cube hosting the compute nodes and archive \& online storage devices and servers.   

FAIR Tier0 will be linked to external facilities based on a {\it federated} computing model, illustrated in Fig.~\ref{fig:comp_model} with the arrows between ``FAIR on campus'' and ``Academic collaborations or commercial procurements''. 
The majority of computing resources for FAIR experiments will be based at FAIR Tier0, specifically the GreenCube, which is managed by the local IT department. We expect FAIR Tier0 to focus on high-bandwidth online processing and large-scale, synchronous data handling. In contrast, the collaborating Tier1 computing centers from other academic institutions will provide resources optimized for less data-intensive tasks, such as Monte Carlo simulations and theoretical studies. Local centers near FAIR, including those at Johannes Gutenberg University Mainz, Technical University Darmstadt, and Goethe University Frankfurt, play a crucial role by forming a Teralink network with the GreenCube. Considering the required bandwidths of the experiments, a sufficiently amount of 100~Gb/s links to the major external centers will be provided. 

Our computing model is more centralized compared to a distributed computing model based on a grid-like infrastructure involving many small compute centers. The choice of our model is primarily driven by the organizational needs of the numerous small nuclear-physics collaborations utilizing FAIR's computing infrastructure. Due to the relatively small size of these collaborations and cultural aspects, their computing teams are in general small and lack the expertise to maintain their local computing facility. Consequently, a grid-like distributed model involving many smaller computing resources would not be feasible and would introduce excessive operational overhead. Instead, we anticipate a centralized IT department at GSI/FAIR that will provide robust support for various experimental computing activities and oversee the local infrastructure.
Taking into consideration that the majority of computing tasks will take place at FAIR Tier0, it would be most advantageous to focus on a distributed computing scenario involving only a few and relatively-large computing and data centers in our model. 

\subsubsection{F.A.I.R. compatible interface}
In preparation towards FAIR, it will be advantageous to work out commonly-used and agreed upon standards with these local centers as a proof-of-principle for a federative model, represented by the box ``F.A.I.R. compatible interface'' in Fig.~\ref{fig:comp_model}. We note that also in the European context, the concept of federation, instead of consolidation, is meanwhile widely accepted for future computing-related activities. During a JENA (Joint ECFA-NuPECC-APPEC) meeting on 12-14th of June 2023~\cite{jena} the theme of federation was placed at the center and discussed in the context of the operation of (future) HPC centers in Europe. This includes the development of software on heterogeneous architectures, federated data management, development of machine learning and artificial intelligence tools, and activities for training, dissemination, and education. This work has led to the development of several White Papers on each of these topics, accompanied by a summary document~\cite{jena-wp}. These were presented at the JENAS 2025 meeting, held from April 8–11, 2025 in Oxfordshire, with participation from funding agencies~\cite{jenas2025}, and with significant contributions from FAIR, representing the major part of the nuclear physics community. We particularly refer to the White Paper of the working group ``Integrating HPC Centers with JENA Computing Infrastructures: Key Findings and Recommendations'' identifying the critical technical and organizational areas that need development and funding to effectively integrate HPC centers with experimental facilities~\cite{wp_hpc}. Similarly, the European Open-Science Cloud (EOSC) association has endorsed this concept for the development of a ``Web of F.A.I.R. data and services''~\cite{eosc}. 

There is a clear and strong push with a tight timeline to a working EOSC federation concept. The EOSC-EU node~\cite{eosc-eu-node} is already in operation (since October 2024) and further build-up nodes are being planned to become online. This includes among 12 organizations, a German national node by National Research Data Infrastructure (NFDI)~\cite{nfdi} with the objective to integrate German research infrastructures, including FAIR. With FAIR as one of the ESFRIs leading the nuclear physics domain and since 2025 a full member of the EOSC association, we will have the opportunity to play an important role in the realization of the federated computing concept, thereby remaining well connected to the computing-related developments and activities, particularly, within the European context. 
During the preparation and write-up of this CDR, the technical implementation of the F.A.I.R. compatible interface is in a continuous development phase. Section~\ref{sec:open-science} describes the planned implementation of the overall resource access in connection to the open science policies. Below, we provide an overview of the strategic steps, from past to future, taken to develop the interface with the aim of adopting the standards outlined in the following chapter.

FAIR computing naturally emerges from the present GSI IT facility and its support. Hence, operational access to the Green IT Cube's compute resources has already been enabled and supported by the GSI-IT department for all GSI, and thereby as well, for FAIR users. Moreover, the development to adopt F.A.I.R. policies has progressed already since the last decade, particularly in the context of projects supported within the German national funding agencies, e.g. BMBF, such as those within the Programmorientierte F\"orderung (POF) scheme for Helmholtz centers. As a key example, GSI is strategically engaged in several NFDI consortia, notably PUNCH4NFDI (Particles, Universe, NuClei and Hadrons for NFDI), which focuses on domain-specific data management for large-scale scientific experiments. This aligns closely with FAIR’s ESFRI ambitions and GSI’s position within the Helmholtz Association, both integral to Germany’s scientific infrastructure. Given Germany’s long-term prioritization of digitization, continued and active collaboration with national initiatives like NFDI is both strategically important and timely. For the next funding period (POFV), GSI will continue to support the further development of F.A.I.R. compatible tools in collaboration with other Helmholtz centers in Germany.

FAIR with a large European community and focus, it advantageously connects with European initiatives in data management and compute access.
GSI/FAIR has been actively and successfully participating in the European Science Cluster of Astronomy \& Particle physics ESFRI research infrastructures (ESCAPE) funded within Horizon 2020 initiative. Specifically, GSI/FAIR contributed to the development of a data infrastructure to enable seamless access and interoperability of data across various research infrastructures (WP2) and establishing a curated repository for open-source scientific software and services (OSSR, WP3). FAIR partner institutes were contributing to develop an ESFRI science analysis platform (ESAP, WP5). Although the project ended its funding, it continued as a collaboration, with GSI/FAIR being an active member. As a successor initiative, GSI/FAIR is now taking part in the Open Science Cluster's Action for Research and Society (OSCARS) project~\cite{oscars} and, e.g., taking a coordinating role in the developing metadata scheme for nuclear physics and beyond with recently funded NAPMIX (Nuclear, Astro, and Particle Metadata Integration for eXperiments) project. Within Horizon Europe initiative, GSI is taking part in the EUROpean Laboratories for Accelerator Based Science (EUROLABS)~\cite{eurolabs} with the goal to integrate three major research communities—nuclear physics, accelerator technology, and detector development for high-energy physics—to enhance collaboration, optimize the use of research infrastructures (RIs), and address both fundamental scientific questions and technological challenges. GSI plays an important role in EUROLABS by providing transnational access to its advanced accelerator facilities and offering support to international research teams (WP2) and by promoting open, diverse, and inclusive science practices (WP5). This project runs till August 2026 and is particularly valuable for the nuclear physics communities both at GSI and FAIR.

FAIR supports scientific collaborations at an international level, thereby going beyond the European scope. Although, FAIR is not directly involved in discussions on Open Science standards and practices for data and compute sharing on a global level, it relies on the ongoing collaborative efforts between EOSC and the Research Data Alliance (RDA)~\cite{rda}.

\subsubsection{Data archival}
For common-sense safety measures, archival data will be duplicated at least once at a different location other than FAIR Tier0. This is indicated in the left-part of Fig.~\ref{fig:comp_model} as ``offsite archive storage''. This will be provided by the FAIR member states, either in-kind or via FAIR procurements. 

At the time of writing the CDR, no formal commitments exist with sites outside GSI/FAIR. The default planning assumes that FAIR member states will host the second copy of data in their data centers, but this contribution remains undefined, as no binding agreements are in place. If a member state provides such resources, they will be treated as an In-Kind contribution to the FAIR Operating Budget, requiring legally binding contracts between FAIR and the respective funding agency. Several countries have expressed interest in hosting computing capabilities, including long-term storage, with further progress toward signed contracts expected after the approval of this document. The FAIR computing model is technically designed to connect with external resources, ensuring flexibility and readiness to leverage future opportunities. Actively exploring partnerships—whether with international institutions or commercial providers is crucial to securing external resources and strengthening the model’s scalability and resilience.

\subsection{FAIR IT support, role of research lines, and scientific collaborations}
\label{sec:FAIR_IT_role}

To identify the responsibilities of FAIR research IT and that of the research lines and scientific collaborations, it is important to understand the challenges and constrains a research facility such as FAIR is facing. 

First of all, there is an large diversity in requirements among the FAIR research lines. Roughly, one may observe two types of categories. The first category are the large-scale monolithic experiments, {\it e.g.}, CBM, PANDA, and partly NUSTAR. These experiments depend strongly on the performance of online data processing and for a large part depend on the HPC farm. They are the main consumers in terms of computing. The second category involves smaller-scale research activities, mostly the APPA community and for a very large part various subcollaborations of NUSTAR. Their computing activities are very diverse and they have a broad user community with a large spectrum of ``use cases'', not necessarily requiring HPC. For an efficient research-IT support, it is of utmost importance to avoid a huge overhead in the IT department for managing the large variety of computing requirements. 

Secondly, financial constraints and the importance of sustainability -- minimizing the CO$_2$ footprint -- particularly in the operation of computing infrastructures, creates additional requirements for the IT facility planned for FAIR. The design of the GreenCube at GSI was already driven by this notion. But also for the conceptual design of the FAIR computing model, it is of utmost importance to maintain this spirit and, hence, minimize operational costs and optimize its efficiency. In this context, we aim to optimize the usage of the computing resources by a capacity-sharing model. This implies that the available computing resources will be shared between online and offline computing and among the research activities, hence, avoiding as-much-as-reasonably-possible the number of dedicated computers only used by small FAIR communities. The main role of the FAIR IT department will be in promoting and supporting such strategy.

The infrastructure part of which the FAIR research IT department is responsible for starts ``at the end of the fibres from the experiments''. Moreover, it is up to the IT department to define and setup the interfaces connecting the experiment/user with the central compute/storage infrastructure. Given the large spectrum of ``use cases'' driven by the diversity at FAIR, the IT department will promote as much as reasonably acceptable common interfaces, and hard- and software infrastructures/frameworks.

To optimize the usage of the centrally operated HPC/HTC and to minimize the number of ``idle'' computers, the usage of virtualization techniques will be further expanded. Already during FAIR Phase Zero, users of the HPC at GSI make successful use of containerization techniques provided by GSI IT. Depending on the needs that may rise during FAIR operations, this can be further expanded by, {\it i.e.}, enabling the usage of virtual machines and cloud services.

An important part of FAIR IT is to support the developments of research-focused software and related services. Therefore, the support of commonly-used services and frameworks, {\it e.g.} FairROOT, FairMQ, CDash, Gitlab, CernVM-FS, etc. will be continued. Moreover, to enhance the synergy between research and IT, the local research-IT group will be expanded in size whereby a large part of their activities will be integrated within the various FAIR research lines. This research IT group will also act as an interface and network with computing experts from outside FAIR. The overall aim is to enable a integrative and prosperous community between FAIR-IT, the research groups, and with experts outside FAIR. Particularly, such construction will be a basis to follow-up the most promising R\&D activities described in Sec.~\ref{sec:randd}.

The dashed boxes and lines in Fig.~\ref{fig:comp_model} indicate not only responsibilities of the FAIR IT department (indicated by ``FAIR IT''), but also the role of the experiments and collaborations. The IT aspects related to the detector-specific data acquisition and control are conceptually integrated in the experiment, hence, in the hands of the corresponding research group or collaboration. The research groups themselves are also responsible for the development of research-specific software and should make use of the common infrastructure. Moreover, they are accountable for the management of their data. The right-hand side of Fig.~\ref{fig:comp_model} depicts the various external infrastructures including HPC centers, cloud infrastructures, laboratories, and IT infrastructures at universities. The accounting and management of this part will be outside the responsibility of FAIR IT and, therefore, be part of the responsibility of the academic collaboration or research group who requested to use these external facilities. FAIR IT will be responsible for the implementation and maintenance of the F.A.I.R. compatible interface as indicated in the figure.

\subsection{FAIR computing model in a nutshell}


In this part, we conclude on the overall computing model foreseen for FAIR. The most important aspects that will be embedded in the FAIR computing model are summarized below.

\begin{itemize}
    \item The majority of computations will take place at FAIR Tier0. To be cost and operation efficient, the overall concept is to ensure that resources are optimally used for the various computing tasks. For this, the available resources will be shared between on- and offline computing and among the various research activities at FAIR.  
    \item Part of, the less-data-intensive, compute tasks can be offloaded to external facilities. We foresee a distributed-computing model that is based on a federation among compute centers from FAIR-experiment countries. The `local' centers in the vicinity of FAIR that are connected via Teralink network will play a central role for federated storage and computing.
    \item To comply with the F.A.I.R. principles without too much administrative overhead, we foresee to centrally store FAIR-related data at Tier0 with an interface to external facilities compatible with these principles.
    \item Containerized approaches and other virtualization methods will be used and supported by FAIR IT to enable flexible compute operations serving the diverse FAIR community and to optimize the overall resource usage.
    \item Usage of hardware architectures largely dedicated for generic use combined with a significant fraction using technologies suited for online data processing, such as accelerator cards.  
    \item Data and compute access will be enabled using common standards compatible with NFDI~\cite{nfdi} (German national), EOSC~\cite{eosc} (European), and, with less of a direct involvement, RDA~\cite{rda} (worldwide).
    \item For common-sense safety measures, archival data will be duplicated at a different location other than FAIR Tier0, provided by the FAIR member states, either in-kind or via FAIR procurements. 
\end{itemize}

\newpage
\section{Resource access, open science and F.A.I.R. policies}
\label{sec:open-science}

\subsection{General open-science policies}
FAIR is committed to support open science practices and recognises the benefits it brings to individual researchers, the scientific community, and to society as a whole. The aim is to support researchers in making their research open whenever possible, and enabling mechanisms for doing this. The GSI policy on research data management~\cite{gsi:rdm} acts as a foundation stone for this, explicitly laying out the expectations of FAIR. 
The large datasets and complex workflows required by scientific projects at FAIR present challenges when it comes to practising open science. The goal is to adhere to the Findable, Accessible, Interoperable, and Reusable (F.A.I.R.) principles~\cite{Wilkinson:2016}. However, achieving this necessitates consideration to the resources required for a given research project to publish and replicate the outputs. The computing power needed to enable open science on such large and complex datasets requires the necessary infrastructure, in particular for access to shared HPC resources on a cross-community basis. Typically, an end user would seek to access the journal article, the research data, the software, and the computational workflow necessary to process the data and replicate the published results, but this is not always feasible for raw datasets, which may involve resource intense pre-processing stages. Additionally, access to the FAIR HPC infrastructure will require some level of authentication to be determined by the principle investigator of the project, and thus full or pre-processed `larger' datasets will only be available to smaller, more involved communities, rather than fully open-access data and workflows. 
To mitigate this, and enable F.A.I.R.-principles, a selection or a smaller volume of pre-processed data can be published open access, including full documentation and workflows. Alternately or in addition, result data from figures or additional data to support results can be published. The former concept is rather preferred, as it will enable the end user to explore (to some degree) the parameter space of a given dataset and can thereby enhance interoperability. The decision on which data to publish rests with the principle investigator, and thus will be evaluated within a project and detailed in a data management plan. The data management plan will allow  researchers to track the life-cycle of research data in a project and identify the requirements to make the data open. 

To aid researchers in practising Open Science, FAIR (GSI) has been and still is involved in a number of external national and international open science projects. On the European level, the facility participated till recently in the ESCAPE (European Science Cluster of Astronomy \& Particle physics ESFRI research infrastructures)~\cite{project:escape} project with successful impact in several work packages. This community-driven project offers a suite of tools and infrastructure geared towards data, software and workflow management in the context of large scale computing clusters with the ambition that these services will eventually contribute to the EOSC (European Open Science Cloud) implementation.
FAIR is a signing member of the follow-up ESCAPE to the future collaboration representing the Nuclear Physics community. Moreover, GSI participates in the EuroLabs project~\cite{project:escape}, which will also offer solutions for data management and dissemination of research output.
To leverage, support, and influence in conceptualising Open Science on the European, FAIR is an observer member of EOSC. FAIR is also involved in the preparation of the NuPECC 2024 long range plan, wherein the requirements are laid out for open science and computing infrastructure for the European nuclear physics community. 

On the German-national level, GSI is a key player in the PUNCH4NFDI project \cite{project:punch4nfdi} and in alignment with the ambitions of EOSC/ESCAPE.
There are also efforts on the institutional level to implement new tools and documentation  (e.g. an institutional data repository, data and software management, open science policies), which are essential to enable researchers to practice open science.

At the institutional level, various open-science initiatives have emerged in recent years. A GSI/FAIR Open Science working group has been established to develop GSI/FAIR-specific use cases, implement prototypes, define best practices, and organize regular teaching and training events. Many of these activities are conducted in collaboration with external partners, aiming to achieve shared objectives.


\subsection{Data, services, and software access}

\subsubsection{Authentication, Authorization and Identity (AAI) services}
\label{sec:aai}


The FAIR computing infrastructure will have a distributed character in which resources, including data, software and services, will be shared among various communities and partner facilities world-wide. A federated computing model in which the various facilities maintain their autonomy is clearly the trend for the upcoming future and, also, the basis of the FAIR computing infrastructure. To ensure interoperability within the distributed infrastructure, it is essential that one follows common standards and interfaces. 

A key service that would be mandatory to commonly agree upon is the Authentication and Authorization, and Identity (AAI) layer. For this we follow the well established AARC blueprint \cite{aarc-blueprint:2019}, was well as the REFEDS \cite{refeds} and SIRTIFI \cite{sirtifi:2022} guidelines where applicable. For the FAIR computing infrastructure, we will require the following AAI features:

\begin{itemize}
    \item Use existing and well-tested solutions for the AAI layer, {\i.e.} based on commonly-used services with high standards in security.
    \item Flexible enough to accommodate various authentication mechanisms (federated identities, X.509 certificates, username \& password, EduGAIN, social logins, ORCID, etc.). Due to increasing security requirements, Multi-Factor-Authentication will be required on GSI accounts.
    \item Enabling integration with legacy services through controlled credential (token) translation.
    \item Provide the abstraction of collaborations and the tools to manage membership, entitlements and access policies that will regulate access to resources.
    \item Easily integrated in existing data access and computing software leveraging standard, off-the-shelf libraries and components, in particular to map collaboration-level authentication and authorization attributes and capabilities to local access mechanisms.
    \item Possibility of multiple identities and credentials in a single account, providing a persistent identifier.
\end{itemize}

\noindent These requirements are well aligned with the wishes of other partners within our science cluster~\cite{indigo} and they are in compliance with EOSC-Hub. For the deployment of an AAI infrastructure for FAIR, we will closely follow-up the developments and, thereby, continue our collaborations with partner institutes and remain compatible with the realization of the EOSC-Hub. An appropriate infrastructure to enhance the cybersecurity of the system using a CERT and SOC will be provided.

\begin{landscape}
    
\begin{figure}[t]
\begin{center}
\vspace*{-0.2cm}
\includegraphics[width=1.8\textwidth]{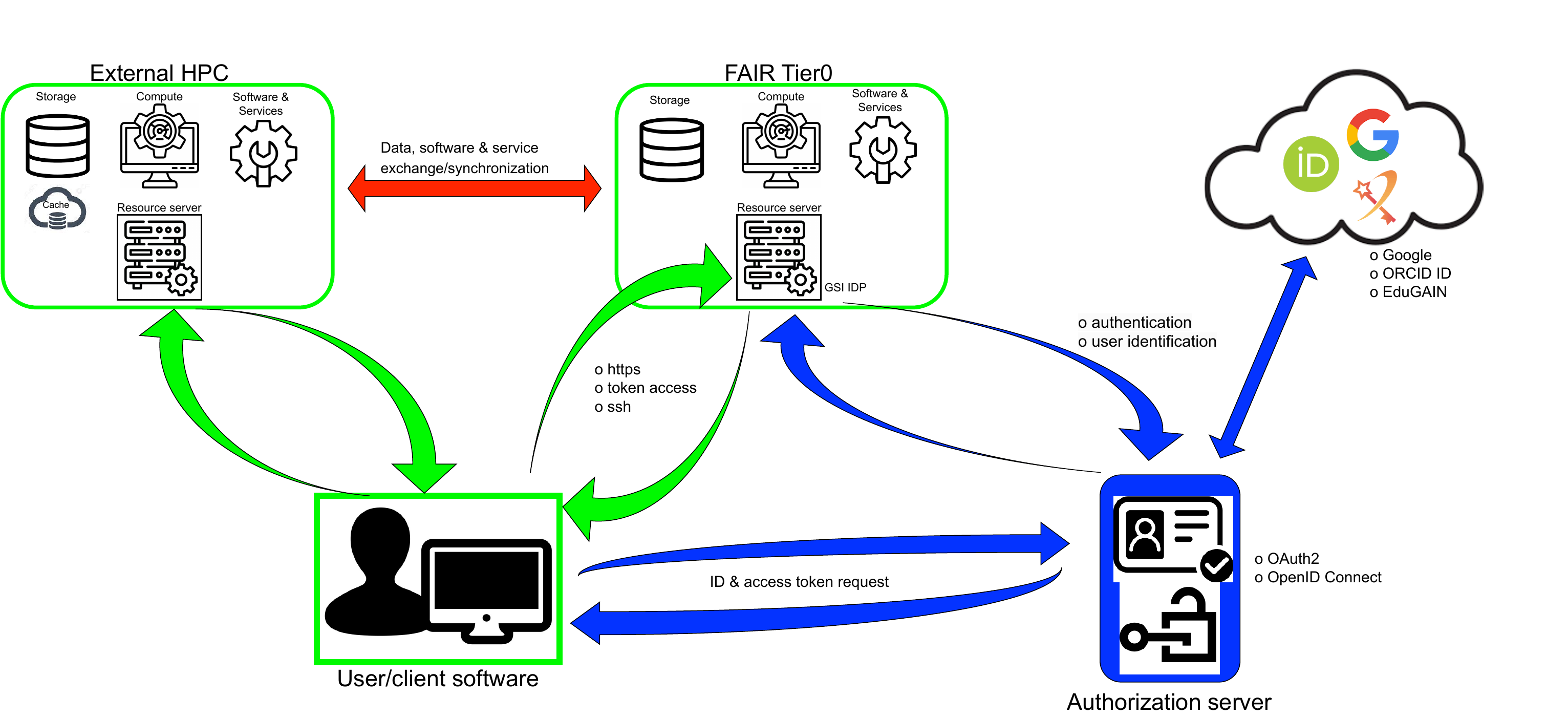}\\
\caption{Sketch of the conceptual model to access FAIR resources. 
A community-wide authorization server will be used to provide tokens for the user/client based on commonly used standards, {\it e.g.} OAuth and OpenID Connect (indicated in blue). The token-based communication with resource servers will follow standard protocols, such as https, and, depending on type of user/client, provide a certain level of access to the compute \& storage resources (marked by green arrows). External HPC facilities and sites will be transparently linked to FAIR Tier0 resources based on a federative compute model (red arrow), whereby the implementation and operational details are hidden to ``common'' researchers and users. Privileged developers and IT experts will be able to use the same AAI mechanism to access lower-level processes, services and data.}
\label{fig:access_model}
\end{center}
\end{figure}
\end{landscape}

FAIR aims to provide a central authorization service for its user community. Accounts are in general tied to a validated identity and an up-to-date affiliation to FAIR, e.g. a valid experiment guest registration or employee status. Some selected services might be open also to other authentication mechanisms, e.g. federated identities or social logins.
We will require that the AAI concept for the FAIR computing infrastructure enables interoperability among our partners within the research domain. For this the foreseen AAI service will be integrated with legacy services where necessary. With this, we are compatible with the F.A.I.R. principle and enable user-friendly and secure mechanisms to access the (distributed) computing resources of FAIR with sufficient flexibility to account for collaboration-specific policies, such as embargoed data. A sketch of the foreseen access model for FAIR resources is shown in Fig.~\ref{fig:access_model}.

\subsubsection{Data management and the ``Data-Lake'' concept}


The amount of experimental, simulated, and derived data that will be produced in the context of FAIR research and made accessible to the user community will reach a scale corresponding to a couple of hundreds of PB. To ensure the accessibility and findability of this data without creating an enormous overhead in the maintenance of services, it will be required that all data are centrally stored at the FAIR Tier0 site and made accessible via a parallel distributed file system such as Lustre. To ease the accessibility of such a centrally orchestrated storage concept, a common and user-friendly interface to store and retrieve data will be provided and maintained by the FAIR IT department. From the outside, we foresee to use standard protocols, such as https, token-based authentication (as mentioned above), possibly, based on open-source protocols
commonly used in HEP and Nuclear Physics.
The data will be duplicated at least twice at another computing site outside FAIR Tier0 as a requirement established by an external audit.
This also provides the possibility to make `easy' use of computing resources at other external sites for higher-level data analysis.  


Besides the objective to centrally administer the data obtained at FAIR, we have the ambition to provide a cloud-like access to parts of the data. This includes primarily higher-level derived data falling outside embargoed periods and authorized by the FAIR collaborations. The choice of what to share and with whom is defined by the policies of each collaboration independently, since the data structure, model, and volume differ largely among the FAIR communities. The infrastructure to enable this will make use of standard technologies presently being developed within the community. The experiences obtained at GSI/FAIR, from various FAIR Phase Zero activities, and in close collaboration with partner facilities, {\it e.g.} ESCAPE-Horizon2020, PUNCH4NFDI, etc., are an excellent basis to realize this so-called ``Data Lake'' concept well before FAIR becomes operational. 


\subsubsection{Software and service repository}

The F.A.I.R. principles will only be of value if the associated software and services used to process, analyze and manage data follow the same principle as well. A central repository for the dissemination and (maximized) use of trusted open-source software, including the communities associated with FAIR, is necessary to accomplish this. GSI has made a significant contribution to the realisation of the Open-source Software and Service Repository (OSSR)~\cite{escape:ossr} within ESCAPE and it has the ambition to further play a leading role in future EOSC calls in support of this. It would, therefore, be natural and advantageous to consider OSSR as a base model for the future FAIR research IT as well. The core of OSSR is implemented as a curated community, named escape2020, of the Zenodo general-purpose repository~\cite{escape:ossr_zenodo}. The design is such to make is easy for researchers, developers, and ESFRIs to find, access, and download the software and services in their community, and to contribute their own tools and services. Its main features are to provide a long-term, safe and secure archive with services that allow various types of contents, the implementation of publication versioning, and to provide a unique DOI for each version. All-in-all, the system is designed to improve the quality of contributions via its curation process ensuring compliance to the F.A.I.R. principles. 

\begin{table}
\begin{center}
\caption{An overview of policies, activities and potential services enabling the F.A.I.R. principles and considered for FAIR computing.}
\vspace*{-0.2cm}
\begin{tabular}{|p{0.15\textwidth}|p{0.85\textwidth}|}
\hline
\vspace{0.2cm}
\centerline{\bf Principles}  & 
\vspace{0.2cm}
\centerline{\bf Policies \& Services} \\
\hline
\vspace{2.5cm}
{\bf F}indable & 
\vspace*{-0.5cm}
\begin{itemize}
\small{
    \setlength\itemsep{-0.3em}
    \item Centrally orchestrated storage and access of data.
    \item Consistent usage of Persistent IDentifiers (PID) such as Digital Object Identifiers (DOI) for data and metadata.
    \item Ensure that all data and software are accompanied by rich, detailed metadata. A dedicated FAIR Open Science Working Group has a team to support researchers in this and to organise and promote projects to develop suitable vocabularies.
    Facilitate cross-platform automated metadata generation. For example, connections between GATE, RDMO and the FAIR publications repository in a dedicated open science ecosystem.
    \item Deploy data catalogues that allow users to locate datasets and software. Examples are Eurolabs OpenNP catalogue and the support of the open-source scientific software and service repository (OSSR) within the ESCAPE collaboration.
    }
\end{itemize}
 \\
 \hline
 \vspace{1.2cm}
{\bf A}ccessible & 
\vspace*{-0.5cm}
\begin{itemize}
\small{
    \setlength\itemsep{-0.3em}
    \item Data and software produced and dedicated for FAIR communities and publications centrally stored. 
    \item Data will be accessible using standard http protocols, possibly, based on open-source protocols commonly used in HEP and Nuclear Physics.
    \item AAI will be token-based and integrated with eduGAIN in line with ongoing concepts introduced within ESCAPE/EOSC. 
    }
\end{itemize}
\\
 \hline
 \vspace{1.5cm}
{\bf I}nteroperable & 
\vspace*{-0.5cm}
\begin{itemize}
\small{
    \setlength\itemsep{-0.3em}
    \item Participate in community-wide open-science initiatives, projects \& programs on institutional, national, and European levels.
    \item Work with partners to develop common metadata standards for data.
    \item Follow-up the ``Datalake'' concept developed within ESCAPE.
    \item Support the OSSR service (see above) to encourage the use of common software and services within the research domain.
    \item Use common AAI services together with partner institutes and facilities (see above).
    }
\end{itemize}
\\
 \hline
 \vspace{1.8cm}
{\bf R}eusable & 
\begin{itemize}
\small{
    \setlength\itemsep{-0.2em}
    \item Data and software available under a suitable open licence (such as GPL or CC BY).
    \item Ensure that all data and software are accompanied by comprehensive and structured documentation including provenance, versioning, and clear usage instructions.
    \item Support the use of controlled domain-specific metadata vocabularies. Currently ongoing projects HELPMI/NAPMIX.
    \item Employ collaborative platforms to facilitate reusable data and software, {\it e.g.} PUNCH4NFDI, ESCAPE ESAP.
    }
\end{itemize}
\\
\hline
\end{tabular}
\label{tab:FAIR}
\end{center}
\end{table}%

\newpage
\subsection{F.A.I.R. principles for FAIR}

Table~\ref{tab:FAIR} summarizes the various aspects and policies that will be incorporated at FAIR in light of the F.A.I.R. principles.
Generally, our objective is to implement F.A.I.R. principles in the computing model by participating in open-science networks (on various levels), by introducing very basic policies, and by making use of existing technologies \& services in collaborative efforts with partner institutes and facilities, primarily, within our domain of research. Conceptually, we concentrate on introducing the very minimum basic elements and policies as a framework for researchers and their data whereby the finer-level details will be up to the collaborations to implement. We, thereby, minimize the overhead in the high-level data management services.  


\newpage
\section{R\&D activities}
\label{sec:randd}
\subsection{Introduction}

The estimates on required computing resources of the research lines at FAIR, specified in Sec.~\ref{sec:requirements}, are to a large extent based on present-day technologies in the computing sector. There are, however, a few promising developments that may potentially increase the performance in terms of reducing the costs or providing more physics output. The most promising research and developments (R\&D) for those purposes will be requiring, on the one hand, additional resources, but, on the other hand, may pay off royally on the long term within the lifetime of an international facility like FAIR. Considering the relatively small scale of FAIR in comparison with that of, {\it i.e.}, CERN, such R\&D will be concentrated towards ongoing activities that are already taking place at FAIR Phase Zero with promising perspectives and results, and which are embedded in larger communities and networks. 

In recent years, R\&D efforts in computing technology at FAIR have yielded significant successes, particularly through strong synergies with other major research facilities. A prime example is the collaboration between FAIR and ALICE at CERN, which has led, e.g., to the development of the ALFA framework~\cite{ref:alfa}, a message-passing layer based on the actor model of concurrency. Originally designed to serve both FAIR experiments and ALICE, ALFA now plays a central role in ALICE’s O2 data processing framework~\cite{ref:o2}, exemplifying the mutual technological benefits of such partnerships. Beyond this, FAIR experiments are actively engaging with other international laboratories, strengthening shared computing-related initiatives that enhance efficiency and innovation across the board. These collaborative efforts underscore FAIR’s commitment to pursuit an integrated, forward-looking research strategy, and reflect the strength of its global scientific network.

It is worth stressing that the role of the next-generation accelerator facility FAIR with its large scientific community has been recognized as a potentially important player in future IT applications. For example, in the context of German-national initiatives, GSI is successfully contributing to the BMBF-funded ErUM-Data~\cite{erum-data} with the aim to address data challenges by exploiting modern digital technologies relevant for our research area. Similarly, and as mentioned before, the contributions of GSI/FAIR within the European ESCAPE project and with its perspectives in the follow-up ESCAPE to the future collaboration have made successful impact and provided a promising basis for future developments, respectively. These are only a few examples demonstrating the important role of GSI and FAIR in technological developments, thereby, not even mentioning the various successful activities that have been initiated by FAIR partners in other countries. In this section, we briefly outline those R\&D aspects particularly relevant for activities that, partly, will take place at FAIR.

To ensure continued alignment with evolving technologies and methodologies, sustained and responsive follow-up R\&D is part of the portfolio of FAIR. Establishing mechanisms such as an advisory board to monitor global trends and guide FAIR’s IT development will be crucial in keeping FAIR at the forefront of data-intensive research.

\subsection{Hard- and software technologies}

The most promising R\&D aspects in the context of the hard- and software developments have to be placed in light of the developments on the market and the plans of other research facilities with similar computing requirements as that of FAIR. From the hardware side, the most noticeably aspect is that Moore's law still holds but single-thread performance clearly has leveled off. Hence, performance gains are now to be made through parallelization~\cite{micro:trend}. As a consequence, the hard- and software trends in our research communities are to a large extent dedicated to the deployment of data processing methodologies based on (massive) parallel computing and compatible architectures. The pre-processed and derived data from experiments and the associated Monte Carlo simulations are often very well parallelizable on standard x86-based computers based on CISC architectures (Intel, AMD) as a consequence of their underlying event-based data structure. Moreover, most of the scientific software, library and compiler support are well tuned towards these general-purpose architectures. Hence, for the upcoming decade the x86 architectures will remain the basis for further developments for FAIR computing as well.

For the online data processing, FAIR will be challenged by huge data rates from the experiments. CBM, for example, will generate 400~GB/s of data that need to be processed with a high throughput capability already during FS+. Similar rates are to be expected from PANDA during MSVc. To cope with such demands and to reduce the data volumes to acceptable volumes, it will become mandatory to perform a nearly-complete data analysis during online data taking for event filtering. A combination of various computing architectures are foreseen to reach these objectives. Particularly the deployment of field programmable gate arrays (FPGAs) for fast feature extraction and graphics processing units (GPUs) for in-situ processing of massive amount of pieces of data simultaneously have become standards in, {\it e.g.}, CERN experiments~\cite{alice:gpu, lhcb:gpu}. 

The adoption of GPUs in FAIR experiments such as CBM and PANDA is driven not only by computational performance but also by architectural advantages demonstrated in large-scale, high-rate experiments like ALICE at CERN. ALICE's transition to a fully continuous readout mode in Run 3, combined with GPU-accelerated online processing, serves as a strong precedent for, f.e., CBM’s strategy. At CBM, which operates at unprecedented interaction rates of up to 10 MHz, GPUs are promising to integrate to the First-Level Event Selection (FLES) system. Drawing on ALICE's experience with TPC track reconstruction on GPUs, CBM employs similar parallel tracking algorithms optimized for GPU execution. This allows real-time event selection and filtering, drastically reducing the data volume without compromising physics sensitivity. ALICE has shown that GPUs offer superior performance-per-Watt and enable online processing at the full input rate, reducing dependence on hardware triggers. For CBM, this may be very beneficial for selecting rare physics events from an immense background. Additionally, PANDA, with its triggerless, continuous data acquisition, benefits from GPU-based parallel processing for real-time event building and reconstruction.

FAIR experiments, particularly CBM and PANDA, will benefit from further exploring AI/ML applications on GPUs — a direction that has proven promising in other large-scale experiments, where ML techniques are being developed for tasks such as online quality monitoring and intelligent filtering. These shared approaches underline a broader trend across modern nuclear and particle physics experiments: leveraging GPUs not only for speed, but as enablers of adaptive, intelligent, and scalable data processing architectures that can meet the demands of high-rate, high-granularity detectors.

Since the foreseen algorithms for experiments like CBM and PANDA require to be rather sophisticated dealing with complex information from the detectors, it remains a challenge to incorporate them on these platforms. With the long-term experiences from operating complex GSI experiments and based on the successful studies performed during FAIR Phase Zero, FAIR has an excellent basis for future operations as well. Particularly, promising developments have been made in the recent past demonstrating the potential to perform triggerless in-situ data processing with FAIR experiments~\cite{kisel:2020,jiang:2023,alicke:2021,gazagnes:2023,heine:2022}. Follow-up R\&D towards placing these algorithms, such as tracking and particle identification, in production mode will become necessary to reach the goals set at FAIR. Preparatory studies at FAIR Phase Zero and planned for the early-science phase play an important role towards this, such as those presently ongoing with the miniCBM setup at GSI.

Besides the promising applications of GPUs for online data processing, they are also very well suited architectures for high-level machine learning operations. For future FAIR-related research, these type of machine-learning-based computations will very likely grow in popularity as outlined in Sec.~\ref{sec:ML}. In general, there is a trend on the market in the design of future computing architectures optimized for machine learning and artificial intelligence applications, such as presented by AMD~\cite{amd:cpu_gpu} and NVIDIA~\cite{nvidia:cpu_gpu} who aim to combine CPU and GPU technologies for that purpose.

In view of minimizing the energy consumption and sustainability in computing, RISC based architectures like ARM and RISC-V show promising perspectives. An active R\&D program is ensuring that a large fraction of our scientific software stack is compatible with especially the ARM architecture. In this context, the usage of chiplets, small and modular chips that can be combined to form a complete system-on-chip (SoC) shows interesting perspectives as well~\cite{chiplets}. 

In general the fast technology evolution demands significant investments in software frameworks acting as an interface between the code developed by the FAIR researcher and the underlying hardware architecture accounting, in a user-friendly and sustainable manner, for changes in hardware technology and supporting libraries. A example of success in this context is the FairROOT framework~\cite{fairroot} developed by the research-IT department at GSI and being used as a standard by a large user community including atomic, nuclear, hadron, heavy-ion, and particle physicists. One of the key features of this framework is that its capabilities can be used for both online, exploiting FairMQ, and offline purposes for data analysis and Monte Carlo simulations. In light of the upcoming computing requirements of FAIR experiments, it would be opportune to follow-up the developments supporting parallelism on multi- or many-core CPU and the execution of algorithms on heterogeneous platforms, {\it e.g.} Alpaka~\cite{alpaka}, Kokos~\cite{kokos}, and oneAPI~\cite{oneapi}. 

The promising hard- and software R\&D aspects illustrated above will be for a large part driven by commercial enterprises and by the large IT community of LHC-scale experiments. The role of the FAIR research community and IT support in this context will mostly be concentrated on those aspects related to the development of software and corresponding services in view of online data processing, data analysis, and simulations. Maintaining tight networks with commercial partners and with IT-driven large-scale experiments world-wide will be of utmost importance to follow-up the developments and to create common projects on a synergy basis.

\subsection{Machine Learning and Artificial Intelligence}
\label{sec:ML}

In the past decades, the application of machine learning (ML) algorithms in the field of subatomic physics has made its appearance in both theory and experiment. 
Particularly, in the past $\sim$5 years its developments were impressive leading to an exponential increase in publications in scientific research connected to the application of ML, Artificial Intelligence (AI), deep learning, and other related terms. That such development is an important step forward in research has been clearly underlined by awarding the Nobel prize in physics in 2024 to John Hopfield and Geoffrey Hinton who helped in laying the foundation for today's powerful machine learning. The applications of ML/AI are promising, particularly for the activities related to the research fields addressed at FAIR.

\begin{figure}[t]
\begin{center}
\includegraphics[width=1\textwidth]{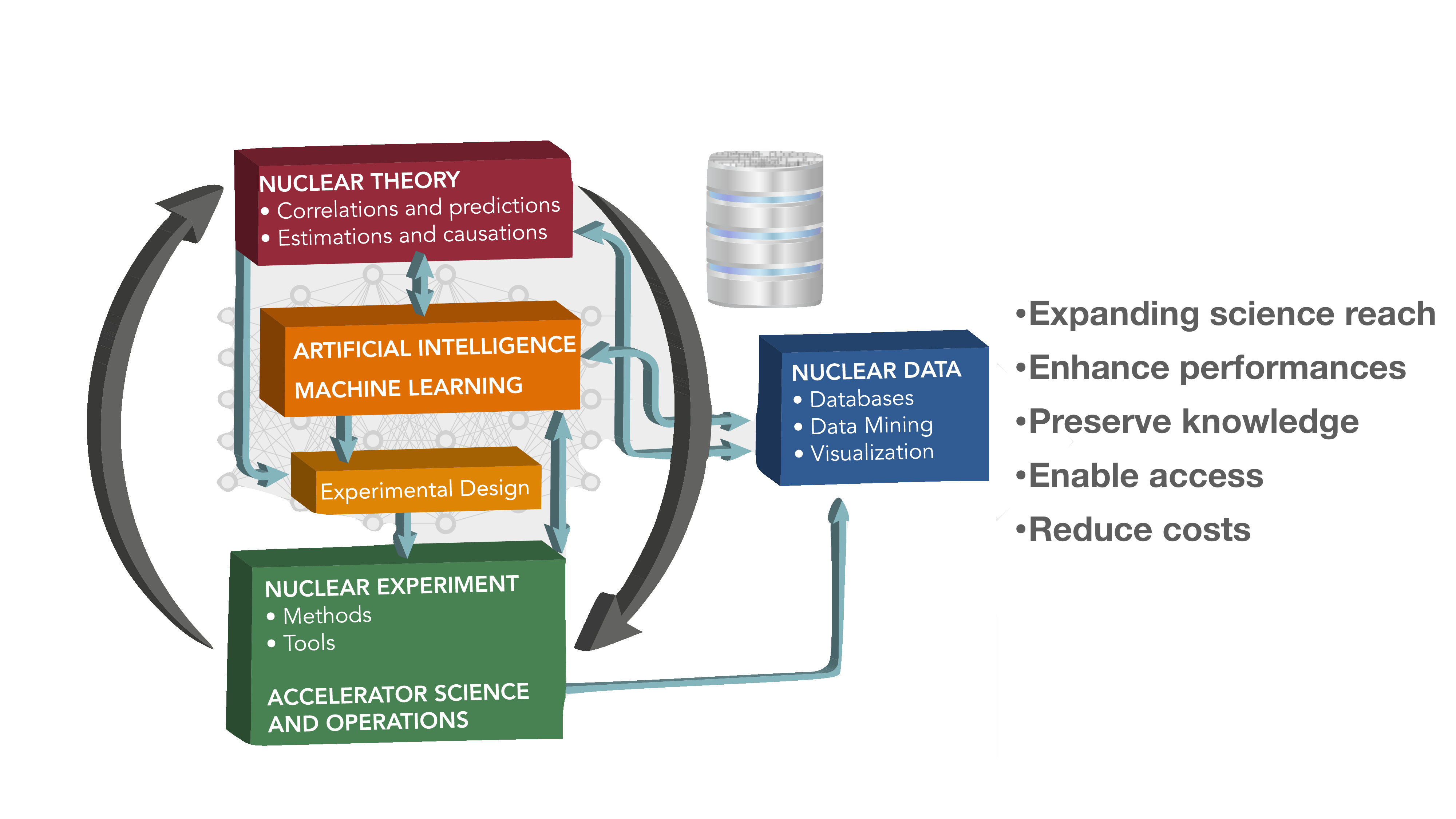} 
\vspace*{-1.2cm}
\caption{Schematic view of the applications of machine learning and artificial intelligence in the field of nuclear physics with the objectives indicated as bullet points. Figure adapted from Ref.~\cite{Boehnlein:2022}.  }
\label{fig:ml_illustration}
\end{center}
\end{figure}

Figure~\ref{fig:ml_illustration} illustrates the various applications in which ML foreseen to play a central role in nuclear physics together with how the elements are coupled together. This figure has been adapted from a recent review article~\cite{Boehnlein:2022} providing a comprehensive overview of the various applications of ML in the field of nuclear physics. In this context, we limit the discussion on those aspects that are primarily related to FAIR computing for the experiments. Additional activities not described in the CDR focus on the facility optimization, e.g. the realization of a "semi-autonomous accelerator operation" as one of the accelerator flagship goals or the energy-efficient operation of the data center. 

The application of ML in the experimental data analysis has become a common standard meanwhile. Particularly, classification and clustering algorithms to perform pattern recognition, {\it e.g.} track finding and cluster finding, particle identification and event selection are extensively used and, to a large extent, these applications demonstrated to outperform ``traditional'' algorithms. In this respect, ML methodologies are meanwhile well integrated in various data analysis tasks and have proven to serve their purposes in terms of improving the data quality by exploiting a high-dimensional feature space. Various software packages have become available as a common basis for users based on Python, {\it e.g.} PyTorch~\cite{pytorch}, TensorFlow~\cite{tensorflow} and scikit-learn~\cite{scikit-learn}, Java, {\it e.g.} DeepLearning4J~\cite{deeplearning4j}, and in C++, {\it e.g.} TMVA integrated within ROOT~\cite{TMVA2007}. As a consequence, the user community has grown enormously which has led to various networking and educational initiatives enabling an environment that provides training for the next generation of physicists in applying ML methods for their research. An example initiative in Germany is the ErUM-Data-Hub~\cite{erumdatahub}, funded by the German Federal Ministry of Education and Research (BMBF) with the mission to act as a central networking and transfer office for the digital transformation in the exploration of universe and matter. 

The spectrum of applications has meanwhile been extended far beyond data analysis and classification problems. A large-scale research laboratory such as FAIR with its complex operation of accelerators, detectors, and data acquisition is a perfect show case to explore novel applications of ML techniques beyond its current applications. A few key examples, in which FAIR already has become a player, are the usage of ML techniques in the design of experiments as proposed by the MODE collaboration~\cite{mode}, the application of Generative Adversarial Networks to significantly boost up the performances of compute-intensive Monte Carlo simulations~\cite{hashemi:2019, khattak:2022}, and the deployment of ML for online, streaming data processing and triggering~\cite{jiang:2023}, accelerator, detector, experiment operation and control~\cite{jeske:2022}. Although these efforts have the potential to make a quantum step in reducing the compute times and/or data rates, thereby improving significantly the efficiency of beam times and preparations, etc., they are still in a pilot stage, often not having fully demonstrated yet their feasibility. Some of these concepts touch even the domain of AI, referring to the broader field of creating machines or software that can perform tasks that typically require human intelligence. These encompass a wide range of techniques and approaches, including machine learning, expert systems, robotics, natural language processing, and more. Clearly, R\&D activities in the direction of ML and AI in the upcoming years are further needed and, hopefully, will provide proof-of-concepts and bring them eventually in production mode. Since it is not clear at this stage how much added value these new developments will bring and on what time scale they may become operational, it is difficult to incorporate their impact in the conceptual design of the FAIR computing infrastructure. 

For the realization of ML/AI within the context of FAIR computing, it is of utmost importance to embed R\&D activities within a larger community that share common challenges such as the one proposed in the context of JENA~\cite{caron:2023} and to connect to ML/AI expertise groups and networks, such as ErUM-Data-Hub~\cite{erumdatahub} and Hessian.AI~\cite{hessian.ai}.
In recent years, a partnership has been formed between Hessian.AI~\cite{hessian.ai} and GSI/FAIR through the Digital Open Lab, with the GreenCube playing a key role. Common training programs with hessian.AI have been successfully established. In addition, several workshops with local partners (ESA/ESOC, Universities, industry) are being organized with the aim to identify common areas of interest and projects.

At the European level, GSI/FAIR is taking a leadership position in a new network called the "European Coalition for AI in Fundamental Physics" (EuCAIF)~\cite{eucaif} with members engaged in research across various fields, including particle/accelerator physics, astroparticle physics, nuclear physics, gravitational wave physics, cosmology, theoretical physics, and simulation and computational infrastructure. 
FAIR is part of its management and represents the nuclear physics community, as mandated by NuPECC. A kick-off conference was recently held in April/May 2024 in Amsterdam~\cite{eucaifconf}, attracting over 250 participants. We also note that GSI and FAIR are the founding members of the ARTIFACT network~\cite{artifact} (with CERN, DESY, GANIL, etc) aiming to establish a European network for AI developments and applications for particle accelerators. One of the goals here is to develop a pan-European AI enabled platform for data and model sharing across the accelerator community.

On the GSI/FAIR institutional level, a ML/AI working group has been established bringing together researchers, engineers from the various fields and departments to gather information about ongoing activities, to develop possible synergies between local research themes and to forge new collaborations/networks, and to formulate a roadmap~\cite{aiworkshop-gsi}.

In addition to our ongoing efforts to enhance intellectual expertise in deploying ML/AI technology for FAIR experiments, we plan to maintain our investments in GPU clusters at GreenCube, which are specifically designed for ML/AI algorithms. Currently, approximately 400 GPUs (including NVIDIA Tesla and AMD Radeon Instinct) are accessible to users engaged in FAIR Phase Zero activities. Further investments for upcoming FAIR initiatives are anticipated, as outlined in the ``maintenance and operation costs'' supplement document.

Of particular interest for the overall infrastructure of FAIR is the deployment of ML algorithms on embedded architectures such as FPGAs, GPUs for online data processing and ML/AI applications with the ambition to enable an intelligent control and operation of experiments and accelerator, optimizing the beam and experiment operation and minimizing the commissioning time. Moreover, the deployment of modern ML and AI, such as deep-learning algorithms and foundation models, to reduce significantly the computational time for simulations and to optimize the workflow for data processing has great potential. We note that the present CDR has not accounted for the impact of such developments.
%

\subsection{Web of FAIR/F.A.I.R. data and services}

The European Open Science Cloud (EOSC) association~\cite{eosc}, a legal entity of the European Union (EU), has the ambition to become the world leader in the realization of a ``web of F.A.I.R. data and related services'' as an additional high-level layer on top of the world-wide web. The purpose is to create a web of scientific insights based upon a federation of relevant existing and future data sources, a virtual space where science producers and consumers come together, an open-ended range of content and services, meeting all European data requirements, and in interaction with other regions in the world. The overarching principle for such development is to place research a the center of EOSC with the main guiding principles based on multi-stakeholder approach, openness (as open as possible, as restricted as necessary), based on F.A.I.R. principles, with a federation of existing and upcoming data- and e-infrastructures, and machine actionable. At present, EOSC is still in its pioneering stage with challenges on both the technical and the social side. On the technical front, the challenge lies primarily in realizing an interoperable system, truly enabling a multidisciplinary approach in the long run. The approach has, therefore, been a more stepwise approach towards interoperability, focusing first on domain specific science clusters, {\it e.g.} ESCAPE~\cite{escape}. Socially, it remains a challenge to converge to common standards in support of the implementation and, hence, ``getting the noses in the same direction''. It is still a long way to go in order to get the ``web of F.A.I.R. data and related services'', but it is also clear that the EU will further pursue the dream in the upcoming years, if not, decades.

FAIR is undoubtfully one of the near-future ESFRI landmarks that will produce an immense amount of valuable data, associated services and derived knowledge. For its own interest, it has endorsed the principle of open science and has, thereby, taken part in EOSC-related activities on the European (ESCAPE) and Germany-national (PUNCH4NFDI, FIDIUM) levels. To participate in follow-up initiatives is foreseen as well. For details, we refer to Sec.~\ref{sec:open-science}. The overall strategy we foresee is to support domain specific efforts in this direction. The research communities associated with FAIR will be encouraged to identify physics-motivated use cases as a source of inspiration and as mid-term objectives. 

With the realization a ``web of F.A.I.R. data and services'', leveraging AI for data mining becomes an appealing prospect. The goal would be to extend the scope of physics research by extracting valuable insights through the integration of data from multiple experiments and facilities, and share knowledge across a broader community, including education and public outreach. Given the remarkable capabilities of modern pre-trained transformers and large language models, such as ChatGPT, the idea of developing an AI-driven platform and infrastructure for data collected from accelerator-based facilities worldwide, including FAIR, could become a possibility to consider.
Whether such initiative will be feasible has to be seen. Discussions in this direction are, however, already informally ongoing and could well end up in long-range plans of, {\it e.g.}, NuPECC. It would be too early to define the role of FAIR in such an endeavour, except that our commitment of providing FAIR-produced data following F.A.I.R. principles is already an important basis for and step towards supporting such vision.


\newpage
\section{Governance model}

\begin{landscape}
\begin{figure}[h]
\begin{center}
\includegraphics[width=1.7\textwidth]{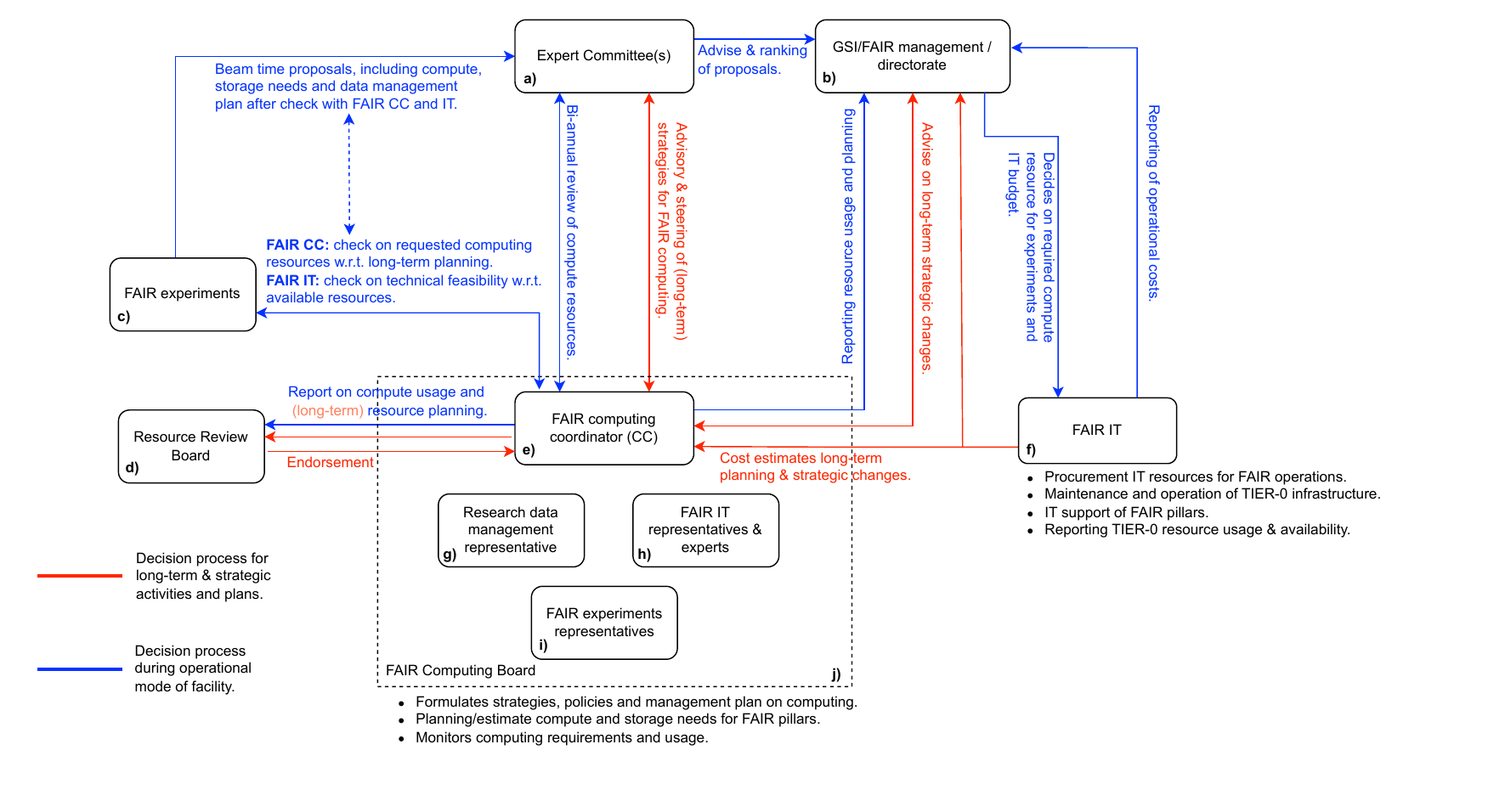}
\caption{A sketch of the decision processes of the usage and developments of the computing infrastructure of FAIR. The roles of the various advisory boards, departments, coordinators, and management board are indicated by boxes. The communication/decision lines are presented by arrows. The decision progresses are subdivided in long-term \& strategic activities and plans (red lines) and for the operational mode of the facility (blue). Each box is alphabetically enumerated and referenced accordingly in the text}.
\label{fig:governance-decision-model}
\end{center}
\end{figure}
\end{landscape}

This section describes the advisory and decision-making processes for resource pledges, allocations, tenders, and strategic developments within the FAIR computing facility. As a guideline, we refer to Fig.~\ref{fig:governance-decision-model}, which illustrates the corresponding flow diagram. The figure depicts the various boards and departments (represented as boxes) and the decision and advisory lines for two different cases, distinguished by color. Each box is alphabetically enumerated and referenced in the descriptive text below.

\subsection{Decision-Making Authority}
The FAIR directorate (b) is the ultimate decision-making body.
Decisions are based on multiple factors, including recommendations from advisory boards and expert committees (a), such as GPAC and ECE, as well as internal boards like the FAIR Computing Board (j).
The evaluation process includes consideration of operational costs and strategic changes, such as unforeseen experimental needs.

\subsection{Decision Processes}
Two types of decision-making processes are identified:

\begin{itemize}
    \item {\bf Operational Model Decision Process (Blue Lines):} This occurs on a timescale of a few years and governs regular operations.
    \item {\bf Strategic Planning Decision Process (Red Lines):}
This applies to long-term developments, such as major facility upgrades.
Examples include introducing new large-scale experimental facilities that were not previously accounted for in planning documents, significantly impacting computational resource needs.

\end{itemize}

\subsection{Role of the FAIR Computing Board}
The FAIR Computing Board (j) plays a central role in:
\begin{itemize}
    \item Strategy formulation
    \item Computing and storage planning
    \item Monitoring resource requirements and usage
\end{itemize}

\noindent The board is chaired by the FAIR Computing Coordinator (CC) (e), appointed by the FAIR management (b).
Members include:
\begin{itemize}
    \item Representatives from FAIR pillars (i) (e.g., computing coordinators for experiments)
    \item Research Data Management representative (g)
    \item FAIR IT Department representative \& experts (h)
\end{itemize}

\noindent The CC organizes board meetings and reports to the directorate (b), expert committees (a), and the Resource Review Board (d).
The Resource Review Board, represented by funding agencies, need to endorse the compute resource planning. The composition and responsibilities of the Resource Review Board can be found at~\cite{rrb-fair}. Reports of advisory boards are forwarded to the Resource Review Board as independent input.

\subsection{Responsibilities of the FAIR IT Department}
The FAIR IT Department (f) oversees:
\begin{itemize}
    \item Development, maintenance, and operation of the TIER-0 IT infrastructure
    \item IT support for FAIR pillars, as outlined in the CDR (Sec.~\ref{sec:FAIR_IT_role})
    \item Monitoring the usage of external computing resources within a distributed computing model
\end{itemize}

\noindent In strategic planning cases, the IT department (f) provides cost estimates to the CC (e).

\subsection{Beam-Time Proposals and Resource Allocation}
FAIR experiments (c) must include compute and storage requirements, along with a data management plan.
Before submission to expert committees (a), such as GPAC, proposals must be:

\begin{itemize}
    \item Verified by the CC (e) for consistency with long-term planning
    \item Assessed by the FAIR IT Department (f) for technical feasibility
\end{itemize}

\noindent Upon approval, expert committees (a) rank and advise the FAIR directorate (b) on proposals.

\subsection{Access and Cost Considerations}
The computing infrastructure is exclusively dedicated to FAIR pillars.
Access is granted solely to members of FAIR collaborations for scientific work related to FAIR.
For details on cost models, refer to the addendum document: Supplemental Information: Maintenance and Operation Costs.

\newpage
\section*{\textcolor{red}{List of Abbreviations}}
\addcontentsline{toc}{section}{List of Abbreviations}

\begin{longtable}[l]{ll} 
  \textbf{Abbreviation} & \textbf{Definition} \\
  \hline
    AAI     & Authentication, Authorization and Identity \\
    AGATA   & Advanced GAmma Tracking Array \\
    AI      & Artificial Intelligence \\
    ALICE   & A Large Ion Collider Experiment \\
    AMD     & Advanced Micro Devices \\
    AOD     & Analyis Object Data \\
    APPA    & Atomic, Plasma Physics and Applications \\
    APPEC   & Astroparticle Physics European Consortium \\
    ARTIFACT& ARTificial Intelligence For Accelerators, \\
     & user Communities and associated Technologies \\
    ASIC    & Application-Specific Integrated Circuit \\
    BMBF    & Bundesministerium f\"ur Bildung und Forschung \\
    BUU     & Boltzmann-Uehling-Uhlenbeck \\
    CA      & Cellular Automaton \\
    CBM     & Compressed Baryonic Matter \\
    CDR     & Conceptual Design Report \\
    CERN    & Conseil Européen pour la Recherche Nucléaire \\
    CISC    & Complex Instruction Set Computer \\
    CPU     & Central Processing Unit \\
    CR      & Collector Ring \\
    CRYRING & Cryogenic ring accelerator \\
    DAQ     & Data Acquisition \\
    DCS     & Distributed Control System \\
    DESPEC  & Detector for Advanced Structure and Decay Properties of Exotic Nuclei \\
    DOI     & Digital Object Identifiers \\
    DSSSD   & Double-Sided Silicon Strip Detector \\
    DST     & Data Summary Tape \\
    EC      & Experimental Collaboration \\
    ECFA    & European Committee for Future Accelerators \\
    ELiSe   & European Laser Initiative for Science and Engineering \\
    EOSC    & European Open Science Cloud \\
    ErUM    & Erforschung von Universum und Materie \\
    ES      & Early Science \\
    ESA     & European Space Agency \\
    ESAP    & ESFRI Science Analysis Platform \\
    ESD     & Event Summary Data \\
    ESFRI   & European Strategy Forum on Research Infrastructures \\
    ESR     & Energy Storage Ring \\
    ESCAPE  & European Science Cluster of Astronomy \& Particle physics ESFRI \\
    EuCAIF  & European Coalition for AI in Fundamental Physics \\
    EUROLABS & EUROpean Laboratories for Accelerator Based Science \\
    EXL     & Exotic nuclei studied with Electromagnetic and Light hadronic probes\\
    FAIR    & Facility for Antiproton and Ion Research \\
    F.A.I.R.& Findable Accessible Interoperable Reusable \\
    FLES    & First Level Event Selection \\
    FPGA    & Field Programmable Gate Arrays \\
    FS(+)   & First Science (+) \\
    GEANT   & GEometry ANd Tracking \\
    GPU     & Graphics Processing Unit \\
    GPGPU   & General Purpose GPU \\
    FLES    & First Level Event Selection \\
    FRS     & FRagment Seperator \\
    FTF     & FRITIOF \\
    GPT     & Generative Pre-trained Transformer \\
    GSI     & Gesellschaft f\"ur Schwerionenforschung \\
    HADES   & High Acceptance DiElectron Spectrometer \\
    HEB     & High Energy Beamline \\
    HELPMI  & Helmholtz Laser-Plasma Metadata Initiative \\
    HEP     & High Energy Physics \\
    HESR    & High Energy Storage Ring\\
    HISPEC  & High-Resolution Inelastic Scattering and Particle Emission Chamber \\
    HITRAP  & Heavy Ion Trap \\
    HPC     & High Performance Computing \\
    HTC     & High Throughput Computing \\
    ILIMA   & Ionic Laser Manipulation of Atoms\\
    I/O     & Input/Output \\
    IT      & Information Technology \\
    JENA    & Joint ECFA NuPECC APPEC\\
    JSON    & JavaScript Object Notation \\
    KF      & Kalman Filter \\
    LDAP    & Lightweight Directory Access Protocol \\
    LHC     & Large Hadron Collider \\
    LLM     & Large Language Model \\
    MC      & Monte Carlo \\
    ML      & Machine Learning \\
    MODE    & Machine-learning Optimized Design of Experiments \\
    MPI     & Message Passing Interface \\
    MSVc    & Modularized Start Version complete \\
    MVD     & Micro Vertex Detector \\
    MUCH    & Muon Detection System \\
    NAPMIX  & Nuclear, Astro, and Particle Metadata Integration for eXperiments \\
    NFDI    & Nationale Forschungsdateninfrastruktur \\
    NuPECC  & Nuclear Physics European Collaboration Committee \\
    NUSTAR  & Nuclear Structure, Astrophysics and Reactions \\
    OpenMP  & Open Message Passing \\
    ORCID   & Open Researcher and Contributor ID \\
    OSCARS  & Open Science Cluster's Action for Research and Society \\
    OSSR    & Open Source Software Repository \\
    PANDA   & antiProton Annihilations at Darmstadt\\
    PIC     & Programmable Interrupt Controller \\
    PID     & Particle IDentification/Persistent IDentifiers  \\
    PSD     & Projectile Spectator Detector \\
    PUE     & Power Usage Effectivenes \\
    PUNCH4NFDI & Particles, Universe, NuClei and Hadrons for NFDI \\
    QCD     & Quantum Chromo Dynamics \\
    QMD     & Quantum Molecular Dynamics \\
    R\&D    & Research \& Development \\
    REANA   & Reusable Analyses \\
    RDA     & Research Data Alliance \\
    RDMO    & Research Data Management Organiser \\
    RICH    & Ring Imaging Cherenkov Detector \\
    RZ      & Rechenzentrum\\
    R3B     & Reactions with Relativistic Radioactive Beams \\
    SIMD    & Single Instruction, Multiple Data \\
    SIS18/100 & Superconducting Ion Synchrotron 18/100\\
    STS     & Silicon Tracking System \\
    TDR     & Technical Design Report \\
    MVA     & Multi-Variate Analysis\\
    TOF     & Time-of-Flight \\
    TRD     & Transition Radiation Detector \\
    UrQMD   & Ultrafast Quantum Molecular Dynamics \\
    VAE     & Virtual Application Environment \\
    WASA    & Wide Angle Shower Apparatus \\
    WLCG    & Worldwide LHC Computing Grid \\
  \hline
\end{longtable}


\begin{thebibliography}{99}

\bibitem{greencube:hpc}
 https://de.wikipedia.org/wiki/Green\_IT\_Cube

\bibitem{gsi:hpc}
https://hpc.gsi.de/virgo/preface.html.

\bibitem{ref:flesnet}
P. Gasik, W. M\"uller, M. Guminski (CBM Collaboration), ``Technical Design Report for the CBM Online Systems - Part I DAQ and FLES Entry Stage'', https://repository.gsi.de/record/340597/files/CBM\%20TDR\%20Online\%20Systems\%20-\%20Part\%20I.pdf.

\bibitem{Albrecht:2019}
J.~Albrecht et al. (HEP Software Foundation),
``A Roadmap for HEP Software and Computing R\&D for the 2020s'',
Computing and Software for Big Science {\bf 3} (2019), https://doi.org/10.1007/s41781-018-0018-8.

\bibitem{jena}
https://agenda.infn.it/event/34738/

\bibitem{jena-wp}
https://nupecc.org/jenaa/index.php?display=computing

\bibitem{jenas2025}
https://indico.global/event/5574/

\bibitem{wp_hpc}
https://nupecc.org/jenaa/docs/report\_wg1.pdf

\bibitem{eosc}
https://eosc.eu

\bibitem{nfdi}
https://www.nfdi.de

\bibitem{eosc-eu-node}
https://open-science-cloud.ec.europa.eu

\bibitem{oscars}
https://eosc.eu/eu-project/oscars/

\bibitem{eurolabs}
https://web.infn.it/EURO-LABS/

\bibitem{rda}
https://www.rd-alliance.org

\bibitem{gsi:rdm}
GSI Helmholtzzentrum f\"{u}r Schwerionenforschung GmbH, Facility for Antiproton and Ion Research GmbH, ``GSI/FAIR Research Data Management (RDM) Policy'' (2023), http://dx.doi.org/10.15120/GSI-2023-00646.

\bibitem{Wilkinson:2016}
Wilkinson, M., Dumontier, M., Aalbersberg, I. et al., ``The FAIR Guiding Principles for scientific data management and stewardship.'' Sci Data {\bf3}, 160018 (2016), https://doi.org/10.1038/sdata.2016.18.

\bibitem{escape:ossr}
https://escape-ossr.gitlab.io/ossr-pages/

\bibitem{escape:ossr_zenodo}
https://zenodo.org/communities/escape2020

\bibitem{cern:rucio}
https://rucio.cern.ch/

\bibitem{project:escape}
https://projectescape.eu/

\bibitem{project:eurolabs}
https://doi.org/10.3030/101057511

\bibitem{project:punch4nfdi}
https://www.punch4nfdi.de/

\bibitem{escape}
https://projectescape.eu/escape-and-eosc-future

\bibitem{indigo}
https://indigo-iam.github.io/v/current/docs/

\bibitem{ndaq:tdr}
  https://fair-center.eu/user/experiments/nustar/documents/technical-design-reports 

\bibitem{ref:agata}
https://www.agata.org/campaign\_gsi\_2012-2014.

\bibitem{ref:wasa}
T.R. Saito et al., ``The WASA-FRS project at GSI and its perspective'', Nucl. Instr. and Meth. in Phys. Res. B {\bf 542}, 22 (2023), https://doi.org/10.1016/j.nimb.2023.05.042; Y.K. Tanaka et al., Acta Phys. Polon. Supp. {\bf 16}, 4 (2023), https://doi.org/10.5506/APhysPolBSupp.16.4-A27.

\bibitem{ref:alfa}
M. Al-Turany et al., ``ALFA: The new ALICE-FAIR software framework'', J. Phys.: Conf. Ser. {\bf 664} 072001 (2015), https://doi.org/10.1088/1742-6596/664/7/072001.

\bibitem{ref:o2}
G. Eulisse et al., ``Evolution of the ALICE Software Framework for Run 3'', EPJ Web of Conferences 214, 05010 (2019), https://doi.org/10.1051/epjconf/201921405010. 

\bibitem{erum-data}
https://www.erum-data.de

\bibitem{micro:trend}
 https://github.com/karlrupp/microprocessor-trend-data

\bibitem{alice:gpu}
David Rohr et al. (ALICE Collaboration),
``Usage of GPUs in ALICE Online and Offline processing during LHC Run 3'',
EPJ Web of Conferences {\bf 251}, 04026 (2021),
https://doi.org/10.1051/epjconf/202125104026.

\bibitem{lhcb:gpu}
https://cds.cern.ch/record/2801723/files/LHCb-PROC-2022-002.pdf

\bibitem{kisel:2020}
I. Kisel, ``Real-time Event Reconstruction and Analysis in CBM and STAR experiments'',
Journal of Physics Conference Series 1602, 012006 (2020),
https://doi.org/10.1088/1742-6596/1602/1/012006.

\bibitem{jiang:2023}
P. Jiang et al.,
``Deep machine learning for the PANDA software trigger'',
Eur. Phys. J. C {\bf 83}, 337 (2023), 
https://doi.org/10.1140/epjc/s10052-023-11494-y.

\bibitem{alicke:2021}
A.~Alicke, T.~Stockmanns, J.~Ritman,
EPJ Web of Conferences 251, 04002 (2021),
https://doi.org/10.1051/epjconf/202125104002.

\bibitem{gazagnes:2023}
S.~Gazagnes et al.,
``Reconstructing charged-particle trajectories in the PANDA Straw Tube Tracker using the LOcal Track Finder (LOTF) algorithm'',
Eur. Phys. J. A {\bf 59}, 100 (2023),
https://doi.org/10.1140/epja/s10050-023-01005-8.

\bibitem{heine:2022}
G.~Heine, https://publish.etp.kit.edu/record/22136.

 \bibitem{amd:cpu_gpu}
 https://www.amd.com/en/newsroom/press-releases/2023-6-13-amd-expands-leadership-data-center-portfolio-with-.html

\bibitem{nvidia:cpu_gpu}
 https://developer.nvidia.com/blog/nvidia-grace-hopper-superchip-architecture-in-depth/

\bibitem{chiplets}
https://www.uciexpress.org/

\bibitem{fairroot}
https://fairroot.gsi.de

\bibitem{alpaka}
https://iopscience.iop.org/article/10.1088/1742-6596/2438/1/012058

\bibitem{kokos}
https://doi.org/10.1051/epjconf/202125103034

\bibitem{oneapi}
https://www.oneapi.io/

\bibitem{Boehnlein:2022}
A.~Boehnlein et al., ``Machine Learning in Nuclear Physics'', Rev. Mod. Phys. {\bf 94}, 031003 (2022), https://doi.org/10.1103/RevModPhys.94.031003.

\bibitem{pytorch}
https://pytorch.org

\bibitem{tensorflow}
https://www.tensorflow.org

\bibitem{deeplearning4j}
https://deeplearning4j.konduit.ai

\bibitem{scikit-learn}
https://scikit-learn.org

\bibitem{TMVA2007} 
        A.~Hoecker, P.~Speckmayer, J.~Stelzer, 
        J.~Therhaag, E.~von Toerne, and H.~Voss,
        ``TMVA: Toolkit for Multivariate Data Analysis,''
        PoS A CAT 040 (2007) [physics/0703039].

\bibitem{erumdatahub}
https://erumdatahub.de/en/

\bibitem{mode}
Atılım Güneş Baydin et al.,
``Toward Machine Learning Optimization of Experimental Design'',
Nuclear Physics News {\bf 31}, 25 (2021),
https://doi.org/10.1080/10619127.2021.1881364.

\bibitem{hashemi:2019}
B. Hashemi et al.,
``LHC analysis-specific datasets with Generative Adversarial Networks'',
arXiv:1901.05282,
https://doi.org/10.48550/arXiv.1901.05282.

\bibitem{khattak:2022}
Gul Rukh Khattak et al.,
``Fast simulation of a high granularity calorimeter by generative adversarial networks'', 
Eur. Phys. J. C {\bf 82}, 386 (2022), 
https://doi.org/10.1140/epjc/s10052-022-10258-4.

\bibitem{jeske:2022}
Torri Jeske et al.,
``AI for Experimental Controls at Jefferson Lab'',
arXiv:2203.05999,
https://doi.org/10.1088/1748-0221/17/03/C03043.

\bibitem{caron:2023}
Elena~Cuoco et al.,
``Expression of Interest for a synergic research plan of potential interest of the JENA group'',
Proposal submitted to JENA computing seminar (https://agenda.infn.it/event/34738/),
private communications Sascha Caron (2023)

\bibitem{hessian.ai}
https://hessian.ai

\bibitem{eucaif}
https://www.eucaif.org

\bibitem{eucaifconf}
https://indico.nikhef.nl/event/4875/

\bibitem{artifact}
https://aghribi1-presentation-20230911-artifact.docs.cern.ch/pres\_artifact\_20231215.html\#/title-slide

\bibitem{aiworkshop-gsi}
https://indico.gsi.de/event/20517/

\bibitem{aarc-blueprint:2019}
AARC Community Members and AppInt Members, 
“AARC Blueprint Architecture 2019 (AARC-G045),” 
Nov. 2019, 
https://doi.org/10.5281/ZENODO.3672784.

\bibitem{refeds}
"The Research and Education Federations (REFEDS)", Retrieved from https://refeds.org

\bibitem{sirtifi:2022}
"Sirtfi: Security Incident Response Trust Framework for Federated Identity.", Retrieved from https://refeds.org/sirtfi, 2022

\bibitem{apptainer:2021}
Apptainer Project,
"Apptainer: A container system for high-performance computing.",
2021
Retrieved from https://apptainer.org

\bibitem{rrb-fair}
https://fair-center.eu/about/organisation/rrb

\end{thebibliography}
\end{document}